\documentclass[preprint,aps]{revtex4}
\usepackage{graphicx,epsf}
\usepackage{lscape}
\usepackage{citesort}

\begin{document}

\title{ {\Large {\bf  Analysis of the semileptonic $B_c\to D_1^0$ transition in QCD sum rules and HQET  } } }

\author{\small R. Khosravi \footnote {e-mail: rezakhosravi@
cc.iut.ac.ir}}
\affiliation{Department of Physics, Isfahan University of
Technology, Isfahan 84156-83111, Iran}

\begin{abstract}
We investigate the structure of the $D_{1}^0(2420[2430])
(J^P=1^+)$ mesons via analyzing the semileptonic $B_{c}\to D_{1}^0
l\nu$ transition  in the frame work of the three--point QCD sum
rules and the heavy quark effective theory. We consider the
$D_{1}^0$ meson in three ways; the pure $|c\bar{u}\rangle$ state,
a mixture of two $|^3P_1\rangle$ and $|^1P_1\rangle$ states with a
mixing angle $\theta$ and a combination of two mentioned states
with mixing angle $\theta=35.3^\circ$ in the heavy quark limit.
Taking into account the gluon condensate contributions, the
relevant form factors are obtained for the three above conditions.
These form factors are numerically calculated for
$|c\bar{u}\rangle$ and the heavy quark limit  cases. The obtained
results for the form factors are used to evaluate the decay rates
and the branching ratios. Also for mixed states, all of the
mentioned physical quantities are plotted with respect to  the
unknown mixing angle $\theta$.
\end{abstract}


\maketitle

\section{Introduction}
There is some difference between the measured and predicted masses
of the even--parity charmed mesons $(J^P=1^+)$, observed in the
laboratories \cite{BaBar1,CLEO1,Belle1,Belle2,Belle3} and considered
in many phenomenological models
\cite{Kolomeitsev,Hofmann,Cheng1,Kim,Szczepaniak,Barnes}. So much
efforts have been dedicated to realize this unexpected disparity
between theory and experiment
\cite{Ebert,Jovanovic,Dmitra,Bracco,Beveren,Vijande}. Therefore the
study of the processes involving these mesons is important for
understanding of the structure and quark content of them. Some
physicists presumed that these discovered states are conventional
$c\bar{u}$ and $c\bar{s}$ mesons
\cite{Godfrey1,Swanson,Rosner,Bardeen,Nowak,Godfrey2,Close,Pierro,Godfrey3}.
Among  these mesons, we focus on the non--strange $D_1^0$ meson.
Sofar the two confirmed $D^0_1$ states, with mass of $2423.4\pm
3.2~MeV$ and $2427\pm 26\pm 25~MeV$, have been observed
\cite{Belle3}. The narrow--width state with lower mass is known as
$D_1^0(2420)$ and the wide--width state with more mass is identified
as $D_1^0(2430)$ \cite{PDG}. Theoretically, the discovered states do
not fit easily into the $c\bar{u}$ spectroscopy \cite{Close}. One of
the proposals, could be the introduction of the $D_1^0$ meson as a
mixture of two $|^1P_1\rangle$ and $|^3P_1\rangle$ states with the
$c\bar{u}$ quark content
\cite{Thomas,Close,Cheng2,Pierro,Godfrey1,Godfrey2}. In this work,
we plan to analyze $D_1^0$ meson as conventional meson with pure
$|c\bar{u}\rangle$ state and also as a combination of
$|^1P_1\rangle$ and $|^3P_1\rangle$ states.

Heavy--light mesons are not charge conjugation eigenstates and so
mixing can occur among states with the same $J^P$ and different
mass that are forbidden for neutral states \cite{Close}. So the
mixing of the physical $D_{1}$ and $D'_{1}$ states can be
parameterized in terms of a mixing angle $\theta$, as follow:
\begin{equation}\label{eq1}
\left[
\begin{array}{c}
|D_1\rangle \\
|D'_1\rangle
\end{array}
\right]=\left[
\begin{array}{cc}
cos\theta & sin\theta\\
-sin\theta & cos\theta
\end{array}
\right]\times\left[
\begin{array}{c}
|^1P_1\rangle\\
|^3P_1\rangle
\end{array}
\right],
\end{equation}
where we used the spectroscopic notation $^{2S+1}L_{J}$ for
introduction of the mixing states. Considering
$|^3P_1\rangle\equiv |D_{1}1\rangle$ and $|^1P_1\rangle\equiv
|D_{1}2\rangle$ with different masses and decay constants
\cite{Close,Thomas}, we can apply these relations, beyond the
heavy quark model, for axial vectors $D_{1}(2420)$ and
$D_{1}(2430)$ mesons with two different masses. i.e.,
\begin{eqnarray}\label{eq2}
|D_{1}(2420)\rangle &=& sin\theta~|D_{1}1\rangle ~+
cos\theta~|D_{1}2\rangle, \nonumber\\
|D_{1}(2430)\rangle &=& cos\theta~|D_{1}1\rangle ~-sin\theta~
|D_{1}2\rangle.
\end{eqnarray}
The masses and decay constant of the $D_{1}1$ and $D_{1}2$ states
are presented in Tables \ref{T1} and \ref{T2}.
\begin{table}[th]
\centering
\begin{tabular}{cccc}
\cline{1-4}Ref&\cite{Godfrey1}&\cite{Pierro}&\cite{Godfrey12}\\
\cline{1-4}\lower0.35cm \hbox{{\vrule width 0pt height 1.0cm }}
$D_{1}1(^3P_{1})$ &2.49 &2.42 &2.47 \\
\cline{1-4}\lower0.35cm \hbox{{\vrule width 0pt height 1.0cm }}
$D_{1}2(^1P_{1})$ &2.44 &2.49 &2.46 \\
\hline
\end{tabular}
\vspace{0.10cm} \caption{Masses of $D_11(^3P_1)$ and $D_12(^1P_1)$
states  in GeV.}\label{T1}
\end{table}

\begin{table}[th]
\centering
\begin{tabular}{cccc}
\cline{1-2}Ref&\cite{Thomas}\\
\cline{1-2}\lower0.35cm \hbox{{\vrule width 0pt height 1.0cm }}
$D_{1}1(^3P_{1})$ &183\\
\cline{1-2}\lower0.35cm \hbox{{\vrule width 0pt height 1.0cm }}
$D_{1}2(^1P_{1})$ &89\\
\hline
\end{tabular}
\vspace{0.10cm} \caption{Decay constant of $D_11(^3P_1)$ and
$D_12(^1P_1)$ states in MeV.}\label{T2}
\end{table}
In the heavy quark limit where the quark mass $m_{c}\to \infty$,
both axial vector $D_1^0(2420)$ and $D_{1}^0(2430)$ mesons can be
produced and identified with $|P_1^{3/2}\rangle$ and
$|P_1^{1/2}\rangle$, respectively. It is useful to change from the
$L-S$ basis $^{2S+1}L_J$ to the $j-j$ coupling basis $L^j_J$,
where $j$ is the total angular momentum of the light quark. The
relationship between these states are given as
\cite{Close,Thomas,Cheng2}:
\begin{equation}\label{eq3}
\left[
\begin{array}{c}
D_{1}^0(2420)\equiv|P_1^{3/2}\rangle \\
D_{1}^0(2430)\equiv|P_1^{1/2}\rangle
\end{array}
\right]=\left[
\begin{array}{cc}
\sqrt{\frac{2}{3}} & \sqrt{\frac{1}{3}}\\
-\sqrt{\frac{1}{3}}& \sqrt{\frac{2}{3}}
\end{array}
\right]\times\left[
\begin{array}{c}
|D_12\rangle\equiv|^1P_1\rangle\\
|D_11\rangle\equiv|^3P_1\rangle
\end{array}
\right].
\end{equation}
These relations  occur for the mixing angle $\theta=35.3^\circ$ in
Eq. (\ref{eq1}). But note that the value of the mixing angle can
be positive equal to $\theta=35.3^\circ$ or negative corresponding
$\theta=-54.7^\circ$ if the expectation of the heavy--quark
spin--orbit interaction is positive or negative, respectively
\cite{Close}.

The $B_c \to D^{*0} l \nu$ \cite{Ghahramany} and $B_c \to D ll/
\nu\bar{\nu}$ \cite{Azizi1} have been studied via three--point QCD
sum rules (3PSR). In this work, we analyze the semileptonic $B_c
\to D_1^0(2420[2430]) l \nu$ decays in 3PSR and heavy quark
effective theory (HQET).  For this aim, we consider the structure
of the $D_1^0$ meson in three conditions:

1- The $D_1^0$ meson as a pure state $(|c\bar{u}\rangle)$.

2- The $D_1^0$ meson as a mixture of two states of the
$|^1P_1\rangle$ and $|^3P_1\rangle$ with a mixing angle $\theta$
(see Eq. (\ref{eq2})).

3- The $D_1^0$ meson as a combination of two $|^1P_1\rangle$ and
$|^3P_1\rangle$ states with the mixing angle $\theta=35.3^\circ$ in
the heavy quark limit (see Eq. (\ref{eq3})).

Taking into account the gluon condensate corrections, as the
important term of the non--perturbative part of the correlation
function, the form factors of the $B_c \to D_1^0$ transition are
obtained within 3PSR for the condition 1, 2 and within HQET
approach for the condition 3. For the condition 1 and 3, the form
factors of the $B_c\to D_1^0(2420[2430])$ transitions are a
function of the transferred momentum square $q^2$. So, we plot
these form factors and decay widthes  of these decays with respect
to $q^2$. Also the branching ratios for these cases are evaluated.
But, it should be remarked, when we consider the $D_1^0$ as a
mixture of two states with mixing angle $\theta$ in the region
$-180\leq\theta\leq180$, the transition form factors of the
$B_c\to D_1^0(2420[2430])$ decays are a function of two variables
$\theta$ and $q^2$. Since the decay width of the $B_c\to D_1^0$
transition is related to the form factors, then it is a function
of mixing angle $\theta$ and $q^2$, too. for a better analysis, we
plot the form factors and the decay widths of the $B_c\to
D_1^0(2420[2430])$ in three dimensions.  In this case, the
branching ratios are shown with respect to the mixing angle
$\theta$. Detection of these channels and their comparison with
the phenomenological models like QCD sum rules could give useful
information about the structure of the $D_{1}^0$ meson and the
unknown mixing angle $\theta$.

This paper is organized as follow. In Section II, we calculate the
form factors for the $B_c\to D_{1}^0$ transition in 3PSR for above
condition 1 and 2.  In Section III, the transition form factors
are evaluated via HQET approach for condition 3.  Finally, Section
IV is devoted to the numeric results and discussions.

\section{Sum rules method }
In this section, we study the transition form factors of the
semileptonic $B_c\to D_1^0 l \nu$ decay by QCD sum rules
mechanism. For this aim, first, we consider the $D_1^0$ meson as a
pure state. The $B_c\to D_1^0 l \nu$ process is governed by the
tree level $b\to u l \nu$ transition and $c$ quark is the
spectator, at quark level (see Fig. 1).
\begin{figure}[th]
\begin{center}
\begin{picture}(160,55)
\centerline{ \epsfxsize=7cm \epsfbox{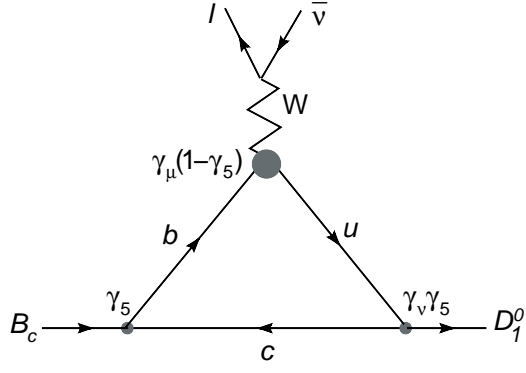}}
\end{picture}
\end{center}
\vspace*{-1cm} \caption{The bare loop diagram for $B_c \rightarrow
D_1^0 l \nu$ transition.}
\end{figure}
The three--point correlation function is considered for the
evaluation of the transition form factors in the framework of the
3PSR. The three--point correlation function is constructed from
the vacuum expectation value of time ordered product of three
currents as follow:
\begin{eqnarray}\label{eq4}
\Pi_{\mu\nu} (p^2,p'^2,q^2) = i^2 \int d^4 x d^4 y ~e^{+i p^\prime
x - i p y} \left\langle 0 \left| \mbox{\rm T} \left\{ J^{D_1^0
}_\nu(x) J^{W}_\mu (0)  {J^{B_c}}^{\dagger}(y)\right\} \right|
0\right\rangle~,
\end{eqnarray}
where $J^{D_1^0}_\nu(x)=\overline{c}\gamma_{\nu}\gamma_{5}u$ and
$J^{B_{c}}(y)=\overline{c}\gamma_{5}b$ are the interpolating
currents of the $ D_1^0 $ and $B_{c} $ mesons.
$J_{\mu}^{W}=\overline{u}\gamma_{\mu}(1-\gamma_{5})b$ is the
current of the weak transition.

\begin{figure}[th]
\vspace*{0.0cm}
\begin{center}
\begin{picture}(160,100)
\centerline{ \epsfxsize=14cm \epsfbox{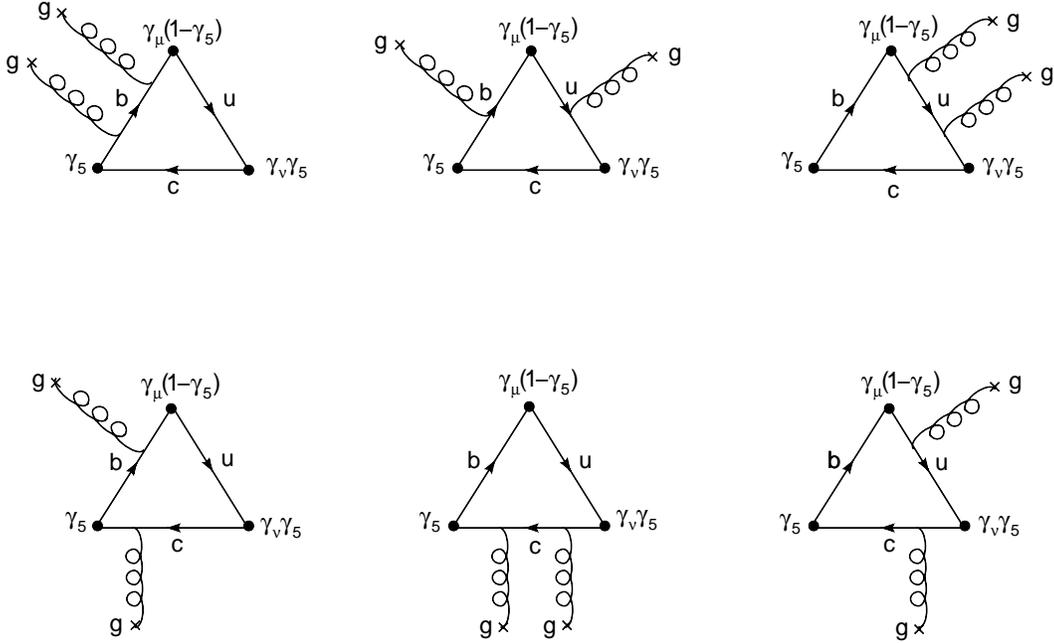}}
\end{picture}
\end{center}
\vspace*{-0.1cm} \caption{Contribution of two gluon condensates
for $B_c \to D^0_1$ transition.} \label{F2}
\end{figure}

We can obtain the correlation function of Eq. (\ref{eq4}) in two
sides. The phenomenological or physical part is calculated
saturating the correlation  by a tower of hadrons with the same
quantum numbers as interpolating currents. The QCD or theoretical
part, on the other side is obtained in terms of the quarks and
gluons interacting in the QCD vacuum. To derive the
phenomenological part of the correlation given in Eq. (\ref{eq4}),
two complete sets of intermediate states with the same quantum
numbers as the currents $J_{D^0_1}$ and $J_{B_{c}}$ are inserted.
This procedure leads to  the following representations of the
above-mentioned correlation:
\begin{eqnarray} \label{eq5}
\Pi _{\mu\nu}(p^2,p'^2,q^2)&=&\frac{\langle0\mid J_{\nu}^{ D_1^0}
\mid D_1^0(p',{\varepsilon})\rangle\langle
D_1^0(p',{\varepsilon})\mid J^{W}_{\mu}\mid B_{c}(p)\rangle\langle
B_{c}(p)\mid {J^{B_c}}^{\dag}\mid
0\rangle}{(p'^2-m_{D_1^0}^2)(p^2-m_{B_c}^2)}
\nonumber\\&+&\mbox{higher resonances and continuum states}~.
\end{eqnarray}
The general expression for the hadronic matrix element of the weak
current with definition of the transition form factors is given by
the formula:
\begin{eqnarray}  \label{eq6}
\langle
D_1^0(p^{\prime},\varepsilon)\mid\overline{u}\gamma_{\mu}(1-\gamma_5)
b\mid B_c(p)\rangle
&=&f^{'}_{V}(q^2)\varepsilon_{\mu\nu\alpha\beta}
\varepsilon^{\ast\nu}p^\alpha
{p^{\prime}}^\beta-i\left[f^{'}_{0}(q^2)\varepsilon_{\mu}^{\ast}
\right.
\nonumber \\
+ f^{'}_{1}(q^2)(\varepsilon^{\ast}p)P_{\mu} &+&
\left. f^{'}_{2}(q^2)(\varepsilon^{\ast}p)q_{\mu}%
\right],
\end{eqnarray}
where:
\begin{eqnarray}  \label{eq7}
f_{V}^{\prime}(q^2)&=&\frac{2f_{V}(q^2)}{(m_{B_c}+m_{D_1^0})}%
~,~~~~~~~~~~~~f_{0}^{\prime}(q^2)=f_{0}(q^2)(m_{B_c} +m_{D_1^0}),
\nonumber
\\
f_{1}^{\prime}(q^2)&=&-\frac{f_{1}(q^2)}{(m_{B_c}+m_{D_1^0})}~,~~~~~~~~~~~~
f_{2}^{\prime}(q^2)=-\frac{f_{2}(q^2)}{(m_{B_c}+m_{D_1^0})},
\end{eqnarray}
and the $f_{V}(q^2)$, $f_{0}(q^2)$, $f_{1}(q^2)$ and $f_{2}(q^2)$
are the transition form factors,  $P_{\mu}=(p+p^{\prime})_{\mu}$, $%
q_{\mu}=(p-p^{\prime})_{\mu}$ and $\varepsilon$ is the
four--polarization vector of the $D_1^0$ meson. Also the following
matrix elements are defined in the standard way in terms of the
leptonic decay constants of the $D_1^0$ and $B_c$ mesons as:
\begin{equation}\label{eq8}
\langle0\mid J^{\nu}_{D_1^0} \mid
D_1^0(p',{\varepsilon})\rangle=f_{D_1^0}m_{D_1^0}\varepsilon^{\nu}~,~~\langle0\mid
J_{B_c}\mid
B_{c}(p)\rangle=i\frac{f_{B_{c}}m_{B_{c}}^2}{m_{b}+m_{c}},
\end{equation}
where $f_{D_1^0}$ and $f_{B_c}$ are the leptonic decay constants
of $D_1^0$ and $B_c$ mesons, respectively. Using Eq. (\ref{eq6})
and Eq. (\ref{eq8}) in Eq. (\ref{eq5}) and performing summation
over the polarization of the  $D_1^0$ meson, we get the following
result for the physical part:
\begin{eqnarray}\label{eq9}
\Pi_{\mu\nu}(p^2,p'^2,q^2)&=&-\frac{f_{B_c}m_{B_c}^2}{(m_{b}+m_{c})}\frac{f_{D_1^0}m_{D_1^0}}
{(p'^2-m_{D_1^0}^2)(p^2-m_{B_c}^2)} \times
\left[if'_V(q^2)\varepsilon_{\mu\nu\alpha\beta}p^{\alpha}p'^{\beta}+f'_{0}(q^2)g_{\mu\nu}
\right. \nonumber \\
&+&\left. f'_{1}(q^2)P_{\mu}p_{\nu} +
f'_{2}(q^2)q_{\mu}p_{\nu}\right]+ \mbox{excited states.}
\end{eqnarray}
The coefficients of Lorentz structures
$i\epsilon_{\mu\nu\alpha\beta}p^{\alpha}p^{'\beta}$, $g_{\mu\nu}$,
$P_{\mu}p_{\nu}$ and $q_{\mu}p_{\nu}$ in the correlation function
$\Pi_{\mu\nu}$ will be chosen  in determination of the form
factors $f_{V}(q^2)$,~$f_{0}(q^2)$,~$f_{1}(q^2)$ and $f_{2}(q^2)$,
respectively. So the Lorentz structures in the correlation
function can be written down as:
\begin{eqnarray}\label{eq10}
\Pi_{\mu\nu}(p^2,p'^2,q^2)&=&i~\Pi_{V}\varepsilon_{\mu\nu\alpha\beta}p^{\alpha}p'^{\beta}
+\Pi_{0}g_{\mu\nu}+\Pi_{1}P_{\mu}p_{\nu}+\Pi_{2}q_{\mu}p_{\nu},
\end{eqnarray}
where, each $\Pi_{i}$ function is defined in terms of the
perturbative and non-perturbative parts as:
\begin{eqnarray} \label{eq11}
\Pi_{i}(p^2,p'^2,q^2) = \Pi_{i}^{per}(p^2,p'^2,q^2)
+\Pi_{i}^{nonper}(p^2,p'^2,q^2)~.
\end{eqnarray}
With the help of the operator product expansion (OPE), in the deep
Euclidean region where $p^2\ll (m_b+m_c)^2$ and $p'^2\ll m_c^2$,
the vacuum expectation value of the expansion of the correlation
function in terms of the local operators, is written as follow
\cite{Ghahramany,Aliev1}:
\begin{eqnarray}\label{eq12}
\Pi_{\mu\nu} (p^2,p'^2,q^2) &=& (C_0)_{\mu\nu} + (C_3)_{\mu\nu}
\langle \bar{q} q \rangle + (C_4)_{\mu\nu} \langle G^2 \rangle+
(C_5)_{\mu\nu} \langle \bar{q}
\sigma_{\alpha\beta} G^{\alpha\beta} q \rangle \nonumber \\
&+& (C_6)_{\mu\nu} \langle \bar{q} \Gamma q \bar{q} \Gamma^\prime
q \rangle~,
\end{eqnarray}
where $(C_i)_{\mu\nu}$ are the Wilson coefficients,
$G_{\alpha\beta}$ is the gluon field strength tensor, $\Gamma $
and $\Gamma^{'}$ are the matrices appearing in the calculations.
The non-perturbative part contains the quark and gluon condensate
diagrams. We consider the condensate terms of dimension $3, 4$ and
$ 5$.  It's found that the heavy quark condensate contributions
are suppressed by inverse of the heavy quark mass and can be
safely omitted. The light $u$ quark condensate contribution is
zero after applying the double Borel transformation with respect
to the both variables $p^2$ and $p'^2$, because only one variable
appears in the denominator. Therefore in this case, we consider
the two gluon condensate diagrams with mass dimension $4$ as a
important term of the non-perturbative corrections, only i.e.,
\begin{eqnarray}\label{eq13}
\Pi_i(p^2,p'^2,q^2) = \Pi_i^{per}(p^2,p'^2,q^2) + \Pi_i^{\langle
G^2 \rangle} (p^2,p'^2,q^2)  \langle \frac{\alpha_s}{\pi} G^2
\rangle~.
\end{eqnarray}
The diagrams for contribution of the gluon condensates are
depicted in Fig. \ref{F2}.
\normalsize To obtain the contributions of these diagrams ,the
Fock-Schwinger fixed-point gauge, $x^\mu A_\mu^a=0$, are used;
where $A_\mu^a$ is the gluon field. The procedure of the
evaluation of such as diagrams in Fig. \ref{F2} has been discussed
in Ref. \cite{Ghahramany}, completely.

Using the double dispersion representation, the bare-loop
contribution is determined:
\begin{eqnarray}\label{eq14}
\Pi_i^{per} = - \frac{1}{(2 \pi)^2} \int \int\frac{\rho_i^{per}
(s,s^\prime, q^2)}{(s-p^2) (s^\prime - p^{\prime 2})} ds ds^\prime
+ \mbox{\rm subtraction terms}~.
\end{eqnarray}
By replacing the propagators with the Dirac-delta
functions (Cutkosky rules):
\begin{eqnarray}\label{eq15}
\frac{1}{k^2-m^2} \rightarrow -2i\pi \delta(k^2-m^2)~,
\end{eqnarray}
the spectral densities $\rho_i^{per}(s,s^{'},q^2)$ are found as:
\begin{eqnarray}\label{eq16}
\rho_V^{per} &=&\frac{
N_c}{\lambda^{1/2}(s,s',q^2)}~\left\{\frac{(2s'\Delta-\Delta'u)(m_b-m_c)}{\lambda(s,s',q^2)}
-\frac{(2s'\Delta-\Delta'u)(m
_{u}+m_c)}{\lambda(s,s',q^2)}-m_c\right\} ~,\nonumber \\ \nonumber\\
\rho_0^{per} &=&\frac{-
N_c}{2\lambda^{1/2}(s,s',q^2)}~\left\{\Delta (m_c+m_{u})-\Delta
^{\prime }(m_b-m_c)+2 m_c^2(m_b-m_c-m_{u}) \right.\nonumber
\\&&\left.+\frac{2(4ss'm_c^2-s\Delta'^2-s'\Delta^2-u^2m_c^2+u\Delta\Delta')(m_b-m_c)}{\lambda(s,s',q^2)}
+m_c(2m_b m_{u}-u)%
\right\}~,
\nonumber \\ \nonumber \\
\rho_1^{per}&=&\frac{N_c}{2\lambda^{1/2}(s,s',q^2)}\left\{\frac{(2s'\Delta-\Delta'u)(m_b-
3m_c)}{\lambda(s,s',q^2)}-\frac{(2s\Delta'-\Delta
u)(m_c+m_{u})}{\lambda(s,s',q^2)}
\right. \nonumber \\
&&\left.
-\frac{2(8ss'^2m_c^2-2ss'\Delta'^2-6s'^2\Delta^2-2u^2s'm_c^2+6s'u\Delta\Delta'
-u^2\Delta'^2)(m_b-m_c)}{\lambda^2(s,s',q^2)}
\right. \nonumber \\
&&\left.
+\frac{2(4ss'um_c^2+4ss'\Delta\Delta'-3su\Delta'^2-3u\Delta^2s'
-u^3m_c^2+2u^2\Delta\Delta')(m_b-m_c)}{\lambda^{2}(s,s',q^2)}-m_c\right\}~,
\nonumber \\ \nonumber \\
\rho_2^{per}&=&\frac{N_c}{2\lambda^{1/2}(s,s',q^2)}\left\{\frac{(2s\Delta'-\Delta
u)(m_c+m_{u})}
{\lambda(s,s',q^2)}-\frac{(2s'\Delta-\Delta'u)(m_b+m_c)}{\lambda(s,s',q^2)}
\right. \nonumber \\
&&\left.
-\frac{2(8ss'^2m_c^2-2ss'\Delta'^2-6s'^2\Delta^2-2u^2s'm_c^2
+6s'u\Delta\Delta'-u^2\Delta'^2)(m_b-m_c)}{\lambda^2(s,s',q^2)}
\right. \nonumber \\
&&\left.
-\frac{2[4ss'um_c^2+4ss'\Delta\Delta'-3su\Delta'^2-3u\Delta^2s'
-u^3m_c^2+2u^2\Delta\Delta'](m_b-m_c)}{\lambda^{2}(s,s',q^2)}+m_c\right\}~,
\nonumber \\
\end{eqnarray}
where $\lambda(a,b,c)=a^2+b^2+c^2-2ac-2bc-2ab$. The $N_c=3$ is the
color factor, $u=s+s^\prime-q^2$, $\Delta=s+m_c^2-m_b^2$ and
$\Delta'=s^\prime +m_c^2-m^2_{u}$.

For the heavy quarkonium $b\bar{c}$, where the relative velocity of
quark movement is small, an essential role is taken by the
Coulomb-like $\alpha_s/v$-corrections \cite{Kiselev1}. It leads to
the finite renormalization for $\rho_i^{per}$, so that:
\begin{equation}
\rho^{c}_i={\cal C} \rho_i^{per},
\end{equation}
with
\begin{equation}
 {\cal C}^2=\frac{4\pi \alpha_s^{\cal
 C}}{3v}\,\frac{1}{\displaystyle 1-\exp\left(-\frac{4\pi\alpha_s^{\cal
 C}}{3v}\right)},
\end{equation}
where $\alpha_s^{\cal C}$ is the coupling constant of effective
coulomb interactions. Also $v$ is the relative velocity of quarks
in the $b\bar{c} $-system,
\begin{equation}
v=\sqrt{1-\frac{4 m_b m_c}{p^2-(m_b-m_c)^2}}.
\end{equation}
The value of $\alpha_s^{\cal C}$ for $B_c$ meson is
\cite{Kiselev1}:
\begin{equation}
\alpha_s^{\cal C}[b\bar{c}]=0.45~.
\end{equation}

By performing the double Borel transformations over the variables
$p^2$ and $p'^2$ on the physical parts of the correlation
functions and bare-loop diagrams and also equating two
representations of the correlation functions, the sum rules for
the $f'_i(q^2)$ are obtained:
\begin{eqnarray}\label{eq17}
f'_{i}(q^2) &=&  \frac{(m_b+m_c)}{f_{B_c} m_{B_c}^2 f_{D_1^0}
m_{D_1^0}}~ e^{\frac{m_{B_c}^2}{M_1^2}}
e^{\frac{m_{D_1}^2}{M_2^2}} \nonumber \\
&\times& \Bigg\{-\frac{1}{4 \pi^2} \int_{m_b^2}^{s_0^\prime}
ds^\prime \int_{s_L}^{s_0} \rho_{i}^{c} (s,s^\prime,q^2)
e^\frac{-s}{M_1^2} e^\frac{-s^{'}}{M_2^2} - iM^{2}_{1}M^{2}_{2}
\left\langle \frac{\alpha_s}{\pi} G^2 \right\rangle
\frac{C_i^4}{6} \Bigg\}~,
\end{eqnarray}
where $i=V,0,1$ and $2$, $s_0$ and $s_0^{'}$ are the continuum
thresholds in pseudoscalar $B_c$ and axial vector $D_1^0$
channels, respectively, and lower bound integration limit of
$s_{L}$ is as follow:
\begin{eqnarray*}
s_L &=& \frac{(m_c^2+q^2-m_b^2-s^\prime) (m_b^2 s^\prime -m_c^2
q^2)} {(m_b^2-q^2) (m_c^2-s^\prime)}~.
\end{eqnarray*}
The explicit expressions for $C_i^{4}$ are presented in Appendix--A.

Now, we would like to consider the form factors related to the
$B_c\to D_1^0$ transition when the $D_1^0$ meson is a mixture of
two $|^1P_1\rangle$ and $|^3P_1\rangle$ states. For this aim,
first the $f_i^{'B_c\to D_11(2)}(q^2)$ are obtained from the above
equations replacing the $f_{D^0_1}$ by decay constant
$f_{{D_1}1(2)}$, and $m_{D^0_1}$ with $m_{{D_1}1(2)}$, i.e.,
\begin{eqnarray}\label{eq18}
f_{i}^{'B_c\to D_11(2)}(q^2)&=&-\frac{(m_b+m_c)}{f_{B_c} m_{B_c}^2
f_{D_11(2)} m_{D_11(2)}} e^\frac{m_{B_c}^2}{M_1^2}
e^\frac{m_{D_11(2)}^2}{M_2^2}\Bigg\{-\frac{1}{4 \pi^2}
\int_{m_c^2}^{s_0^\prime} ds^\prime \int_{s_L}^{s_0} \rho_{i}^{c}
(s,s^\prime,q^2) e^\frac{-s}{M_1^2}
e^\frac{-s^{'}}{M_2^2} \nonumber \\
&-& iM^{2}_{1}M^{2}_{2} \left \langle \frac{\alpha_s}{\pi} G^2
\right\rangle \frac{C_{i}^{4}}{6} \Bigg\}~.
\end{eqnarray}
where $f_{D_11} = (183\pm 25) MeV$, and $f_{D_12} = (89\pm 7) MeV$ \cite{Thomas}. Then using the straight forward calculations, the
$f_{i}^{(2420)}(q^2)$ form factors of $B_{c}\to D^0_1(2420)$
transition are found as follows:
\begin{eqnarray}\label{eq19}
f_0^{(2420)}(q^2)&=&\left(\frac{m_{B_c}+m_{D_11}}{m_{B_c}+m_{D^0_1}}\right)f_0^{B_{c}\to
D_11}(q^2)~sin\theta+\left(\frac{m_{B_c}+m_{D_12}}{m_{B_c}+m_{D^0_1}}\right)f_0^{B_{c}\to
D_12}(q^2)~cos\theta~,\nonumber\\
f_{i'}^{(2430)}(q^2)&=&\left(\frac{m_{B_c}+m_{D^0_1}}{m_{B_c}+m_{D_11}}\right)~f_{i'}^{B_{c}\to
D_11}(q^2)~sin\theta+\left(\frac{m_{B_c}+m_{D^0_1}}{m_{B_c}+m_{D_12}}\right)~f_{i'}^{B_{c}\to
D_12}(q^2)~cos\theta~,\nonumber\\
\end{eqnarray}
where $i'=V, 1, 2$. Note that, the $f_{i}^{(2430)}(q^2)$ form
factors of the $B_{c}\to D^0_1(2430)$ decay are obtained from the
above equations by replacing the $sin{\theta}\rightarrow
cos{\theta}$ and $cos{\theta}\rightarrow -sin{\theta}$.

\section{Heavy quark effective theory}
In this section, we apply the heavy quark effective theory (HQET)
to analyze the form factors of $B_c\to D_1^0 l \nu$ calculated by
3PSR. As it was mentioned in the introduction, in the heavy quark
mass limit, when $m_c\to \infty$, the $D_1^0$ meson can be
considered
 as Eq. (\ref{eq3}). Therefore, to estimate the $f_i^{HQ}$
form factors in this approach, first, we present the dependence of
the $f_i^{HQ(B_c\to D_1k)}$, $(k=1, 2)$ on $y$ where
\begin{equation}\label{eq20}
y=\nu\nu'=\frac{m_{B_{c}}^2+m_{D_1k}^2-q^2}{2m_{B_{c}}m_{D_1k}}.
\end{equation}
Here, $\nu$ and $\nu'$ are the four velocities of $B_c$ and $D_1k$
mesons, respectively (for some details see Refs.
\cite{Huang,Neubert}). After some complicated calculations, the
y--dependent expressions of the $f_i^{HQ(B_c\to D_1k)}(y)$ are
obtained as follows:
\begin{eqnarray}\label{eq21}
f_{V}^{HQ(B_c\to D_1k)}(y) &=&\frac{1}{\hat{f}_{B_{c}}\hat{f}_{D_1k}}e^{\frac{%
\Lambda }{T_{1}}}e^{\frac{\overline{{\Lambda }}}{T_{2}}}\Bigg
\{\frac{1}{{( 1+\sqrt {Z} )
{Z}^{\frac{7}{4}}\sqrt {{ \frac {-1+{y}^{2}}{Z}}}}}\nonumber\\
&&\times\left[-3+ 3( 3y+1 ) \sqrt {Z}-6\,(y^2+y ) Z  \right] \nonumber\\
&&\times\frac{-1}{(2\pi)^2}\int_{0}^{\nu _{0}}d\nu \int_{0}^{\nu
_{0}^{^{\prime }}}d{\nu }^{^{\prime }}e^{-\frac{\nu
}{2T_{1}}}e^{-\frac{{\nu }^{_{^{\prime }}}}{2T_{2}}}\theta
(2y\nu \nu ^{^{\prime }}-\nu ^{2}-{{\nu }^{_{^{\prime }}}}^{2})\nonumber \\
&&+\left(i~\frac{2Z^{\frac{1}{4}}}{3(1+\sqrt{Z})}T_1T_2~%
\left\langle \frac{\alpha _{s}}{\pi }G^{2}\right\rangle\right)
\times \lim_{m_{b}\rightarrow \infty }C_{V}^{HQET}\Bigg \}~,
\end{eqnarray}

\begin{eqnarray}\label{eq22}
f_{0}^{HQ(B_c\to D_1k)}(y) &=&\frac{1}{\hat{f}_{B_{c}}\hat{f}_{D_1k}}e^{\frac{%
\Lambda }{T_{1}}}e^{\frac{\overline{{\Lambda
}}}{T_{2}}}\Bigg\{\frac{1}{{2( 1+\sqrt {Z} )
^{3}{Z}^{\frac{7}{4}}\sqrt {{ \frac {-1+{y}^{2}}{Z}}}}}\nonumber\\
&&\times\left[ -3-3(4y-1)\sqrt {Z} -6(3y^2-2y-1)Z +12(y^3+y^2-y)Z^{\frac{3}{2}}\right] \nonumber\\
&&\times\frac{-1}{(2\pi)^2}\int_{0}^{\nu _{0}}d\nu \int_{0}^{\nu
_{0}^{^{\prime }}}d{\nu }^{^{\prime }}e^{-\frac{\nu
}{2T_{1}}}e^{-\frac{{\nu }^{_{^{\prime }}}}{2T_{2}}}\theta
(2y\nu \nu ^{^{\prime }}-\nu ^{2}-{{\nu }^{_{^{\prime }}}}^{2})\nonumber \\
&&+\left(i~\frac{2Z^{\frac{5}{4}}}{3{(1+\sqrt{Z})^{3}}}T_1T_2~%
\left\langle \frac{\alpha _{s}}{\pi }G^{2}\right\rangle\right)
\times \lim_{m_{b}\rightarrow \infty }C_{0}^{HQET}\Bigg\}~,
\end{eqnarray}

\begin{eqnarray}\label{eq23}
f_{1}^{HQ(B_c\to D_1k)}(y) &=&\frac{1}{\hat{f}_{B_{c}}\hat{f}_{D_1k}}e^{\frac{%
\Lambda }{T_{1}}}e^{\frac{\overline{{\Lambda
}}}{T_{2}}}\Bigg\{\frac{1}{4(1+\sqrt{Z})({\frac{-1+{y}^{2}}{Z}})^{\frac{3}{2}}{Z}^{{\frac{15}{4}}}}\nonumber\\
&&\times\left[-9+9(5y+1)\sqrt{Z}-3(26y^2+15y-2)Z \right.\nonumber\\
&&\left.+6(9y^3+11y^2-3y+1){Z}^{\frac{3}{2}}-12(y^4+2y^3-y^2+y){Z}^{2}\right] \nonumber\\
&&\times\frac{-1}{(2\pi)^2}\int_{0}^{\nu _{0}}d\nu \int_{0}^{\nu
_{0}^{^{\prime }}}d{\nu }^{^{\prime }}e^{-\frac{\nu
}{2T_{1}}}e^{-\frac{{\nu }^{_{^{\prime }}}}{2T_{2}}}\theta
(2y\nu \nu ^{^{\prime }}-\nu ^{2}-{{\nu }^{_{^{\prime }}}}^{2})\nonumber \\
&&+\left(i~\frac{2Z^{\frac{1}{4}}}{3{(1+\sqrt{Z})}}T_1T_2~%
\left\langle \frac{\alpha _{s}}{\pi }G^{2}\right\rangle\right)
\times\lim_{m_{b}\rightarrow \infty }C_{1}^{HQET}\Bigg\}~,
\end{eqnarray}

\begin{eqnarray}\label{eq24}
f_{2}^{HQ(B_c\to D_1k)}(y) &=&\frac{1}{\hat{f}_{B_{c}}\hat{f}_{D_1k}}e^{\frac{%
\Lambda }{T_{1}}}e^{\frac{\overline{{\Lambda
}}}{T_{2}}}\Bigg\{\frac{1}{4(1+\sqrt{Z})({\frac{-1+{y}^{2}}{Z}})^{\frac{3}{2}}{Z}^{{\frac{15}{4}}}}\nonumber\\
&&\times\left[-9+9(3y+1)\sqrt{Z}-9(2y^2+3y-2)Z\right. \nonumber\\
&&\left.-6(y^3-y^2+5y+1){Z}^{\frac{3}{2}}+12(y^4+2y^3-y^2+y){Z}^{2} \right] \nonumber\\
&&\times\frac{-1}{(2\pi)^2}\int_{0}^{\nu _{0}}d\nu \int_{0}^{\nu
_{0}^{^{\prime }}}d{\nu }^{^{\prime }}e^{-\frac{\nu
}{2T_{1}}}e^{-\frac{{\nu }^{_{^{\prime }}}}{2T_{2}}}\theta
(2y\nu \nu ^{^{\prime }}-\nu ^{2}-{{\nu }^{_{^{\prime }}}}^{2})\nonumber \\
&&+\left(i~\frac{2Z^{\frac{1}{4}}}{3{(1+\sqrt{Z})}}T_1T_2~%
\left\langle \frac{\alpha _{s}}{\pi }G^{2}\right\rangle\right)
\times\lim_{m_{b}\rightarrow \infty }C_{2}^{HQET}\Bigg\}~.
\end{eqnarray}
In these heavy quark limit expressions $\Lambda=m_{B_c}-m_{b}$,
$\bar{\Lambda}=m_{D_1k}-m_c$, $\sqrt{Z}=y+\sqrt{y^2-1}$,
$\hat{f}_{B_c}=\sqrt{m_b}f_{B_c}$,
$\hat{f}_{D_1k}=\sqrt{m_c}f_{D_1k}$. The continuum thresholds
$\nu_{0}$, $\nu_{0}^{'}$ and integration variables $\nu$,
$\nu^{'}$ are defined as:
\begin{equation}\label{eq25}
\nu_{0}=\frac{s_{0}-m_{b}^2}{m_{b}},~~~~~~
\nu'_{0}=\frac{s'_{0}-m_{c}^2}{m_{c}},
\end{equation}
\begin{equation}\label{eq26}
\nu=\frac{s-m_{b}^2}{m_{b}},~~~~~~ \nu'=\frac{s'-m_{c}^2}{m_{c}}.
\end{equation}
Also we apply $T_1=M_1^2/2m_b$, $T_2=M_2^2/2m_c$ and
$m_c=m_b/\sqrt{Z}$.

The explicit expressions of the coefficients $C_{i}^{HQET}$ are
given in Appendix--B. In the expressions of the $C_{i}^{HQET}$,
$\bar{I}_0(a,b,c)$, $\bar{I}_{1(2)}(a,b,c)$, $\bar{I}_j(a,b,c)$;
$j=3,4,5$ and $\bar{I}_6(a,b,c)$ are defined as:
\begin{eqnarray}  \label{eq27}
\bar{I}_0(a,b,c)\!\!\! &=& \!\!\! \frac{(-1)^{a+b+c}}{16
\pi^2\,\Gamma(a)
\Gamma(b) \Gamma(c)} (\frac{1}{\sqrt{Z}})^{2-a-c}(2m_b)^{4-2a-b-c}~T_1^{2-a-b} ~T_2^{2-a-c} \,\nonumber\\
&&\mathcal{U}%
_0^{HQET}(a+b+c-4,1-c-b)~,  \nonumber \\
\bar{I}_{1(2)}(a,b,c) \!\!\! &=& \!\!\! i \frac{(-1)^{a+b+c+1}}{16
\pi^2\,\Gamma(a) \Gamma(b)
\Gamma(c)}(\frac{1}{\sqrt{Z}})^{4-a-c-1(2)}(2m_b)^{5-2a-b-c}
~T_1^{1-a-b+1(2)}~T_2^{4-a-c-1(2)} \,\nonumber\\
&&\mathcal{U}_0^{HQET}(a+b+c-5,1-c-b)~,  \nonumber \\
\nonumber \\
\bar{I}_j(a,b,c)\!\!\! &=& \!\!\! i \frac{(-1)^{a+b+c}}{16
\pi^2\,\Gamma(a) \Gamma(b) \Gamma(c)}
(\frac{1}{\sqrt{Z}})^{7-a-c-j}(2m_b)^{6-2a-b-c}~T_1^{-a-b-1+j}
~T_2^{7-a-c-j} \,\nonumber\\
&&\mathcal{U}_0^{HQET}(a+b+c-6,1-c-b)~,  \nonumber \\
\nonumber \\
\bar{I}_6(a,b,c)\!\!\! &=& \!\!\! i \frac{(-1)^{a+b+c+1}}{32
\pi^2\,\Gamma(a) \Gamma(b) \Gamma(c)}
(\frac{1}{\sqrt{Z}})^{3-a-c}(2m_b)^{6-2a-b-c}~T_1^{3-a-b}
~T_2^{3-a-c} \,\nonumber\\
&&\mathcal{U}_0^{HQET}(a+b+c-6,2-c-b)~.
\end{eqnarray}
The function $\mathcal{U}_0^{HQET}(m,n)$ takes the following form
\begin{eqnarray}\label{eq28}
\mathcal{U}_0^{HQET}(m,n)=\int _{0}^{\infty }\!
(2m_b)^{m}~(\frac{x}{2m_b}+T_1+\frac{T_2}{\sqrt{Z}})^{m}
~x^{n}[-\frac{\overline{B}_{-1}}{x}-\overline{B}_{0}-\overline{B}_{1}x]{dx},
\end{eqnarray}
with
\begin{eqnarray}\label{eq29}
\overline{B}_{-1}&=&\frac{\sqrt{Z}}{T_1T_2}~[\frac{mb^2}{Z}~T_2^2+\frac{1}{\sqrt{Z}}T_1T_2(m_b^2-q^2)],%
\nonumber\\
\overline{B}_{0}&=&\frac{\sqrt{Z}}{2m_bT_1T_2}~[m_c^2T_1+\frac{T_2}{\sqrt{Z}}(m_b^2+m_c^2)],%
\nonumber \\
\overline{B}_{1}&=&\frac{1}{4\sqrt{Z}T_1T_2}.
\end{eqnarray}
Then by considering $D_1^0$ meson as a combination of two states
$|D_1k\rangle$, $(k=1, 2)$ the $f_i^{HQ(2420)}(y)$ and
$f_i^{HQ(2430)}(y)$ form factors for the $B_c\to D_1^0(2420) l
\nu$ and $B_c\to D_1^0(2430) l \nu$ decays are obtained as:
\begin{eqnarray}\label{eq30}
f_i^{HQ(2420)}(y)&=&\sqrt{\frac{1}{3}}f_0^{HQ(B_{c}\to
D_11)}(y)+\sqrt{\frac{2}{3}}f_0^{HQ(B_{c}\to
D_12)}(y)~,\nonumber\\
f_{i}^{HQ(2430)}(y)&=&\sqrt{\frac{2}{3}}~f_{i'}^{HQ(B_{c}\to
D_11)}(y)-\sqrt{\frac{1}{3}}~f_{i'}^{HQ(B_{c}\to D_12)}(y)~,
\end{eqnarray}
where $i=V, 0, 1, 2$.

\section{Numerical analysis}
\begin{figure}
\includegraphics[width=6cm,height=5cm]{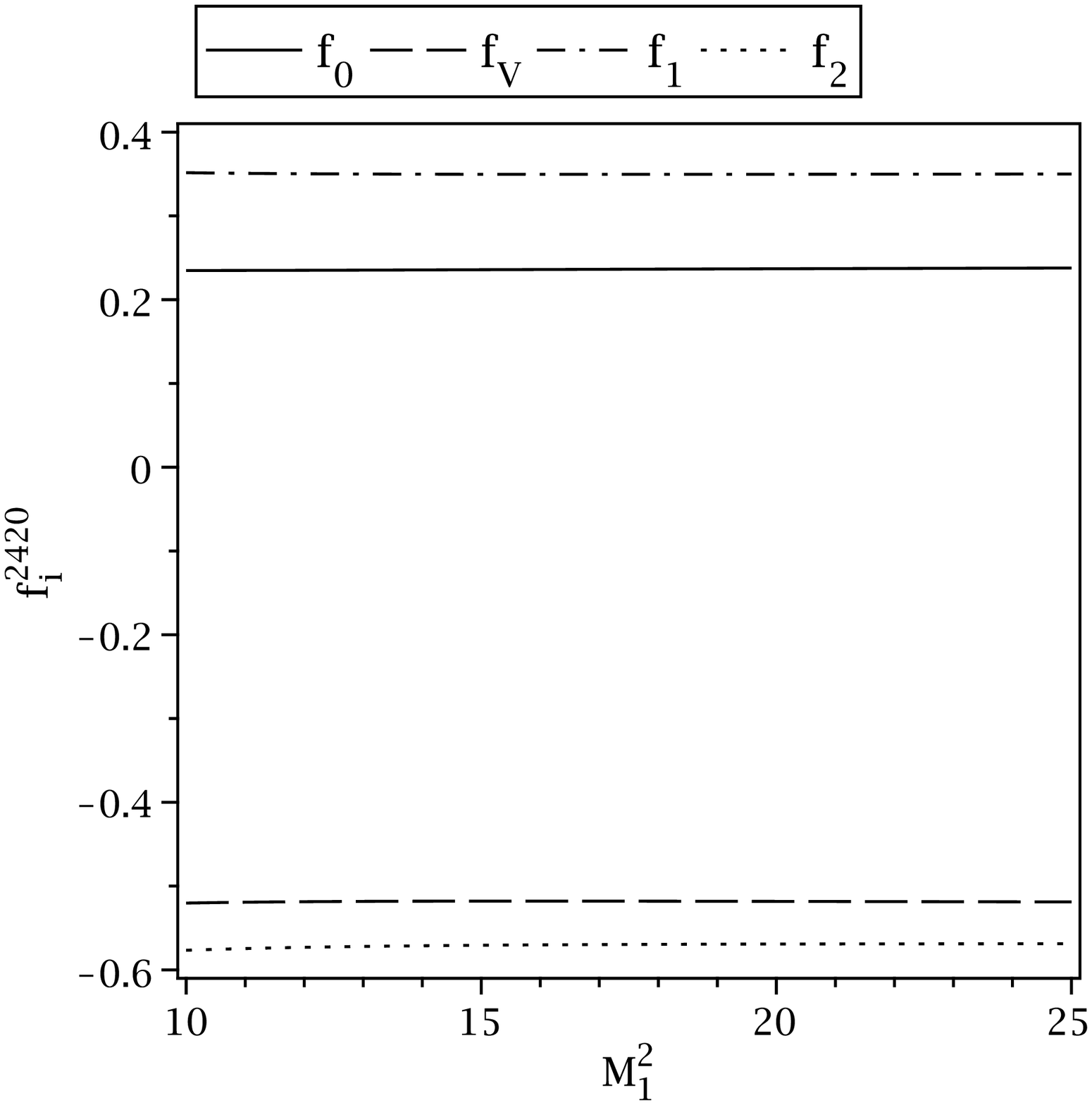}
\includegraphics[width=6cm,height=5cm]{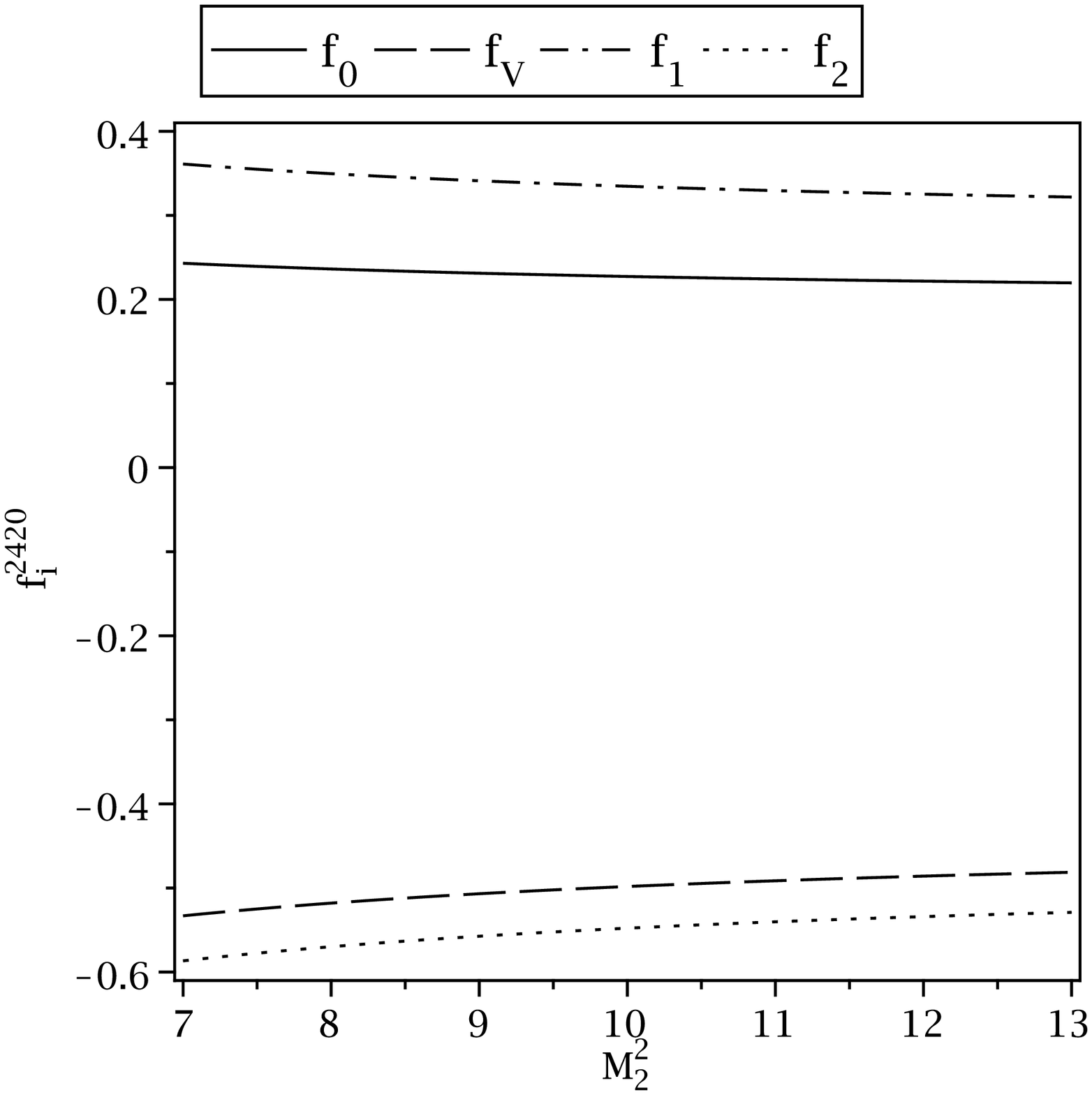}
\caption{The dependence of the transition form
factors on the Borel parameters for the $B_c\to D_{1}^0(2420)$
transition.}\label{F0}
\end{figure}

Now, we present our numerical analysis of the form factors $%
f_{i}(q^2)~,(i=V, 0, 1, 2)$ via 3PSR and HQET. From the sum rules
expressions of the form factors, it is clear that the main input
parameters entering the expressions are gluon condensates, element
of the CKM matrix $V_{ub}$, leptonic decay constants $f_{B_c}$,
$f_{D_{1}^0}$, $f_{D_{1}1}$ and $f_{D_{1}2}$, Borel parameters
$M_{1}^2$ and $M_{2}^2$ as well as the continuum thresholds
$s_{0}$ and $s'_{0}$.
\begin{table}[th]
\begin{center}
\caption{Input values in numerical calculations.}\label{T4}
\begin{tabular} {|c|c|}
\hline
$\langle\frac{\alpha_s}{\pi}G^2\rangle$ & $0.044 \pm 0.007~GeV^4$\cite{Dominguez} \\
$\mid V_{ub}\mid$ & $(3.8\pm0.5)\times 10^{-3}$\cite{Crivellin} \\
$f_{D_{1}}$ & $220\pm12~MeV$ \cite{Bazavov} \\
$f_{B_c}$ & $395\pm15~MeV$ \cite{Kiselev2} \\
\hline
\end{tabular}
\end{center}
\end{table}
The sum rules for the form factors contain also four auxiliary
parameters: Borel mass squares $M_{1}^2$ and $M_{2}^2$ and continuum
thresholds $s_{0}$ and $s'_{0}$. These are not physical quantities,
so the the form factors as physical quantities should be independent
of them. The parameters $s_0$ and $s_0^\prime$, which are the
continuum thresholds of $B_c$ and $D_{1}^0$ mesons, respectively,
are determined from the condition that guarantees the sum rules to
practically be stable in the allowed regions for $M_1^2$ and
$M_2^2$. The values of the continuum thresholds calculated from the
two--point QCD sum rules are taken to be $s_0=(45-50)~GeV^2$ \cite{Kiselev2} and
$s_0^\prime=(6-8)~GeV^2$ \cite{Colangelo2}. We search for the intervals of the Borel mass parameters so that our
results are almost insensitive to their variations. One more
condition for the intervals of these parameters is the fact that the
aforementioned intervals must suppress the higher states, continuum
and contributions of the highest-order operators. In other words,
the sum rules for the form factors must converge (for more details, see \cite{Colangelo}).  As a result, we
get $10~{GeV}^{2}\leq M_{1}^{2} \leq 25~{GeV}^2$ and
$7~{GeV}^2\leq M_{2}^{2}\leq 13~{GeV}^2$. To show how the
form factors depend on the Borel mass parameters, for example, we
depict the variations of the form factors for $B_{c}\rightarrow D_{1}^0(2420)$ at $q^{2} = 0$ with
respect to the variations of the $M_1^2$ and $M_2^2$ parameters in
their working regions in Fig. \ref{F0}. From these figures, it revealed
that the form factors weakly depend on these parameters in their
working regions.

For analysis the form factors of the semileptonic $B_c\to
D_1^0(2420[2430]) l \nu$ decays, first, we consider the $D_1^0$
meson as a pure state, i.e., $|c\bar{u}\rangle$ and analyze the
form factors and the value of the branching ratios of the $B_c\to
D_1^0(2420[2430])$ transitions in 3PSR. Then the form factors,
decay widths and branching ratios of these decays are plotted when
$D_1^0$ meson is a combination of two states with mixing angle
$\theta$. Finally, considering the $D_1^0$ meson as Eq.
(\ref{eq3}), we investigate and estimate this mentioned physical
quantities via HQET approach in $m_b(m_c)\to \infty$ limit.
Therefore, there are three conditions for the study of the form
factors of the $B_c\to D_1^0(2420[2430]) l \nu$ decays, related to
the structure of the $D_1^0$ meson as following:

\subsection*{$\bullet$ Pure state or $|c\bar{u}\rangle$ state}
If the $D_{1}^0$ meson is the pure $|c\bar{u}\rangle$ state, using
Eqs. (\ref{eq7}) and (\ref{eq17}), the values of the form factors
at $q^2=0$ are presented in Table \ref{T5}.
\begin{table}[th]
\caption{The value of the form factors of the
conventional $B_{c}\to D_{1}^0(2420,2430)$ transitions at $q^2=0$,
$M_1^2=15~GeV^2$ and $M_2^2=8~GeV^2$ .}\label{T5}
\begin{ruledtabular}
\begin{tabular}{cc}
Form Factor &Value\\
\hline
$f_{V}^{B_c\to D_{1}^0(2420,2430)}(0)$ &$-0.51\pm0.12$\\
$f_{0}^{B_c\to D_{1}^0(2420,2430)}(0)$ &$0.23\pm0.07$\\
$f_{1}^{B_c\to D_{1}^0(2420,2430)}(0)$ &$0.33\pm0.09$\\
$f_{2}^{B_c\to D_{1}^0(2420,2430)}(0)$ &$-0.55\pm0.14$\\
\end{tabular}
\end{ruledtabular}
\end{table}
In this case, the values of the transition form factors at $q^2=0$
for $B_c\to D_1^0(2420) l \nu$ decay are the same as those for
$B_c\to D_1^0(2430) l \nu$. Our calculations show, the other
physical quantities of these decays are the same, nearly.

The sum rules for the form factors are truncated at about $9
~GeV^2$, so to extend our results to the full physical region, we
look for a parametrization of the form factors in such a way that
in the region $0 \leq q^2 \leq (m_{B_c}-m_{D_{1}^0})^2~ GeV^2$,
this parametrization coincides with the sum rules predictions. Our
numerical calculations show that the sufficient parametrization of
the form factors with respect to $q^2$ is as follows \cite{Patricia Ball}:
\begin{equation}\label{eq31}
f_{i}(q^2)=\frac{a}{(1- \frac{q^2}{m^{2}_{fit}})}+ \frac{b}{(1-\frac{q^2}{m^{2}_{fit}})^2}.
\end{equation}
The values of the parameters $a, b$ and  $m_{fit}$ are given in
Table \ref{T6}. Fig. \ref{F3} depict the fit function of the form
factors $f_i^{(2420,2430)}(q^2)$, (i=V, 0, 1, 2) of the $B_c\to
D^0_1(2420,2430) l\nu$ decays with respect to the transferred
momentum square $q^2$. This figure also contains the form factors
obtained via 3PSR (see Eq.(\ref{eq17})). The form factors and
their fit functions coincide well in the interval $0 \leq q^2 \leq
9~ GeV^2$.
\begin{table}[th]
\caption{Parameters appearing in the fit function
for the form factors of the $B_{c}\to D_{1}(2420,2430)$  at
$M_1^2=15~GeV^2$ and $M_2^2=8~GeV^2$ .}\label{T6}
\begin{ruledtabular}
\begin{tabular}{cccc}
Form Factor &$a$&$b$&$m_{fit}$\\
\hline
$f_{V}^{B_c\to D_{1}^0(2420,2430)}(q^2)$ &-0.34&-0.17&4.94\\
$f_{0}^{B_c\to D_{1}^0(2420,2430)}(q^2)$ &0.19&0.04&6.88\\
$f_{1}^{B_c\to D_{1}^0(2420,2430)}(q^2)$ &0.24&0.09&5.91\\
$f_{2}^{B_c\to D_{1}^0(2420,2430)}(q^2)$ &-0.35&-0.20&4.82\\
\end{tabular}
\end{ruledtabular}
\end{table}

By using the expressions for the form factors, the differential
decay width ${d\Gamma}/{dq^2}$ for the process $B_c\rightarrow
D_1^0 l \nu$ in terms of $H_{i}$ is presented as follow:
\begin{eqnarray}\label{eq32}
\frac{d\Gamma _{\pm }(B_{c}\rightarrow D_1^0 l \nu)}{dq^2}
&=&\frac{G^{2}\left| V_{ub}\right| ^{2}}{192\pi
^{3}m^3_{B_c}}~q^{2}\lambda
^{1/2}(m^2_{B_c},m^2_{D_1^0},q^{2})\left|
H_{\pm }\right| ^{2}~,  \nonumber \\
&&  \nonumber \\
\frac{d\Gamma _{0}(B_{c}\rightarrow D_1^0 l \nu)}{dq^2}
&=&\frac{G^{2}\left| V_{ub}\right| ^{2}}{192\pi ^{3}m^3_{B_c}}~
q^{2}\lambda ^{1/2}(m^2_{B_c},m^2_{D_1^0},q^{2})\left|
H_{0}\right| ^{2}~,\nonumber \\
&&  \nonumber \\
\frac{d\Gamma _{tot}(B_{c}\rightarrow D_1^0 l \nu)}{dq^2}
&=&\frac{d\Gamma _{\pm}(B_{c}\rightarrow D_1^0 l
\nu)}{dq^2}+\frac{d\Gamma _{0}(B_{c}\rightarrow D_1^0 l
\nu)}{dq^2}~,
\end{eqnarray}
$H_{\pm}$ and $H_{0}$ are defined as:
\begin{eqnarray*}
H_{\pm }(q^{2})
&=&(m_{B_c}+m_{D_1^0})f_{0}(q^{2})\mp \frac{%
\lambda ^{1/2}(m^2_{B_c},m^2_{D_1^0},q^{2})}{%
m_{B_c}+m_{D_1^0}}f_{V}(q^{2})~, \\
\\
H_{0}(q^{2}) &=&\frac{1}{2m_{D_1^0}\sqrt{q^2}}\Bigg[%
(m^2_{B_c}-m^2_{D_1^0}-q^{2})(m_{B_c}+m_{D_1^0}%
)f_{0}(q^{2}) -\frac{\lambda
(m^2_{B_c},m^2_{D_1^0},q^{2})}{%
m_{B_c}+m_{D_1^0}}f_{1}(q^{2})\Bigg]~,
\end{eqnarray*}
where $\pm, 0$ refer to the $D_1^0$ helicities. Note that in the
limit of vanishing lepton mass (in our case electron and muon) the
$f_2(q^2)$ form factor does not contribute to the decay width
formula.

To calculate the branching ratios of the $B_c\to D_{1}(2420,2430)l
\nu$ decays, we integrate Eq. (\ref{eq32}) over $q^2$ in the whole
physical region and use the total mean life time
$\tau_{B_c}=(0.46\pm0.07)~ps$. Our numerical analysis
shows that the contribution of the non-perturbative part (the
gluon condensate diagrams ) is about $13\%$ of the total and the
main contribution comes from the perturbative part of the form
factors. The value for the branching ratio of these decays is
obtained as presented in Table \ref{T7}. The function of decay
width of $B_c\to D_1^0(2420,2430) l \nu$ decays with respect to
$q^2$ is shown in Fig. \ref{F4}.
\begin{table}[th]
\caption{The branching ratio value of the
semileptonic $B_c\to D_{1}^0(2420,2430)l \nu$ decays.}\label{T7}
\begin{ruledtabular}
\begin{tabular}{cc}
MOD& BR \\
\hline
$B_c\to D_{1}^0(2420,2430) l {\nu}$ &$(0.71\pm 0.18) \times 10^{-4}$\\
\end{tabular}
\end{ruledtabular}
\end{table}

\subsection*{$\bullet$ Mixture of $|^3P_1\rangle$ and $|^1P_1\rangle$ states}
Now, we would like to analyze the form factors of the $B_c\to
D_1^0$ transition when we consider the $D_{1}^0$ meson as a
mixture of two $|^3P_1\rangle$ and $|^1P_1\rangle$ states with
mixing angle $\theta$ (see Eq. (\ref{eq19})). The transition form
factors of the $B_c\to D_{1}^0(2420[2430]) l \nu$ at $q^2=0$ in
the interval $-180^\circ\leq\theta\leq180^\circ$ are shown in
Figs. \ref{F5}, \ref{F6}. The dependence of the form factors of
the $B_c\to D_{1}^0(2420)$ and $B_c\to D_{1}^0(2430)$ decays on
mixing angle $\theta $ and the transferred momentum square $q^2$
are plotted in Figs. \ref{F7}, \ref{F8} in the regions $0 \leq q^2
\leq (m_{B_c}-m_{D_{1}^0})^2~ GeV^2$ and  $\theta=\pm N\pi/6,~ N= 1, 2, 3$. Using Eq.
(\ref{eq32}), we denote the variation of the decay widths with
respect to $q^2$ and $\theta$ in the same regions for each decay
in Fig. \ref{F9}. Also the branching ratios only in terms of
mixing angle $\theta$ are shown in Fig. \ref{F10}.

\subsection*{$\bullet$ Compound state in the heavy quark limit}
Eventually, we study the structure of the $D_1^0$ meson as a
mixture of two $|^3P_1\rangle$ and $|^1P_1\rangle$ states with the
mixing angle $\theta=35.3^\circ$ in the heavy quark limit. The
HQET form factors of the $B_c\to D_1^0$ transition were evaluated
in Eq. (\ref{eq30}). Figs. \ref{F11}, \ref{F12} depict the
$f_i^{HQ(2420)}$ and $f_i^{HQ(2430)}$ with respect to the $q^2$
for $B_c\to D_1^0(2420) l \nu$ and $B_c\to D_1^0(2430) l \nu$,
respectively. It is noted, at  $y=1$ in Eq. (\ref{eq20}) called
the zero recoil limit (corresponding to
$q^2=(m_{B_c}-m_{D^0_1})^2$), the HQET limit of the form factors
are not finite. For other values of y and corresponding $q^2$, the
behavior of the $f_i^{(2420,2430)}$ form factors shown in Fig.
\ref{F3} and their HQET form factors $f_i^{HQ(2420)}$ and
$f_i^{HQ(2430)}$ in Figs. \ref{F11}, \ref{F12} are the same, i.e.,
when $q^2$ increases (y decreases) both the form factors and their
HQET values increases.

The values of the HQET form factors at $q^2=0$ are presented in
Table \ref{T8}.
\begin{table}[th]
\caption{The value of the form factors of the
 $B_{c}\to D_{1}^0(2420)$ and $B_{c}\to D_{1}^0(2430)$
transitions via HQET at $q^2=0$.}\label{T8}
\begin{ruledtabular}
\begin{tabular}{cccc}
Form Factor& Value& Form Factor& Value\\
\hline
$f_{V}^{HQ(2420)}(0)$ &$-0.70\pm0.15$&$f_{V}^{HQ(2430)}(0)$&$-0.66\pm0.14$\\
$f_{0}^{HQ(2420)}(0)$ &$0.16\pm0.05$&$f_{0}^{HQ(2430)}(0)$&$0.16\pm0.04$\\
$f_{1}^{HQ(2420)}(0)$ &$0.38\pm0.10$&$f_{1}^{HQ(2430)}(0)$&$0.36\pm0.10$\\
$f_{2}^{HQ(2420)}(0)$ &$-0.10\pm0.03$&$f_{2}^{HQ(2430)}(0)$&$-0.10\pm0.03$\\
\end{tabular}
\end{ruledtabular}
\end{table}

Also using Eq. (\ref{eq32}) and HEQT form factors, we evaluated
the branching ratios of the $B_c\to D_1^0(2420[2430]) l \nu$
decays as given in Table \ref{T9}. The Fig. \ref{F13} depict the
dependence of the decay widths of these decays on the $q^2$ in
HQET approach.
\begin{table}[th]
\caption{The branching ratio values of the
semileptonic $B_c\to D_{1}^0(2420)l \nu$ and $B_c\to
D_{1}^0(2430)l \nu$ decays via HQET.}\label{T9}
\begin{ruledtabular}
\begin{tabular}{ccc}
MOD& $B_c\to D_{1}^0(2420) l {\nu}$& $B_c\to D_{1}^0(2430) l {\nu}$\\
\hline
BR &$(1.56\pm 0.41) \times 10^{-4}$&
$(1.05\pm 0.33) \times 10^{-4}$\\
\end{tabular}
\end{ruledtabular}
\end{table}

\section*{Conclusion }
In summary, We analyzed the semileptonic $B_{c}\to D_{1}^0
(2420[2430])l\nu$ decays in the framework of the three--point QCD
sum rules and HQET approach. First, we assumed the $D_{1}^0(2420)$
and $D_{1}^0(2430)$ axial vector mesons as the pure
$|c\bar{u}\rangle$ state. In this case, the  related form factors
were computed. The branching ratios of these decays were also
estimated. Second, $D_{1}^0(2420[2430])$ mesons were considered as
a combination of two states $|^3P_1\rangle\equiv |D_{1}1\rangle$
and $|^1P_1\rangle\equiv |D_{1}2\rangle$ with different masses and
decay constants. We evaluated the transition form factors and the
decay widths of these decays with respect to mixing angle $\theta$
and the transferred momentum square $q^2$. The dependence of the
branching ratios on $\theta$ was also presented. Finally, we
obtained all of the mentioned physical quantity in HQET approach.
Any future experimental measurement on these form factors as well
as decay rates and branching fractions and their comparison with
the obtained results in the present work can give considerable
information about the structure of these mesons and unknown mixing
angle $\theta$.

\section*{Acknowledgments}
Partial support of the Isfahan university of
technology  research council is
appreciated.
\newpage

\appendix
\begin{center}
{\Large \textbf{Appendix--A}}
\end{center}
\setcounter{equation}{0} \renewcommand{\theequation}

In this appendix, the explicit expressions of the coefficients of
the gluon condensate  entering  the sum rules of the form factors
$f_i(q^2)$, $(i=V, 0, 1, 2)$ are given.

\begin{eqnarray*}
C_{V}^{4} &=&-10\,\hat{I}_{{1}}(3,2,2){m_{{b}}}^{3}{m_{{c}}}^{2}+10\,\hat{I}_{{1}}(3,2,2)%
{m_{{b}}}^{2}{m_{{c}}}^{3}+10\,\hat{I}_{{2}}(3,2,2){m_{{b}}}^{2}{m_{{c}}}%
^{3}+10\,\hat{I}_{{0}}(3,2,2){m_{{b}}}^{2}{m_{{c}}}^{3} \\
&&+60\,\hat{I}_{{2}}(1,4,1){m_{{b}}}^{2}m_{{c}}-20\,\hat{I}_{{2}}(3,2,1){m_{{b}}}^{2}m_{{%
c}}+10\,\hat{I}_{2}^{[0,1]}(3,2,2){m_{{b}}}^{2}m_{{c}}-20\,\hat{I}_{{0}}(3,2,1){m_{{b}}}%
^{2}m_{{c}} \\
&&+10\,\hat{I}_{{1}}(3,2,1)m_{{b}}{m_{{c}}}^{2}+40\,\hat{I}_{{2}}(2,3,1)m_{{b}}{m_{{c}}}%
^{2}-10\,\hat{I}_{{0}}(3,2,1)m_{{b}}{m_{{c}}}^{2}+20\,\hat{I}_{{1}}(2,3,1)m_{{b}}{m_{{c}}%
}^{2} \\
&&-10\,\hat{I}_{{0}}(3,2,2){m_{{c}}}^{5}+20\,\hat{I}_{{1}}(3,2,1){m_{{b}}}^{3}+10\,\hat{I}_{{1}%
}(2,2,2){m_{{b}}}^{3}-20\,\hat{I}_{{1}}(2,3,1){m_{{b}}}^{3} \\
&&+10\,\hat{I}_{{0}}(3,2,1){m_{{c}}}^{3}-10\,\hat{I}_{{2}}(3,1,2){m_{{c}}}^{3}-20\,\hat{I}_{{0}%
}(2,2,2){m_{{c}}}^{3}-20\,\hat{I}_{{2}}(2,2,2){m_{{c}}}^{3} \\
&&-10\,\hat{I}_{{0}}(3,1,2){m_{{c}}}^{3}+20\,\hat{I}_{0}^{[0,1]}(3,2,2){m_{{c}}}%
^{3}-50\,\hat{I}_{{1}}(2,2,1)m_{{b}}+20\,\hat{I}_{1}^{[0,1]}(2,3,1)m_{{b}} \\
&&-20\,\hat{I}_{1}^{[0,1]}(3,1,2)m_{{b}}-20\,\hat{I}_{{0}}(2,2,1)m_{{b}}+30\,\hat{I}_{{1}%
}(2,1,2)m_{{b}}+100\,\hat{I}_{{2}}(1,3,1)m_{{b}} \\
&&+30\,\hat{I}_{{0}}(2,2,1)m_{{c}}+30\,\hat{I}_{2}^{[0,1]}(3,1,2)m_{{c}%
}+20\,\hat{I}_{2}^{[0,1]}(3,2,1)m_{{c}}+10\,\hat{I}_{0}^{[0,1]}(3,2,1)m_{{c}} \\
&&+20\,\hat{I}_{{2}}(2,2,1)m_{{c}}-30\,\hat{I}_{{2}}(2,1,2)m_{{c}}+10\,\hat{I}_{{0}}(3,1,1)m_{{%
c}}+20\,\hat{I}_{0}^{[0,1]}(2,2,2)m_{{c}} \\
&&+20\,\hat{I}_{2}^{[0,1]}(2,2,2)m_{{c}}-10\,\hat{I}_{{2}}(3,1,1)m_{{c}}-20\,\hat{I}_{{1}%
}(2,1,2)m_{{c}}-30\,\hat{I}_{{0}}(2,1,2)m_{{c}}. \\
\end{eqnarray*}
\begin{eqnarray*}
C_{0}^{4} &=&-20\,\hat{I}_{{6}}(3,2,2){m_{{c}}}^{5}-40\,\hat{I}_{{6}}(3,2,1){m_{{c}}}%
^{3}-20\,\hat{I}_{{6}}(3,1,2){m_{{c}}}^{3}+40\,\hat{I}_{6}^{[0,6]}(3,2,2){m_{{c}}}^{3} \\
&&+20\,\hat{I}_{{6}}(2,2,2){m_{{b}}}^{3}+5\,\hat{I}_{{0}}(2,2,1){m_{{b}}}^{3}-120\,\hat{I}_{{6}%
}(1,4,1){m_{{b}}}^{3}+40\,\hat{I}_{{6}}(2,3,1){m_{{b}}}^{3} \\
&&+10\,\hat{I}_{0}^{[0,1]}(2,2,2){m_{{b}}}^{3}-5\,\hat{I}_{{0}}(1,2,2){m_{{b}}}%
^{3}-20\,\hat{I}_{6}^{[0,1]}(3,2,2){m_{{b}}}^{3}+20\,\hat{I}_{6}^{[0,1]}(3,1,2)m_{{c}} \\
&&+5\,\hat{I}_{0}^{[0,1]}(3,1,1)m_{{c}}+5\,\hat{I}_{{0}}(1,1,2)m_{{c}}+20\,\hat{I}_{{6}%
}(2,1,2)m_{{c}}+40\,\hat{I}_{{6}}(3,1,1)m_{{c}} \\
&&-10\,\hat{I}_{0}^{[0,1]}(1,3,1)m_{{b}}-15\,\hat{I}_{{0}}(1,2,1)m_{{b}}-40\,\hat{I}_{{6}%
}(2,2,1)m_{{b}}+15\,\hat{I}_{0}^{[0,1]}(2,2,1)m_{{b}} \\
&&-20\,\hat{I}_{6}^{[0,1]}(2,2,2)m_{{b}}+20\,\hat{I}_{6}^{[0,2]}(3,2,2)m_{{b}%
}-40\,\hat{I}_{6}^{[0,1]}(3,1,2)m_{{b}}-15\,\hat{I}_{{0}}(1,1,2)m_{{b}} \\
&&+10\,\hat{I}_{0}^{[0,1]}(3,1,1)m_{{b}}-15\,\hat{I}_{0}^{[0,2]}(3,2,1)m_{{b}}-20\,\hat{I}_{{6}%
}(1,2,2)m_{{b}}-40\,\hat{I}_{6}^{[0,1]}(2,3,1)m_{{b}} \\
&&-10\,\hat{I}_{{0}}(2,3,1){m_{{c}}}^{4}m_{{b}}+15\,\hat{I}_{0}^{[0,1]}(3,2,2){m_{{c}}}%
^{4}m_{{b}}+20\,\hat{I}_{{6}}(3,2,2){m_{{c}}}^{4}m_{{b}}-15\,\hat{I}_{{0}}(2,2,2){m_{{c}}%
}^{4}m_{{b}} \\
&&+5\,\hat{I}_{{0}}(3,2,2){m_{{c}}}^{5}{m_{{b}}}^{2}-30\,\hat{I}_{{0}}(1,4,1)m_{{c}}{m_{{%
b}}}^{4}-5\,\hat{I}_{0}^{[0,1]}(3,2,2)m_{{c}}{m_{{b}}}^{4}+10\,\hat{I}_{{0}}(3,2,1)m_{{c}%
}{m_{{b}}}^{4} \\
&&-10\,\hat{I}_{0}^{[0,1]}(3,2,2){m_{{c}}}^{3}{m_{{b}}}^{2}+5\,\hat{I}_{{0}}(3,2,1){m_{{c%
}}}^{3}{m_{{b}}}^{2}+15\,\hat{I}_{{0}}(4,1,1){m_{{c}}}^{3}{m_{{b}}}^{2}+20\,\hat{I}_{{6}%
}(2,2,2){m_{{c}}}^{2}m_{{b}} \\
&&+10\,\hat{I}_{{0}}(1,3,1){m_{{c}}}^{2}m_{{b}}+20\,\hat{I}_{0}^{[0,1]}(3,2,1){m_{{c}}}%
^{2}m_{{b}}-20\,\hat{I}_{{0}}(1,2,2){m_{{c}}}^{2}m_{{b}}-15\,\hat{I}_{{0}}(2,1,2){m_{{c}}%
}^{2}m_{{b}} \\
&&-10\,\hat{I}_{{0}}(3,1,1){m_{{c}}}^{2}m_{{b}}+20\,\hat{I}_{{6}}(3,1,2){m_{{c}}}^{2}m_{{%
b}}+15\,\hat{I}_{{0}}(2,2,1){m_{{c}}}^{2}m_{{b}}+20\,\hat{I}_{0}^{[0,1]}(2,3,1){m_{{c}}}%
^{2}m_{{b}} \\
&&+15\,\hat{I}_{{0}}(2,1,2)m_{{c}}{m_{{b}}}^{2}+5\,\hat{I}_{{0}}(3,1,1)m_{{c}}{m_{{b}}}%
^{2}-20\,\hat{I}_{0}^{[0,1]}(3,1,2)m_{{c}}{m_{{b}}}^{2}-20\,\hat{I}_{{6}}(2,2,2)m_{{c}}{%
m_{{b}}}^{2} \\
&&-10\,\hat{I}_{{0}}(2,2,1)m_{{c}}{m_{{b}}}^{2}+5\,\hat{I}_{0}^{[0,2]}(3,2,2)m_{{c}}{m_{{%
b}}}^{2}.
\end{eqnarray*}
\begin{eqnarray*}
C_{1}^{4} &=&-40\,\hat{I}_{4}^{[0,1]}(2,3,1)m_{{b}}+20\,\,\hat{I}_{4}^{[0,2]}(3,2,2)m_{{b%
}}-40\,\hat{I}_{{3}}(2,2,1)m_{{b}}-20\,\hat{I}_{{1}}(1,2,2)m_{{b}} \\
&&-20\,\hat{I}_{3}^{[0,1]}(2,2,2)m_{{b}}-20\,\hat{I}_{{3}}(1,2,2)m_{{b}}-20\,\hat{I}_{{4}%
}(1,2,2)m_{{b}}-10\,\hat{I}_{1}^{[0,1]}(2,3,1)m_{{b}} \\
&&-15\,\hat{I}_{{1}}(3,2,2){m_{{c}}}^{5}-45\,\hat{I}_{{1}}(3,2,1){m_{{c}}}^{3}-20\,\hat{I}_{{4}%
}(3,1,2){m_{{c}}}^{3}-20\,\hat{I}_{{2}}(3,2,1){m_{{c}}}^{3} \\
&&-45\,\hat{I}_{{1}}(4,1,1){m_{{c}}}^{3}-20\,\hat{I}_{{4}}(3,2,2){m_{{c}}}^{5}-5\,\hat{I}_{{0}%
}(3,1,2){m_{{c}}}^{3}-40\,\hat{I}_{{3}}(2,2,2){m_{{c}}}^{3} \\
&&-15\,\hat{I}_{{0}}(4,1,1){m_{{c}}}^{3}-5\,\hat{I}_{{2}}(3,1,2){m_{{c}}}^{3}-10\,\hat{I}_{{0}%
}(2,2,2){m_{{c}}}^{3}-20\,\hat{I}_{{1}}(3,1,2){m_{{c}}}^{3} \\
&&-20\,\hat{I}_{{3}}(3,2,2){m_{{c}}}^{2}{m_{{b}}}^{3}+20\,\hat{I}_{{4}}(2,2,2){m_{{c}}}%
^{2}m_{{b}}-20\,\hat{I}_{{0}}(2,3,1){m_{{c}}}^{2}m_{{b}}+40\,\hat{I}_{{4}}(3,2,1){m_{{c}}%
}^{2}m_{{b}} \\
&&+20\,\hat{I}_{{3}}(3,1,2){m_{{c}}}^{2}m_{{b}}+20\,\hat{I}_{{3}}(2,2,2){m_{{c}}}^{2}m_{{%
b}}+5\,\hat{I}_{{1}}(3,2,2){m_{{c}}}^{4}m_{{b}}+20\,\hat{I}_{{3}}(3,2,2){m_{{c}}}^{4}m_{{%
b}} \\
&&+15\,\hat{I}_{{1}}(3,2,2){m_{{c}}}^{3}{m_{{b}}}^{2}+20\,\hat{I}_{{3}}(3,2,2){m_{{c}}}%
^{3}{m_{{b}}}^{2}-20\,\hat{I}_{{4}}(3,2,2){m_{{c}}}^{2}{m_{{b}}}^{3}-5\,\hat{I}_{{1}%
}(3,2,2){m_{{c}}}^{2}{m_{{b}}}^{3} \\
&&-50\,\hat{I}_{{1}}(2,3,1){m_{{c}}}^{2}m_{{b}}-10\,\hat{I}_{1}^{[0,1]}(3,2,2){m_{{c}}}%
^{2}m_{{b}}+35\,\hat{I}_{{1}}(3,2,1){m_{{c}}}^{2}m_{{b}}+20\,\hat{I}_{{4}}(3,1,2){m_{{c}}%
}^{2}m_{{b}} \\
&&+40\,\hat{I}_{{3}}(2,3,1){m_{{b}}}^{3}+20\,\hat{I}_{{3}}(2,2,2){m_{{b}}}%
^{3}-5\,\hat{I}_{1}^{[0,1]}(3,2,2){m_{{b}}}^{3}+40\,\hat{I}_{{4}}(2,3,1){m_{{b}}}^{3} \\
&&+20\,\hat{I}_{{4}}(3,2,1){m_{{b}}}^{3}+10\,\hat{I}_{{1}}(2,3,1){m_{{b}}}^{3}-30\,\hat{I}_{{1}%
}(1,4,1){m_{{b}}}^{3}-40\,\hat{I}_{{4}}(2,2,2){m_{{c}}}^{3} \\
&&-20\,\hat{I}_{{4}}(2,2,2)m_{{c}}{m_{{b}}}^{2}-30\,\hat{I}_{{1}}(3,2,1)m_{{c}}{m_{{b}}}%
^{2}+90\,\hat{I}_{{1}}(1,4,1)m_{{c}}{m_{{b}}}^{2}+120\,\hat{I}_{{3}}(1,4,1)m_{{3}}{m_{{b}%
}}^{2} \\
&&+40\,\hat{I}_{{3}}(3,1,1)m_{{c}}-5\,\hat{I}_{{0}}(2,2,1)m_{{c}}+10%
\,\hat{I}_{0}^{[0,1]}(2,2,2)m_{{c}}+20\,\hat{I}_{{4}}(2,1,2)m_{{c}}+40%
\,\hat{I}_{3}^{[0,1]}(3,2,1)m_{{c}} \\
&&+40\,\hat{I}_{4}^{[0,1]}(2,2,2)m_{{c}}+5\,\hat{I}_{{2}}(3,1,1)m_{{c}}+20\,\hat{I}_{{4}%
}(2,2,2){m_{{b}}}^{3}+30\,\hat{I}_{{0}}(1,4,1)m_{{c}}{m_{{b}}}^{2} \\
&&+30\,\hat{I}_{{2}}(1,4,1)m_{{c}}{m_{{b}}}^{2}-20\,\hat{I}_{{4}}(3,2,1)m_{{c}}{m_{{b}}}%
^{2}+15\,\hat{I}_{1}^{[0,1]}(3,2,2)m_{{c}}{m_{{b}}}^{2}-10\,\hat{I}_{{1}}(2,2,2)m_{{c}}{%
m_{{b}}}^{2} \\
&&-5\,\hat{I}_{2}^{[0,2]}(3,2,2)m_{{c}}+5\,\hat{I}_{{1}}(2,2,1)m_{{c}}+40%
\,\hat{I}_{4}^{[0,1]}(3,2,1)m_{{c}}+10\,\hat{I}_{2}^{[0,1]}(3,2,1)m_{{c}} \\
&&-5\,\hat{I}_{0}^{[0,2]}(3,2,2)m_{{c}}+40\,\hat{I}_{3}^{[0,1]}(2,2,2)m_{{c}}+20\,\hat{I}_{{3}%
}(2,1,2)m_{{c}}-15\,\hat{I}_{{0}}(2,1,2)m_{{c}} \\
&&+20\,\hat{I}_{3}^{[0,2]}(3,2,2)m_{{b}}-40\,\hat{I}_{{3}}(1,3,1)m_{{b}%
}-40\,\hat{I}_{4}^{[0,1]}(3,1,2)m_{{b}}+10\,\hat{I}_{{1}}(1,3,1)m_{{b}} \\
&&+10\,\hat{I}_{{0}}(1,3,1)m_{{b}}-20\,\hat{I}_{4}^{[0,1]}(3,2,1)m_{{b}%
}-20\,\hat{I}_{3}^{[0,2]}(3,2,2)m_{{c}}-10\,\hat{I}_{{0}}(3,1,1)m_{{c}}.
\end{eqnarray*}
\begin{eqnarray*}
C_{2}^{4} &=&15\,\hat{I}_{{2}}(4,1,1){m_{{c}}}^{2}m_{{b}}-40\,\hat{I}_{3}^{[0,1]}(3,2,2){%
m_{{c}}}^{2}m_{{b}}-40\,\hat{I}_{{4}}(3,2,1){m_{{c}}}^{2}m_{{b}}-10\,\hat{I}_{{2}}(2,3,1)%
{m_{{c}}}^{2}m_{{b}} \\
&&+40\,\hat{I}_{4}^{[0,1]}(3,2,2){m_{{c}}}^{2}m_{{b}}-60\,\hat{I}_{{4}}(4,1,1){m_{{c}}}%
^{2}m_{{b}}+40\,\hat{I}_{{4}}(2,3,1){m_{{c}}}^{2}m_{{b}}+20\,\hat{I}_{{3}}(2,2,2){m_{{c}}%
}^{2}m_{{b}} \\
&&-20\,\hat{I}_{{4}}(3,1,2){m_{{c}}}^{2}m_{{b}}+10\,\hat{I}_{2}^{[0,1]}(3,2,2){m_{{c}}}%
^{3}+60\,\hat{I}_{{4}}(4,1,1){m_{{c}}}^{3}-20\,\hat{I}_{{3}}(3,1,2){m_{{c}}}^{3} \\
&&-15\,\hat{I}_{{2}}(4,1,1){m_{{c}}}^{3}-5\,\hat{I}_{{2}}(3,2,1){m_{{c}}}%
^{3}+10\,\hat{I}_{1}^{[0,1]}(3,2,2){m_{{c}}}^{3}+5\,\hat{I}_{{0}}(3,1,2){m_{{c}}}^{3} \\
&&-5\,\hat{I}_{{1}}(3,1,2){m_{{c}}}^{3}+15\,\hat{I}_{{0}}(4,1,1){m_{{c}}}^{3}-20\,\hat{I}_{{1}%
}(3,2,1){m_{{c}}}^{3}+20\,\hat{I}_{4}^{[0,1]}(3,2,2){m_{{b}}}^{3} \\
&&-20\,\hat{I}_{4}^{[0,1]}(3,2,2)m_{{c}}{m_{{b}}}^{2}+5\,\hat{I}_{{2}}(3,1,2)m_{{c}}{m_{{%
b}}}^{2}-20\,\hat{I}_{{3}}(3,2,1)m_{{c}}{m_{{b}}}^{2}+20\,\hat{I}_{{4}}(2,2,2)m_{{c}}{m_{%
{b}}}^{2} \\
&&-10\,\hat{I}_{{2}}(2,2,2)m_{{c}}{m_{{b}}}^{2}-30\,\hat{I}_{{0}}(1,4,1)m_{{c}}{m_{{b}}}%
^{2}+120\,\hat{I}_{{3}}(1,4,1)m_{{c}}{m_{{b}}}^{2}+5\,\hat{I}_{1}^{[0,1]}(3,2,2)m_{{c}}{%
m_{{b}}}^{2} \\
&&+20\,\hat{I}_{{4}}(3,2,1)m_{{c}}{m_{{b}}}^{2}-10\,\hat{I}_{{1}}(3,2,1)m_{{c}}{m_{{b}}}%
^{2}+30\,\hat{I}_{{2}}(1,4,1)m_{{c}}{m_{{b}}}^{2}+20\,\hat{I}_{{3}}(2,2,2){m_{{b}}}^{3}\\
&&-20\,\hat{I}_{{4}}(3,2,1){m_{{b}}}^{3}+10\,\hat{I}_{{2}}(2,3,1){m_{{b}}}^{3}+10\,\hat{I}_{{2}%
}(3,2,1){m_{{b}}}^{3}-120\,\hat{I}_{{3}}(1,4,1){m_{{b}}}^{3} \\
&&+5\,\hat{I}_{{0}}(2,2,1)m_{{c}}+15\,\hat{I}_{1}^{[0,1]}(3,1,2)m_{{c}}-5\,\hat{I}_{2}^{[0,2]%
}(3,2,2)m_{{c}}-15\,\hat{I}_{0}^{[0,1]}(3,2,1)m_{{c}} \\
&&+40\,\hat{I}_{{3}}(3,1,1)m_{{c}}+5\,\hat{I}_{{1}}(3,1,1)m_{{c}}-20%
\,\hat{I}_{4}^{[0,1]}(3,1,2)m_{{c}}+10\,\hat{I}_{{0}}(3,1,1)m_{{c}} \\
&&+10\,\hat{I}_{1}^{[0,1]}(2,2,2)m_{{c}}+20\,\,\hat{I}_{3}^{[0,1]}(3,1,2)m_{{c}}-20\,\hat{I}_{{%
4}}(2,1,2)m_{{c}}+20\,\hat{I}_{{3}}(2,1,2)m_{{c}} \\
&&+5\,\hat{I}_{{2}}(2,2,1)m_{{c}}+10\,\hat{I}_{2}^{[0,1]}(2,2,2)m_{{c}}-15\,\hat{I}_{{1}%
}(2,1,2)m_{{c}}-15\,\hat{I}_{0}^{[0,1]}(3,1,2)m_{{c}} \\
&&-40\,\hat{I}_{3}^{[0,1]}(3,1,2)m_{{b}}+40\,\hat{I}_{4}^{[0,1]}(3,1,2)m_{{b}}-20\,\hat{I}_{{2}%
}(2,1,2)m_{{b}}+20\,\hat{I}_{4}^{[0,1]}(3,2,1)m_{{b}} \\
&&+5\,\hat{I}_{2}^{[0,2]}(3,2,2)m_{{b}}+10\,\hat{I}_{{0}}(2,2,1)m_{{b}}-20\,\hat{I}_{{3}%
}(2,1,2)m_{{b}}-10\,\hat{I}_{{1}}(2,2,1)m_{{b}} \\
&&-20\,\hat{I}_{{2}}(1,2,2)m_{{b}}+20\,\hat{I}_{{4}}(2,1,2)m_{{b}}-40\,\hat{I}_{{3}}(3,1,1)m_{{%
b}}-10\,\hat{I}_{{2}}(1,3,1)m_{{b}} \\
&&+20\,\hat{I}_{3}^{[0,2]}(3,2,2)m_{{b}}-40\,\hat{I}_{{3}}(2,2,1)m_{{b}%
}+20\,\hat{I}_{4}^{[0,1]}(2,2,2)m_{{b}}+40\,\hat{I}_{4}^{[0,1]}(2,3,1)m_{{b}} \\
&&+10\,\hat{I}_{{2}}(3,1,2){m_{{c}}}^{2}m_{{b}}+20\,\hat{I}_{{4}}(3,2,2){m_{{c}}}%
^{5}+5\,\hat{I}_{{0}}(3,2,2){m_{{c}}}^{5}-5\,\hat{I}_{{2}}(3,2,2){m_{{c}}}^{5} \\
&&+40\,\hat{I}_{{4}}(3,2,1){m_{{c}}}^{3}+40\,\hat{I}_{3}^{[0,1]}(3,2,2){m_{{c}}}%
^{3}+20\,\hat{I}_{{4}}(3,1,2){m_{{c}}}^{3}+5\,\hat{I}_{{2}}(3,2,2){m_{{c}}}^{4}m_{{b}} \\
&&+20\,\hat{I}_{{3}}(3,2,2){m_{{c}}}^{3}{m_{{b}}}^{2}-20\,\hat{I}_{{4}}(3,2,2){m_{{c}}}%
^{3}{m_{{b}}}^{2}+5\,\hat{I}_{{2}}(3,2,2){m_{{c}}}^{3}{m_{{b}}}^{2}-5\,\hat{I}_{{2}%
}(3,2,2){m_{{c}}}^{2}{m_{{b}}}^{3}.
\end{eqnarray*}
where
\begin{eqnarray}
\hat{I}_n^{[i,j]} (a,b,c) = \left( M_1^2 \right)^i \left( M_2^2
\right)^j
\frac{d^i}{d\left( M_1^2 \right)^i} \frac{d^j}{d\left( M_2^2 \right)^j} %
\left[\left( M_1^2 \right)^i \left( M_2^2 \right)^j \hat{I}_n(a,b,c) \right]%
~.  \nonumber
\end{eqnarray}
\newpage

\appendix
\begin{center}
{\Large \textbf{Appendix--B}}
\end{center}

\setcounter{equation}{0} \renewcommand{\theequation}

In this appendix, the explicit expressions of the coefficients of
the gluon condensate  entering  the HQET limit of the  form
factors $f_{V}^{HQ}, f_{0}^{HQ}, f_{1}^{HQ}$ and $f_{2}^{HQ}$ are
given.
\begin{eqnarray*}
C_{V}^{HQET} &=&10\,{\frac{\bar{I}_{2}^{[0,1]}(3,2,1)}{\sqrt{Z}}}-10\,{\frac{%
\bar{I}_{2}^{[0,2]}(3,2,2)}{\sqrt{Z}}}+20\,{\frac{\bar{I}_{{0}}(2,2,1)}{\sqrt{Z}}}-20\,{%
\frac{\bar{I}_{{2}}(1,2,2)}{\sqrt{Z}}} \\
&&+20\,{\frac{\bar{I}_{0}^{[0,1]}(2,2,2)}{\sqrt{Z}}}-20\,{\frac{\bar{I}_{{1}}(2,1,2)}{%
\sqrt{Z}}}+30\,{\frac{\bar{I}_{0}^{[0,1]}(3,1,2)}{\sqrt{Z}}}-30\,{\frac{\bar{I}_{{2}%
}(2,1,2)}{\sqrt{Z}}} \\
&&+20\,{\frac{\bar{I}_{1}^{[0,2]}(3,2,1)}{\sqrt{Z}}}-30\,{\frac{\bar{I}_{{0}}(2,1,2)}{%
\sqrt{Z}}}-10\,{\frac{\bar{I}_{1}^{[0,2]}(3,2,2)}{\sqrt{Z}}}-10\,{\frac{%
\bar{I}_{0}^{[0,2]}(3,2,2)}{\sqrt{Z}}} \\
&&+20\,{\frac{\bar{I}_{2}^{[0,1]}(2,2,2)}{\sqrt{Z}}}+20\,{\frac{%
\bar{I}_{1}^{[0,1]}(3,1,2)}{\sqrt{Z}}}+20\,{\frac{\bar{I}_{1}^{[0,1]}(2,2,2)}{\sqrt{Z}}}%
+20\,{\frac{\bar{I}_{0}^{[0,1]}(3,2,1)}{\sqrt{Z}}} \\
&&+30\,{\frac{\bar{I}_{{2}}(2,2,1)}{\sqrt{Z}}}+20\,{\frac{\bar{I}_{{1}}(2,2,1)}{\sqrt{Z}}%
}-10\,{\frac{\bar{I}_{{1}}(3,1,1)}{\sqrt{Z}}}-20\,{\frac{\bar{I}_{{0}}(1,2,2)}{\sqrt{Z}}}
\\
&&+10\,{\frac{\bar{I}_{{2}}(3,1,1)}{\sqrt{Z}}}-20\,{\frac{\bar{I}_{{1}}(1,2,2)}{\sqrt{Z}}%
}+30\,{\frac{\bar{I}_{2}^{[0,1]}(3,1,2)}{\sqrt{Z}}}-10\,{\frac{\bar{I}_{{0}}(3,1,1)}{%
\sqrt{Z}}} \\
&&+60\,\bar{I}_{{2}}(1,3,1)+20\,\bar{I}_{{1}}(1,3,1)-20\,\bar{I}_{1}^{[0,1]}(3,2,1)+10%
\,\bar{I}_{1}^{[0,2]}(3,2,2) \\
&&+20\,\bar{I}_{{1}}(1,2,2)+20\,\bar{I}_{1}^{[0,1]}(2,3,1)-20\,\bar{I}_{1}^{[0,1]}(2,2,2)+100%
\,\bar{I}_{{0}}(1,3,1) \\
&&-50\,\bar{I}_{{1}}(2,2,1)-20\,\bar{I}_{{2}}(2,2,1)-20\,\bar{I}_{{0}}(2,2,1)-20%
\,\bar{I}_{1}^{[0,1]}(3,1,2) \\
&&+30\,\bar{I}_{{1}}(2,1,2)+40\,\bar{I}_{2}^{[0,1]}(2,3,1).
\end{eqnarray*}
\begin{eqnarray*}
C_{0}^{HQET}&=&-5\,{\frac{\bar{I}_{0}^{[0,1]}(3,2,1)}{\sqrt{Z}}}-20\,{\frac{%
\bar{I}_{0}^{[0,1]}(3,1,2)}{\sqrt{Z}}}+120\,{\frac{\bar{I}_{6}^{[0,1]}(1,4,1)}{\sqrt{Z}}}%
+15\,{\frac{\bar{I}_{{0}}(2,1,2)}{\sqrt{Z}}} \\
&&+10\,{\frac{\bar{I}_{{0}}(1,2,2)}{\sqrt{Z}}}-10\,{\frac{\bar{I}_{0}^{[0,1]}(2,2,2)}{%
\sqrt{Z}}}+5\,{\frac{\bar{I}_{{0}}(3,1,1)}{\sqrt{Z}}}+5\,{\frac{%
\bar{I}_{0}^{[0,2]}(3,2,2)}{\sqrt{Z}}} \\
&&-10\,{\frac{\bar{I}_{{0}}(2,2,1)}{\sqrt{Z}}}-20\,{\frac{\bar{I}_{{6}}(3,2,1)}{\sqrt{Z}}%
}-30\,{\frac{\bar{I}_{{0}}(1,3,1)}{\sqrt{Z}}}+20\,{\frac{\bar{I}_{6}^{[0,1]}(3,2,2)}{%
\sqrt{Z}}} \\
&&-20\,{\frac{\bar{I}_{{6}}(2,2,2)}{\sqrt{Z}}}+15\,{\frac{\bar{I}_{{0}}(2,2,1)}{Z}}+30\,{%
\frac{\bar{I}_{0}^{[0,1]}(2,2,2)}{Z}}+20\,{\frac{\bar{I}_{0}^{[0,1]}(2,3,1)}{Z}} \\
&&-20\,{\frac{\bar{I}_{{0}}(1,2,2)}{Z}}-15\,{\frac{\bar{I}_{{0}}(2,1,2)}{Z}}-40\,{\frac{%
\bar{I}_{{6}}(2,3,1)}{Z}}+20\,{\frac{\bar{I}_{0}^{[0,1]}(3,2,1)}{Z}} \\
&&-15\,{\frac{\bar{I}_{0}^{[0,2]}(3,2,2)}{Z}}+20\,{\frac{\bar{I}_{{6}}(3,1,2)}{Z}}+40\,{%
\frac{\bar{I}_{{6}}(3,2,1)}{Z}}+10\,{\frac{\bar{I}_{{0}}(1,3,1)}{Z}} \\
&&-10\,{\frac{\bar{I}_{{0}}(3,1,1)}{Z}}+60\,{\frac{\bar{I}_{{6}}(4,1,1)}{Z}}+15\,{\frac{%
\bar{I}_{0}^{[0,1]}(4,1,1)}{Z}}+10\,{\frac{\bar{I}_{0}^{[0,1]}(3,1,2)}{Z}} \\
&&-40\,{\frac{\bar{I}_{6}^{[0,1]}(3,2,2)}{Z}}+20\,{\frac{\bar{I}_{{6}}(2,2,2)}{Z}}-40\,{%
\frac{\bar{I}_{{6}}(2,2,2)}{{Z}^{3/2}}}+40\,{\frac{\bar{I}_{6}^{[0,1]}(3,2,2)}{{Z}^{3/2}}%
} \\
&&-40\,{\frac{\bar{I}_{{6}}(3,2,1)}{{Z}^{3/2}}}-20\,{\frac{\bar{I}_{{6}}(3,1,2)}{{Z}%
^{3/2}}}+5\,{\frac{\bar{I}_{{0}}(3,1,1)}{{Z}^{3/2}}}-60\,{\frac{\bar{I}_{{6}}(4,1,1)}{{Z}%
^{3/2}}} \\
&&-5\,\bar{I}_{{0}}(1,2,2)+20\,\bar{I}_{{6}}(2,2,2)+20\,\bar{I}_{{6}}(3,2,1)-30%
\,\bar{I}_{0}^{[0,1]}(1,4,1) \\
&&-120\,\bar{I}_{{6}}(1,4,1)-5\,\bar{I}_{0}^{[0,2]}(3,2,2)+5\,\bar{I}_{{0}}(2,2,1)-20%
\,\bar{I}_{6}^{[0,1]}(3,2,2) \\
&&+10\,\bar{I}_{0}^{[0,1]}(2,3,1)+40\,\bar{I}_{{6}}(2,3,1)-20\,\bar{I}_{{0}}(1,3,1)+10%
\,\bar{I}_{0}^{[0,1]}(2,2,2)+15\,\bar{I}_{0}^{[0,1]}(3,2,1).
\end{eqnarray*}
\begin{eqnarray*}
C_{1}^{HQET} &=&10\,{\frac{\bar{I}_{2}^{[0,1]}(2,2,2)}{\sqrt{Z}}}+10\,{\frac{%
\bar{I}_{0}^{[0,1]}(2,2,2)}{\sqrt{Z}}}+45\,{\frac{\bar{I}_{1}^{[0,1]}(3,2,1)}{\sqrt{Z}}}%
+20\,{\frac{\bar{I}_{{4}}(2,2,1)}{\sqrt{Z}}} \\
&&+40\,{\frac{\bar{I}_{{4}}(3,1,1)}{\sqrt{Z}}}-5\,{\frac{\bar{I}_{{0}}(2,2,1)}{\sqrt{Z}}}%
+40\,{\frac{\bar{I}_{4}^{[0,1]}(2,2,2)}{\sqrt{Z}}}+40\,{\frac{\bar{I}_{3}^{[0,1]}(3,2,1)%
}{\sqrt{Z}}} \\
&&-5\,{\frac{\bar{I}_{0}^{[0,2]}(3,2,2)}{\sqrt{Z}}}+40\,{\frac{\bar{I}_{3}^{[0,1]}(2,2,2)%
}{\sqrt{Z}}}+20\,{\frac{\bar{I}_{{3}}(2,1,2)}{\sqrt{Z}}}+20\,{\frac{%
\bar{I}_{4}^{[0,1]}(3,1,2)}{\sqrt{Z}}} \\
&&+20\,{\frac{\bar{I}_{{3}}(2,2,1)}{\sqrt{Z}}}+40\,{\frac{\bar{I}_{4}^{[0,1]}(3,2,1)}{%
\sqrt{Z}}}+20\,{\frac{\bar{I}_{{4}}(2,1,2)}{\sqrt{Z}}}+5\,{\frac{\bar{I}_{{2}}(3,1,1)}{%
\sqrt{Z}}} \\
&&+5\,{\frac{\bar{I}_{{1}}(2,2,1)}{\sqrt{Z}}}-20\,{\frac{\bar{I}_{1}^{[0,2]}(3,2,2)}{%
\sqrt{Z}}}-10\,{\frac{\bar{I}_{{0}}(3,1,1)}{\sqrt{Z}}}+20\,{\frac{\bar{I}_{{1}}(3,1,1)}{%
\sqrt{Z}}} \\
&&-20\,{\frac{\bar{I}_{3}^{[0,2]}(3,2,2)}{\sqrt{Z}}}-20\,{\frac{%
\bar{I}_{4}^{[0,2]}(3,2,2)}{\sqrt{Z}}}+20\,{\frac{\bar{I}_{3}^{[0,1]}(3,1,2)}{\sqrt{Z}}}%
+15\,{\frac{\bar{I}_{2}^{[0,1]}(3,1,2)}{\sqrt{Z}}} \\
&&+10\,{\frac{\bar{I}_{2}^{[0,1]}(3,2,1)}{\sqrt{Z}}}-15\,{\frac{\bar{I}_{{2}}(2,1,2)}{%
\sqrt{Z}}}+40\,{\frac{\bar{I}_{1}^{[0,1]}(2,2,2)}{\sqrt{Z}}}-5\,{\frac{%
\bar{I}_{2}^{[0,2]}(3,2,2)}{\sqrt{Z}}} \\
&&-25\,{\frac{\bar{I}_{{1}}(2,1,2)}{\sqrt{Z}}}+15\,{\frac{\bar{I}_{0}^{[0,1]}(3,1,2)}{%
\sqrt{Z}}}+40\,{\frac{\bar{I}_{{3}}(3,1,1)}{\sqrt{Z}}}+15\,{\frac{%
\bar{I}_{0}^{[0,1]}(3,2,1)}{\sqrt{Z}}} \\
&&+45\,{\frac{\bar{I}_{1}^{[0,1]}(3,1,2)}{\sqrt{Z}}}-15\,{\frac{\bar{I}_{{0}}(2,1,2)}{%
\sqrt{Z}}}-40\,\bar{I}_{3}^{[0,1]}(2,3,1)-20\,\bar{I}_{4}^{[0,1]}(3,2,1) \\
&&+10\,\bar{I}_{{2}}(1,3,1)-40\,\bar{I}_{{4}}(2,2,1)-20\,\bar{I}_{{3}}(1,2,2)-20\,\bar{I}_{{4}%
}(1,2,2) \\
&&+10\,\bar{I}_{{0}}(1,3,1)-10\,\bar{I}_{{0}}(2,2,1)-40\,\bar{I}_{{4}}(1,3,1)-20\,\bar{I}_{{1}%
}(1,2,2) \\
&&-40\,\bar{I}_{{3}}(1,3,1)-20\,\bar{I}_{3}^{[0,1]}(3,2,1)+20\,\bar{I}_{{1}}(1,3,1)-40%
\,\bar{I}_{4}^{[0,1]}(3,1,2) \\
&&-10\,\bar{I}_{{2}}(2,2,1)-40\,\bar{I}_{4}^{[0,1]}(2,3,1)-20\,\bar{I}_{4}^{[0,1]}(2,2,2)-10%
\,\bar{I}_{1}^{[0,1]}(2,3,1) \\
&&-40\,\bar{I}_{{1}}(2,2,1)-20\,\bar{I}_{3}^{[0,1]}(2,2,2)+5\,\bar{I}_{1}^{[0,2]}(3,2,2)-40%
\,\bar{I}_{3}^{[0,1]}(3,1,2) \\
&&-40\,\bar{I}_{{3}}(3,1,1)-40\,\bar{I}_{{4}}(3,1,1)-20\,\bar{I}_{{1}}(2,1,2)-10%
\,\bar{I}_{1}^{[0,1]}(3,1,2) \\
&&-10\,\bar{I}_{1}^{[0,1]}(3,2,1)-20\,\bar{I}_{{4}}(2,1,2)+20\,\bar{I}_{3}^{[0,2]}(3,2,2)+20%
\,\bar{I}_{4}^{[0,2]}(3,2,2) \\
&&-20\,\bar{I}_{{3}}(2,1,2)-40\,\bar{I}_{{3}}(2,2,1).
\end{eqnarray*}
\begin{eqnarray*}
C_{2}^{HQET} &=&20\,{\frac{\bar{I}_{{3}}(2,1,2)}{\sqrt{Z}}}+15\,{\frac{%
\bar{I}_{1}^{[0,1]}(3,1,2)}{\sqrt{Z}}}-20\,{\frac{\bar{I}_{{4}}(2,1,2)}{\sqrt{Z}}}-5\,{%
\frac{\bar{I}_{{1}}(2,2,1)}{\sqrt{Z}}} \\
&&+10\,{\frac{\bar{I}_{0}^{[0,1]}(2,2,2)}{\sqrt{Z}}}-15\,{\frac{\bar{I}_{{0}}(2,1,2)}{%
\sqrt{Z}}}-5\,{\frac{\bar{I}_{0}^{[0,2]}(3,2,2)}{\sqrt{Z}}}+20\,{\frac{%
\bar{I}_{3}^{[0,1]}(3,1,2)}{\sqrt{Z}}} \\
&&-10\,{\frac{\bar{I}_{{2}}(3,1,1)}{\sqrt{Z}}}+40\,{\frac{\bar{I}_{3}^{[0,1]}(2,2,2)}{%
\sqrt{Z}}}-10\,{\frac{\bar{I}_{{0}}(3,1,1)}{\sqrt{Z}}}-40\,{\frac{%
\bar{I}_{4}^{[0,1]}(2,2,2)}{\sqrt{Z}}} \\
&&-5\,{\frac{\bar{I}_{{0}}(2,2,1)}{\sqrt{Z}}}-5\,{\frac{\bar{I}_{1}^{[0,1]}(3,2,1)}{%
\sqrt{Z}}}-40\,{\frac{\bar{I}_{4}^{[0,1]}(3,2,1)}{\sqrt{Z}}}-20\,{\frac{%
\bar{I}_{3}^{[0,2]}(3,2,2)}{\sqrt{Z}}} \\
&&-35\,{\frac{\bar{I}_{{1}}(2,1,2)}{\sqrt{Z}}}+15\,{\frac{\bar{I}_{0}^{[0,1]}(3,1,2)}{%
\sqrt{Z}}}+30\,{\frac{\bar{I}_{{2}}(2,1,2)}{\sqrt{Z}}}-20\,{\frac{\bar{I}_{{4}}(2,2,1)}{%
\sqrt{Z}}} \\
&&+20\,{\frac{\bar{I}_{{3}}(2,2,1)}{\sqrt{Z}}}+40\,{\frac{\bar{I}_{3}^{[0,1]}(3,2,1)}{%
\sqrt{Z}}}-20\,{\frac{\bar{I}_{2}^{[0,1]}(2,2,2)}{\sqrt{Z}}}+40\,{\frac{\bar{I}_{{3}%
}(3,1,1)}{\sqrt{Z}}} \\
&&-20\,{\frac{\bar{I}_{2}^{[0,1]}(3,2,1)}{\sqrt{Z}}}-40\,{\frac{\bar{I}_{{4}}(3,1,1)}{%
\sqrt{Z}}}+15\,{\frac{\bar{I}_{0}^{[0,1]}(3,2,1)}{\sqrt{Z}}}+20\,{\frac{%
\bar{I}_{4}^{[0,2]}(3,2,2)}{\sqrt{Z}}} \\
&&-30\,{\frac{\bar{I}_{2}^{[0,1]}(3,1,2)}{\sqrt{Z}}}-20\,{\frac{%
\bar{I}_{4}^{[0,1]}(3,1,2)}{\sqrt{Z}}}+10\,{\frac{\bar{I}_{2}^{[0,2]}(3,2,2)}{\sqrt{Z}}}%
+10\,\bar{I}_{1}^{[0,1]}(3,2,1) \\
&&+40\,\bar{I}_{{4}}(3,1,1)+40\,\bar{I}_{4}^{[0,1]}(2,3,1)+40\,\bar{I}_{{4}}(1,3,1)+40%
\,\bar{I}_{4}^{[0,1]}(3,1,2) \\
&&-10\,\bar{I}_{{0}}(2,2,1)-40\,\bar{I}_{{3}}(3,1,1)+20\,\bar{I}_{{1}}(1,2,2)-40\,\bar{I}_{{3}%
}(1,3,1) \\
&&+10\,\bar{I}_{1}^{[0,1]}(3,1,2)-40\,\bar{I}_{3}^{[0,1]}(3,1,2)+20\,\bar{I}_{{2}%
}(2,2,1)+20\,\bar{I}_{{1}}(2,1,2) \\
&&-20\,\bar{I}_{{3}}(2,1,2)+20\,\bar{I}_{{4}}(2,1,2)+10\,\bar{I}_{{0}}(1,3,1)-20\,\bar{I}_{{3}%
}(1,2,2) \\
&&-20\,\bar{I}_{3}^{[0,1]}(2,2,2)+20\,\bar{I}_{4}^{[0,1]}(2,2,2)+40\,\bar{I}_{{4}%
}(2,2,1)-40\,\bar{I}_{{3}}(2,2,1) \\
&&+20\,\bar{I}_{{4}}(1,2,2)+10\,\bar{I}_{1}^{[0,1]}(2,3,1)-20\,\bar{I}_{{2}}(1,3,1)+20%
\,\bar{I}_{3}^{[0,2]}(3,2,2) \\
&&-20\,\bar{I}_{4}^{[0,2]}(3,2,2)+20\,\bar{I}_{{1}}(1,3,1)-5\,\bar{I}_{1}^{[0,2]}(3,2,2)-20%
\,\bar{I}_{3}^{[0,1]}(3,2,1) \\
&&+20\,\bar{I}_{4}^{[0,1]}(3,2,1)-40\,\bar{I}_{3}^{[0,1]}(2,3,1).
\end{eqnarray*}
where
\begin{eqnarray}
\bar{I}_n^{[i,j]} (a,b,c) =\frac{2^{i+j}}{(\sqrt{Z})^{j}} \left(
T_1 \right)^i \left( T_2 \right)^j
\frac{d^i}{d\left( T_1 \right)^i} \frac{d^j}{d\left( T_2 \right)^j} %
\left[\left( T_1 \right)^i \left( T_2 \right)^j \bar{I}_n(a,b,c) \right]%
~.  \nonumber
\end{eqnarray}
\newpage
\begin{figure}[th]
\begin{center}
\begin{picture}(0,0)
\put(-80,-90){ \epsfxsize=8cm \epsfbox{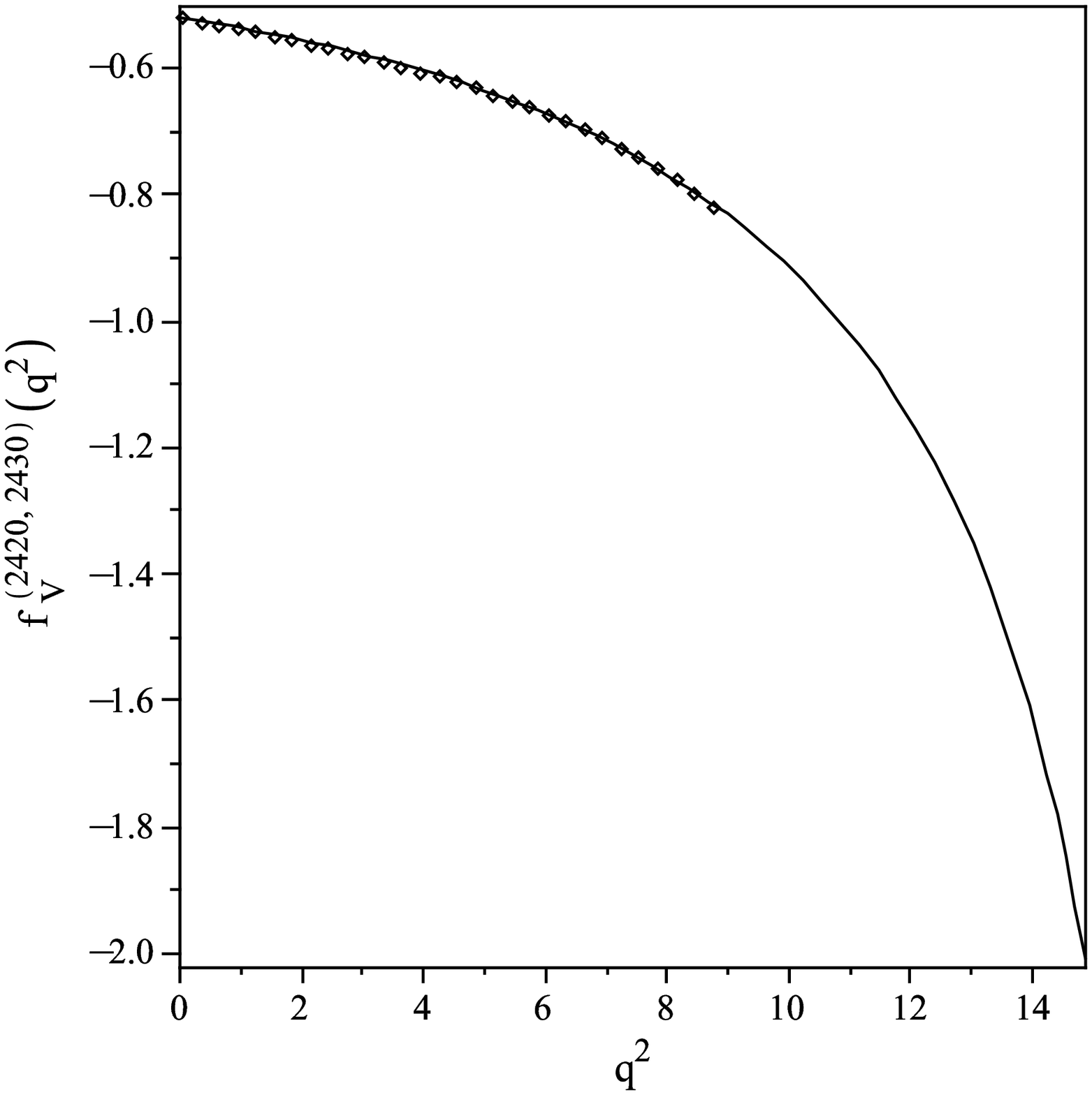} \epsfxsize=8cm
\epsfbox{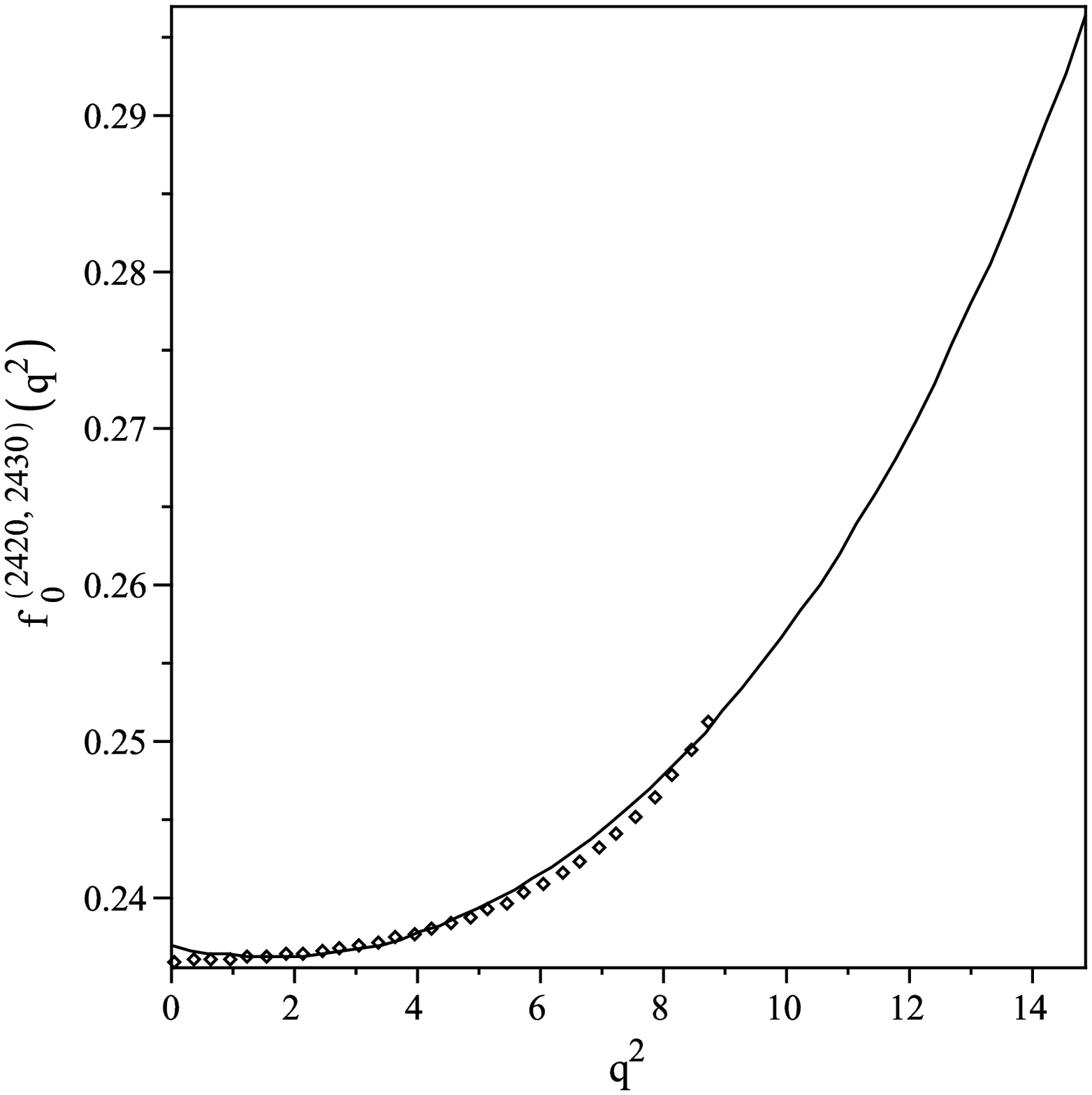} } \put(-80,-175){\epsfxsize=8cm
\epsfbox{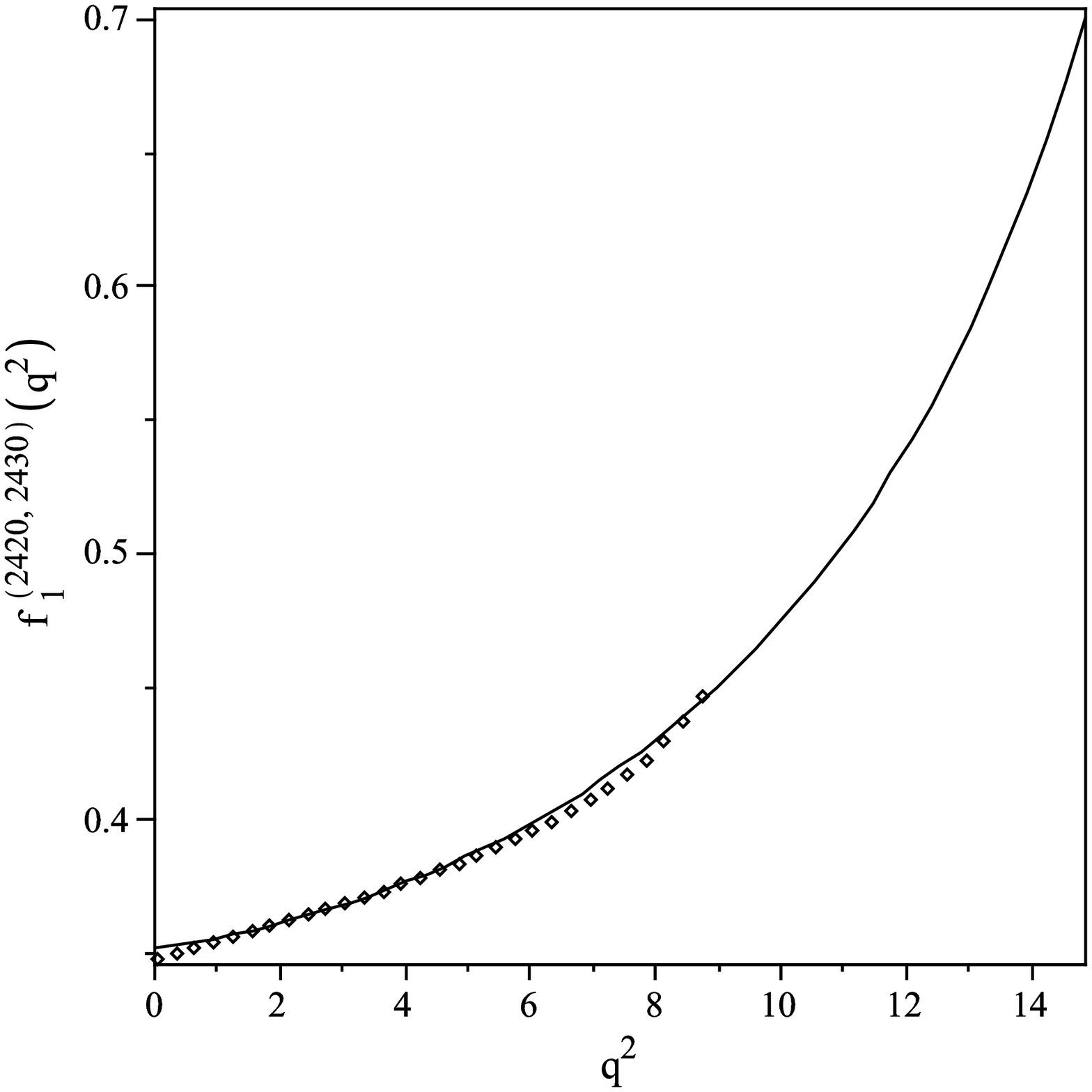} \epsfxsize=8cm \epsfbox{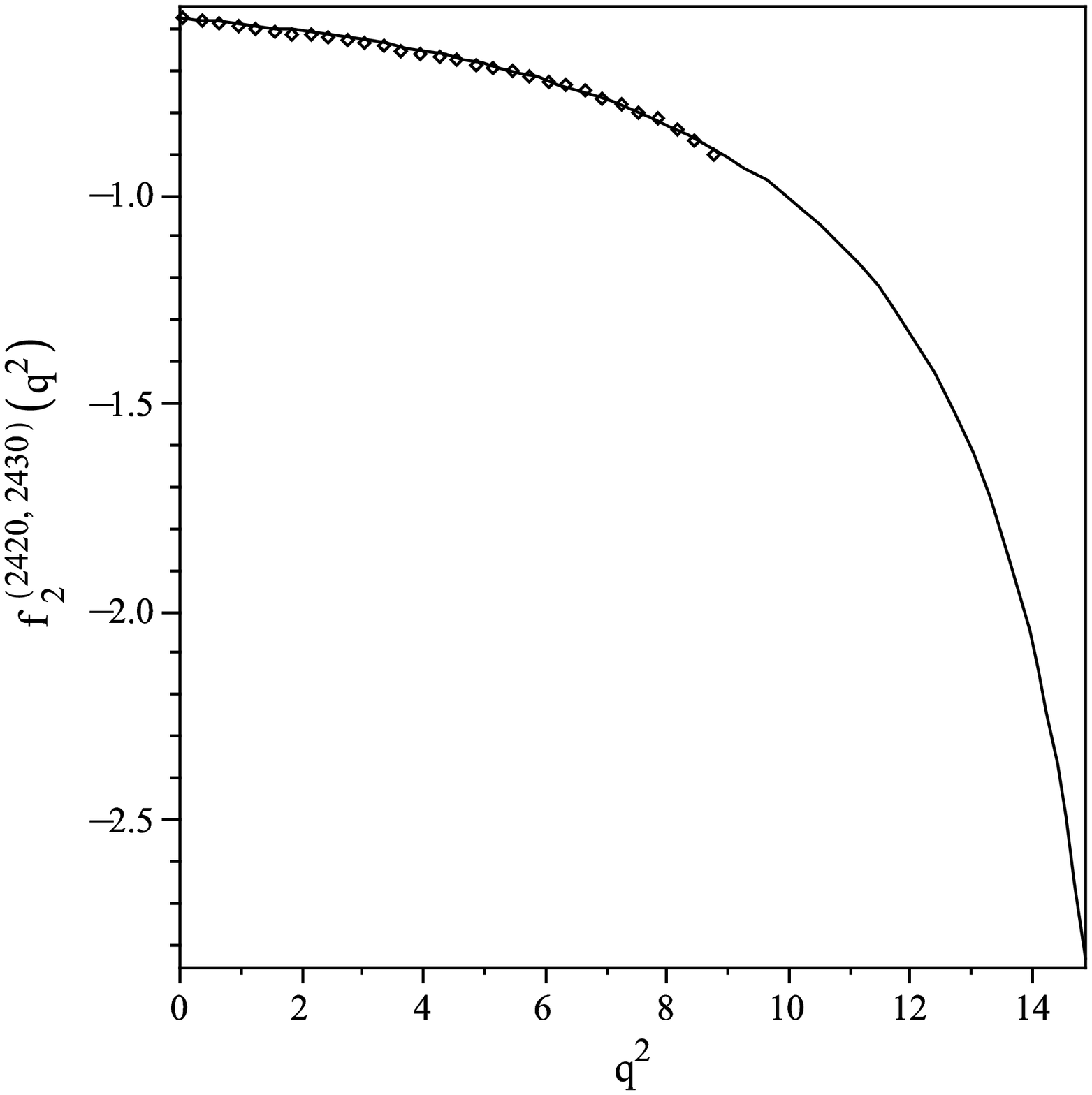} }
\end{picture}
\end{center}
\vspace*{17cm}\caption{The dependence of the form factors as well
as the fit parametrization of the form factors on $q^2$. The small
boxes correspond to the form factors, the solid lines belong to
the fit parametrization of the form factors.}\label{F3}
\end{figure}
\newpage
\begin{figure}[th]
\begin{center}
\begin{picture}(160,100)
\centerline{ \epsfxsize=10cm \epsfbox{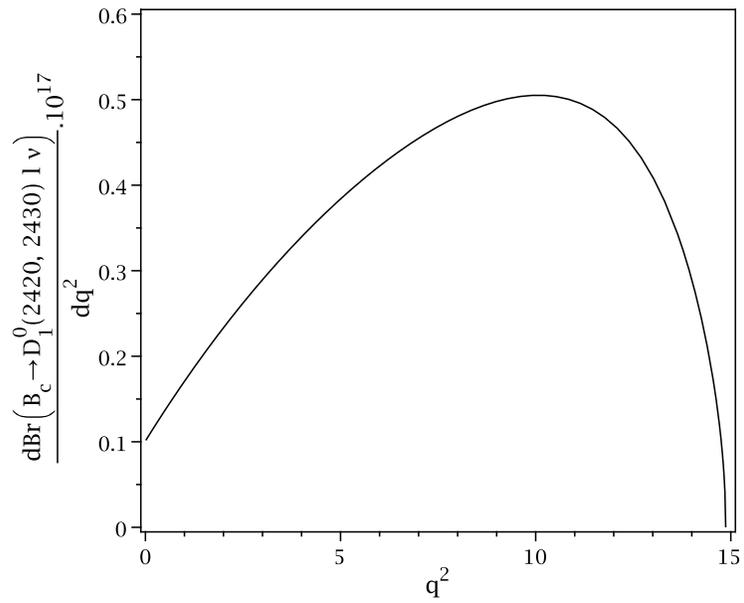}}
\end{picture}
\end{center}
\vspace*{0cm}\caption{The dependence of the decay width of the
$B_c\to D_1^0(2420,2430)$ decays on $q^2$. }\label{F4}
\end{figure}
\normalsize
\newpage
\begin{figure}[th]
\vspace*{0.4cm}
\begin{center}
\begin{picture}(160,150)
\centerline{ \epsfxsize=14cm \epsfbox{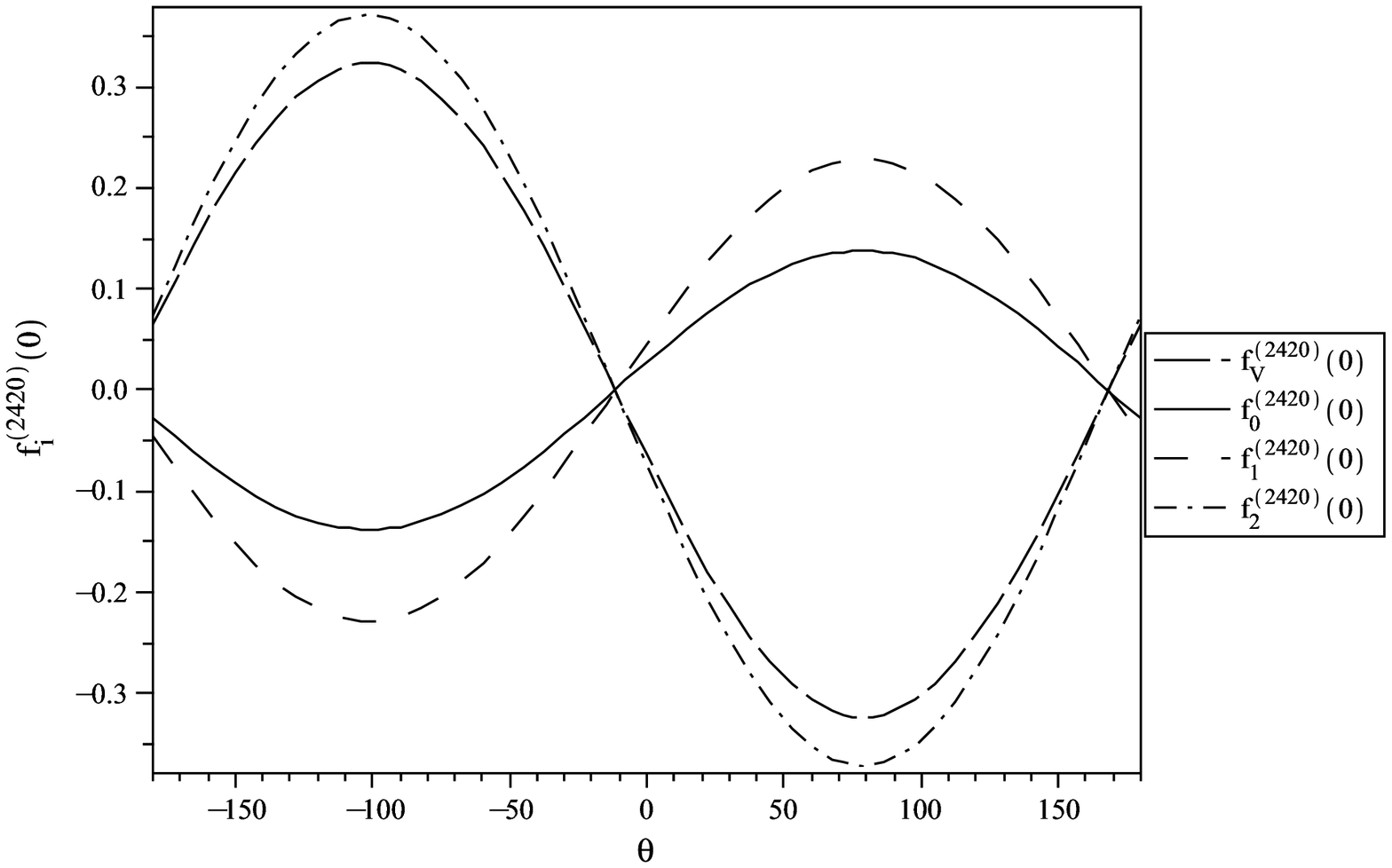}}
\end{picture}
\end{center}
\vspace*{-0.1cm} \caption{The dependence of the form factors on
$\theta$ at $q^2=0$ for the $B_c\to D_{1}^0(2420)$ transition.}
\label{F5}
\end{figure}
\newpage
\begin{figure}[th]
\vspace*{0.4cm}
\begin{center}
\begin{picture}(160,150)
\centerline{ \epsfxsize=14cm \epsfbox{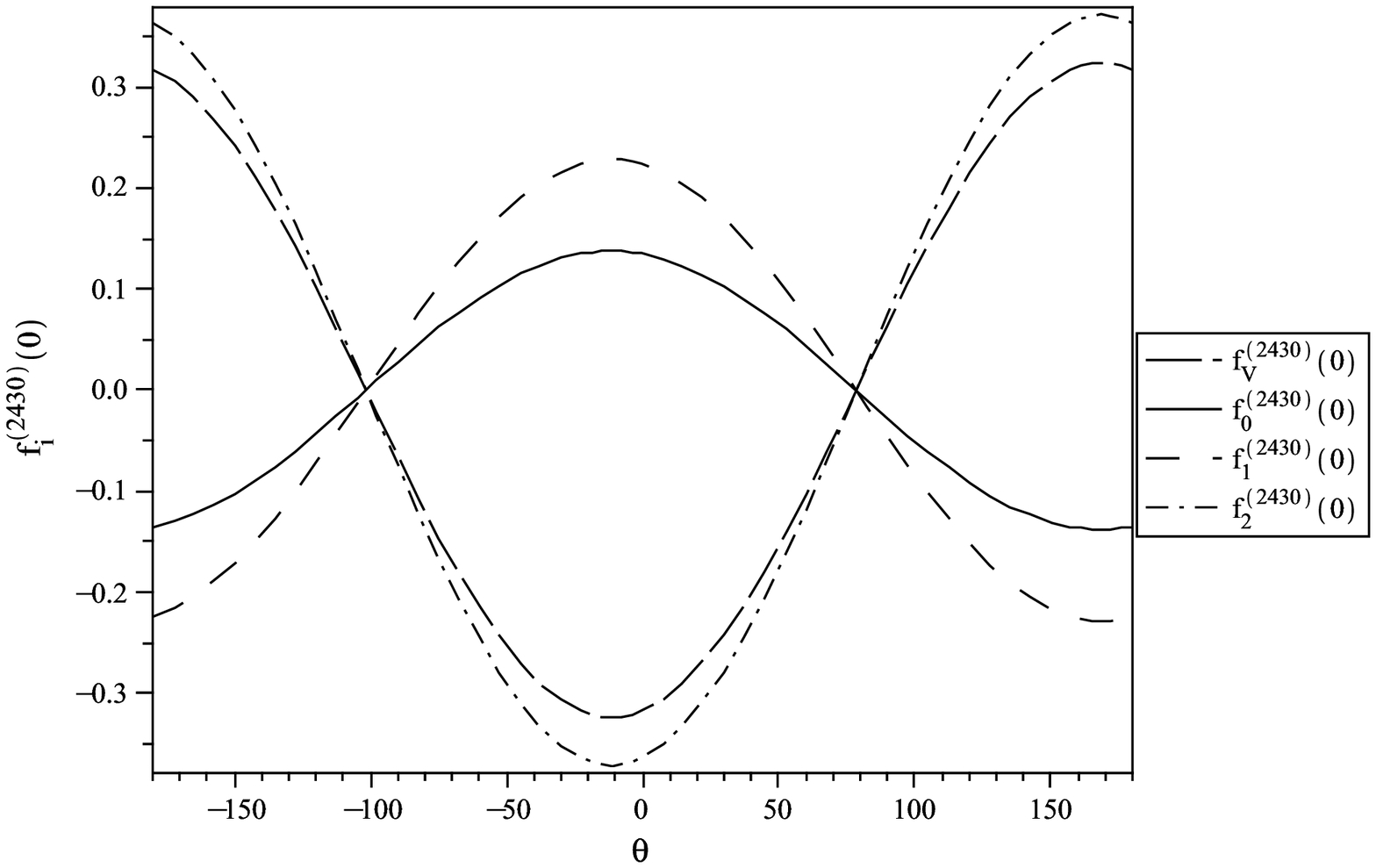}}
\end{picture}
\end{center}
\vspace*{-0.1cm} \caption{The dependence of the form factors on
$\theta$ at $q^2=0$ for the $B_c\to D_{1}^0(2430)$ transition.}
\label{F6}
\end{figure}
\newpage
\begin{figure}[th]
\begin{center}
\begin{picture}(0,0)
\put(-78,-90){ \epsfxsize=5cm \epsfbox{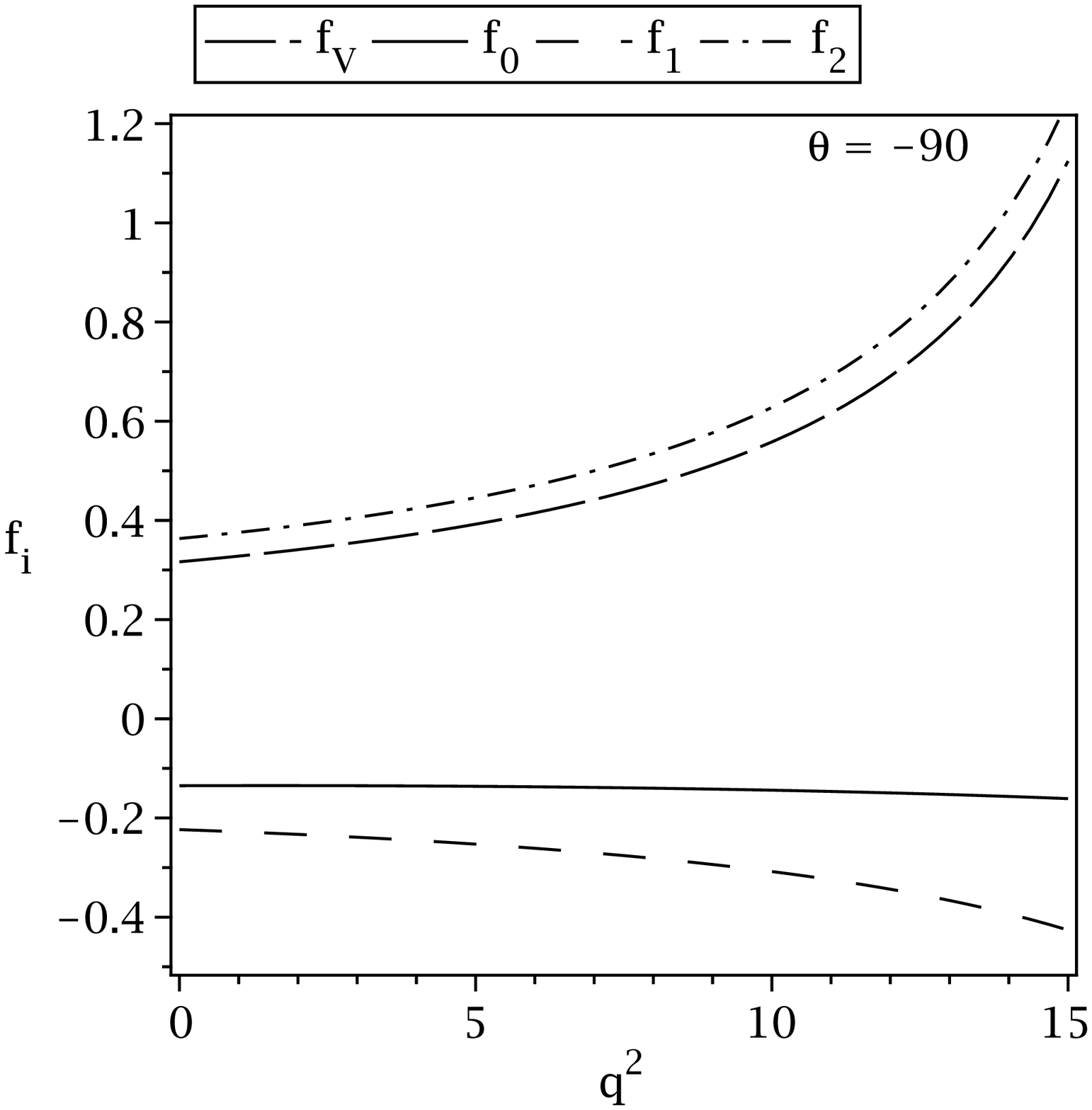} \epsfxsize=5cm
\epsfbox{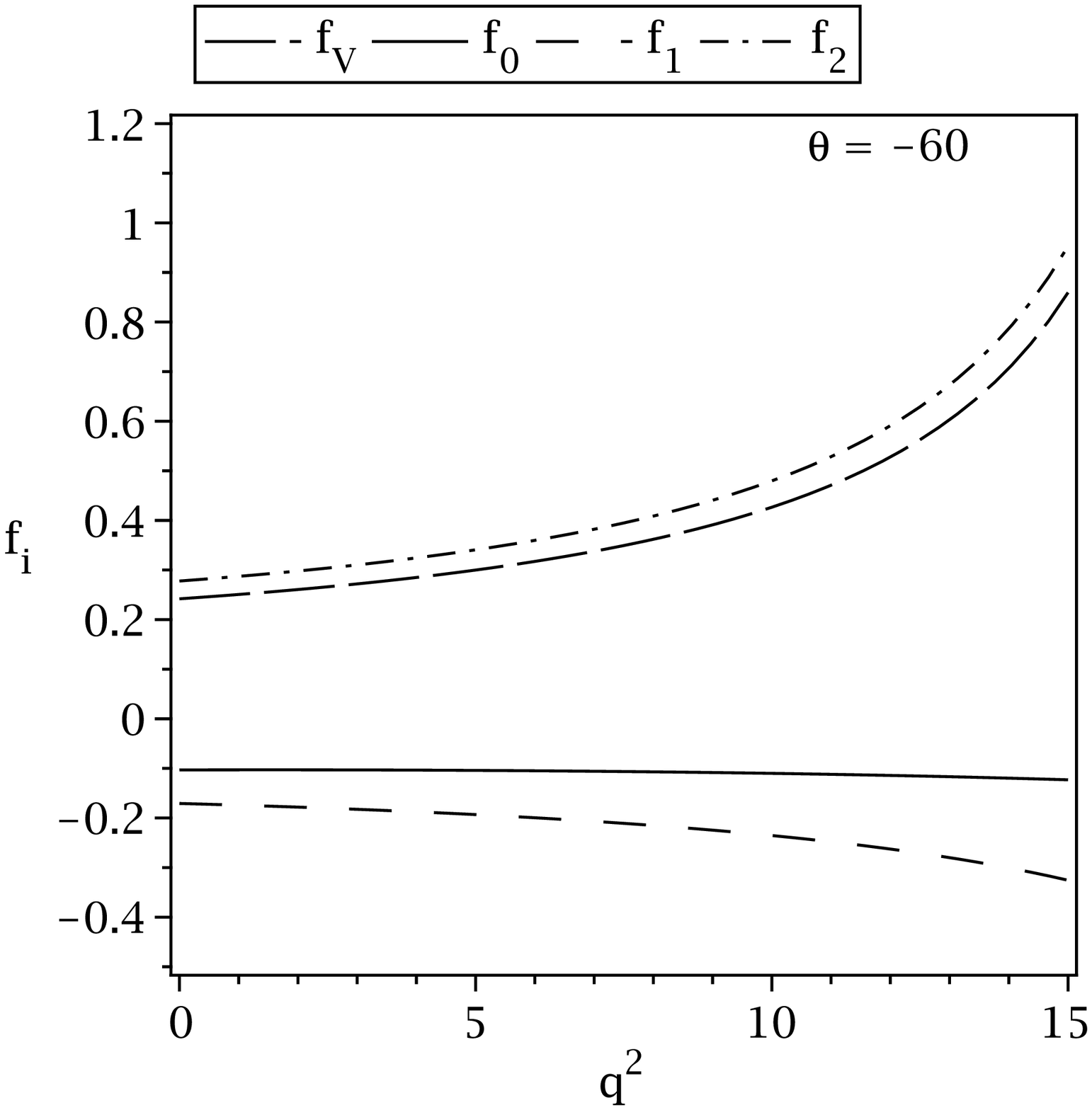}\epsfxsize=5cm \epsfbox{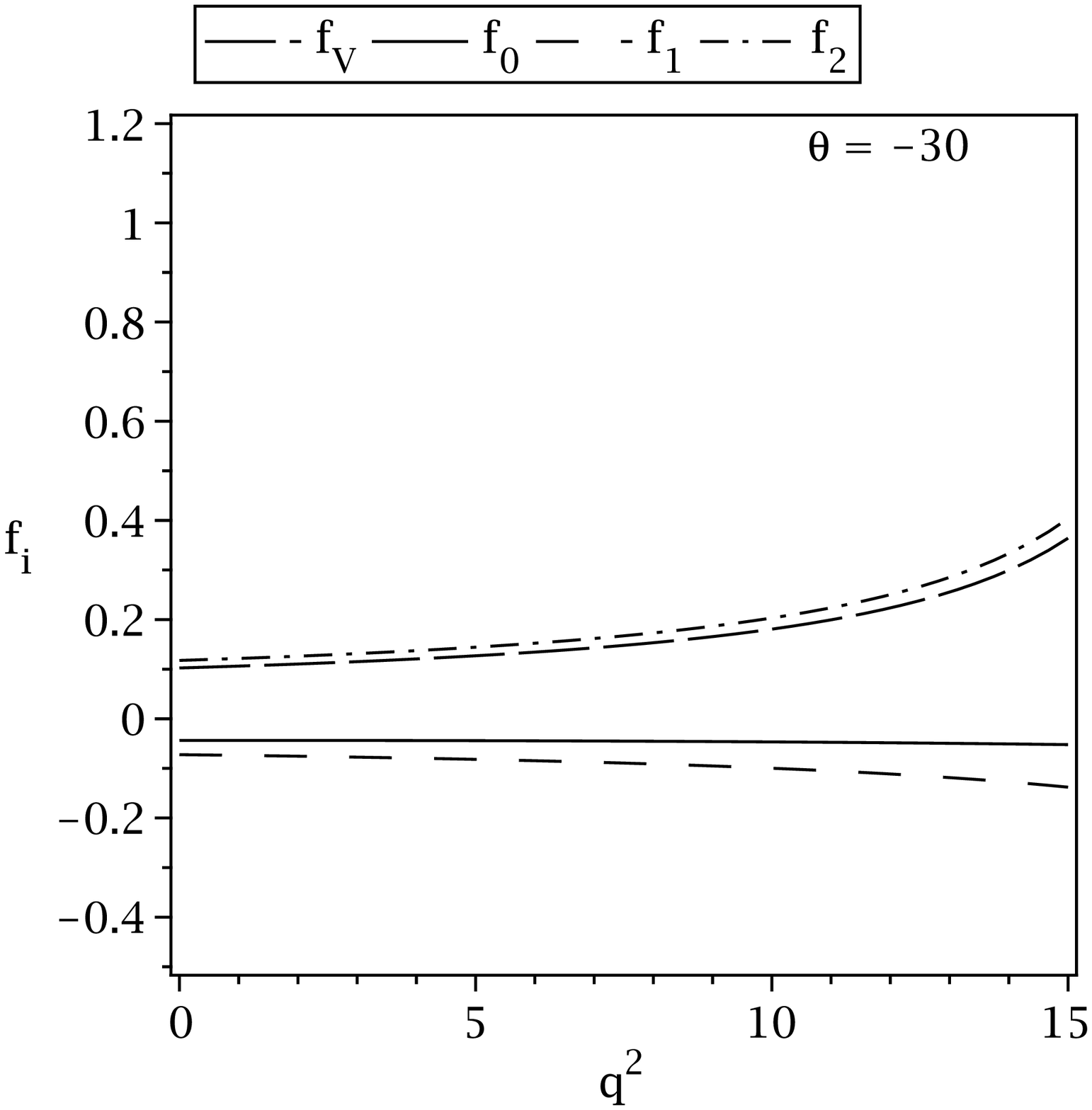} } \put(-77,-140){\epsfxsize=5cm
\epsfbox{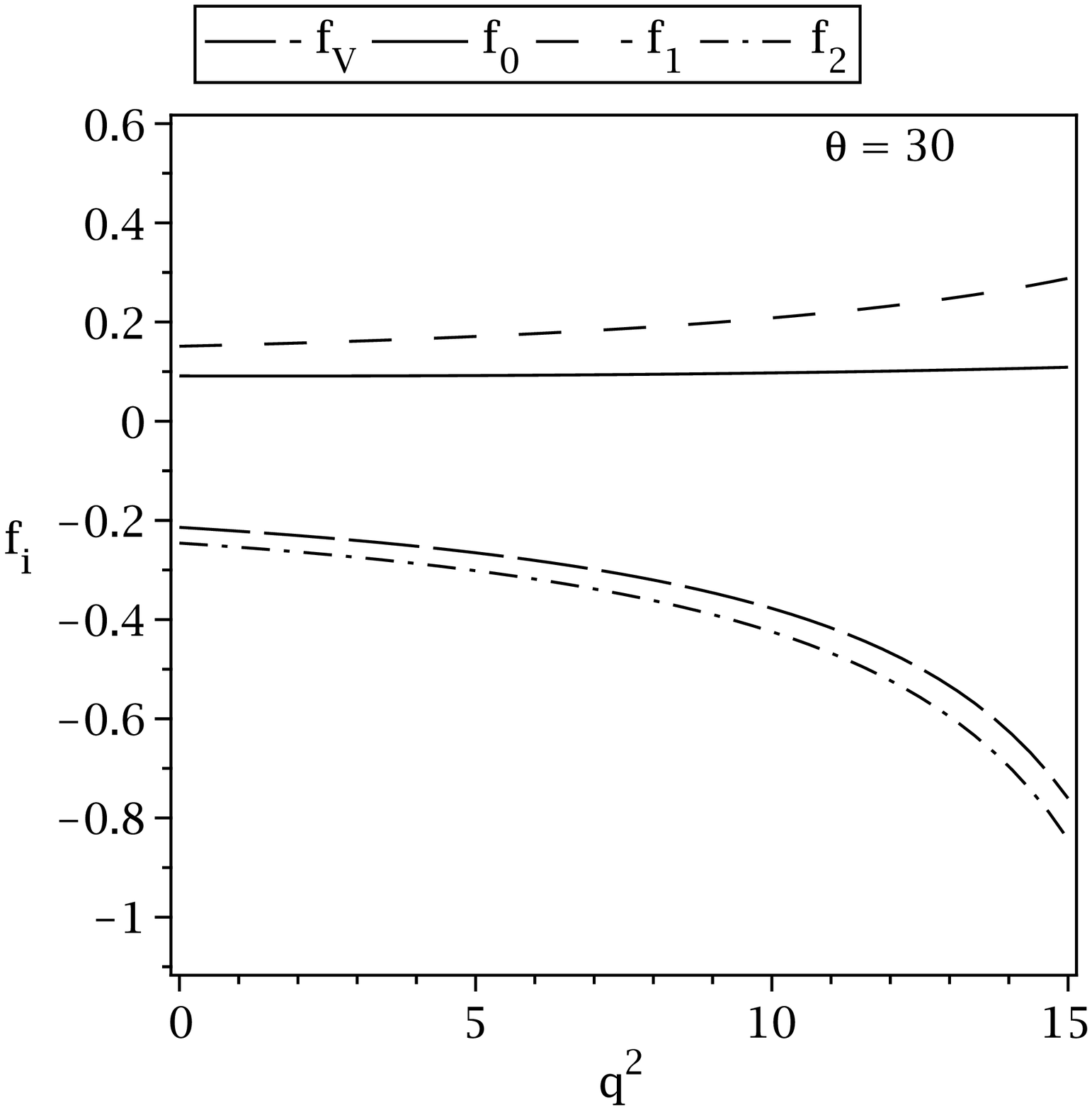} \epsfxsize=5cm \epsfbox{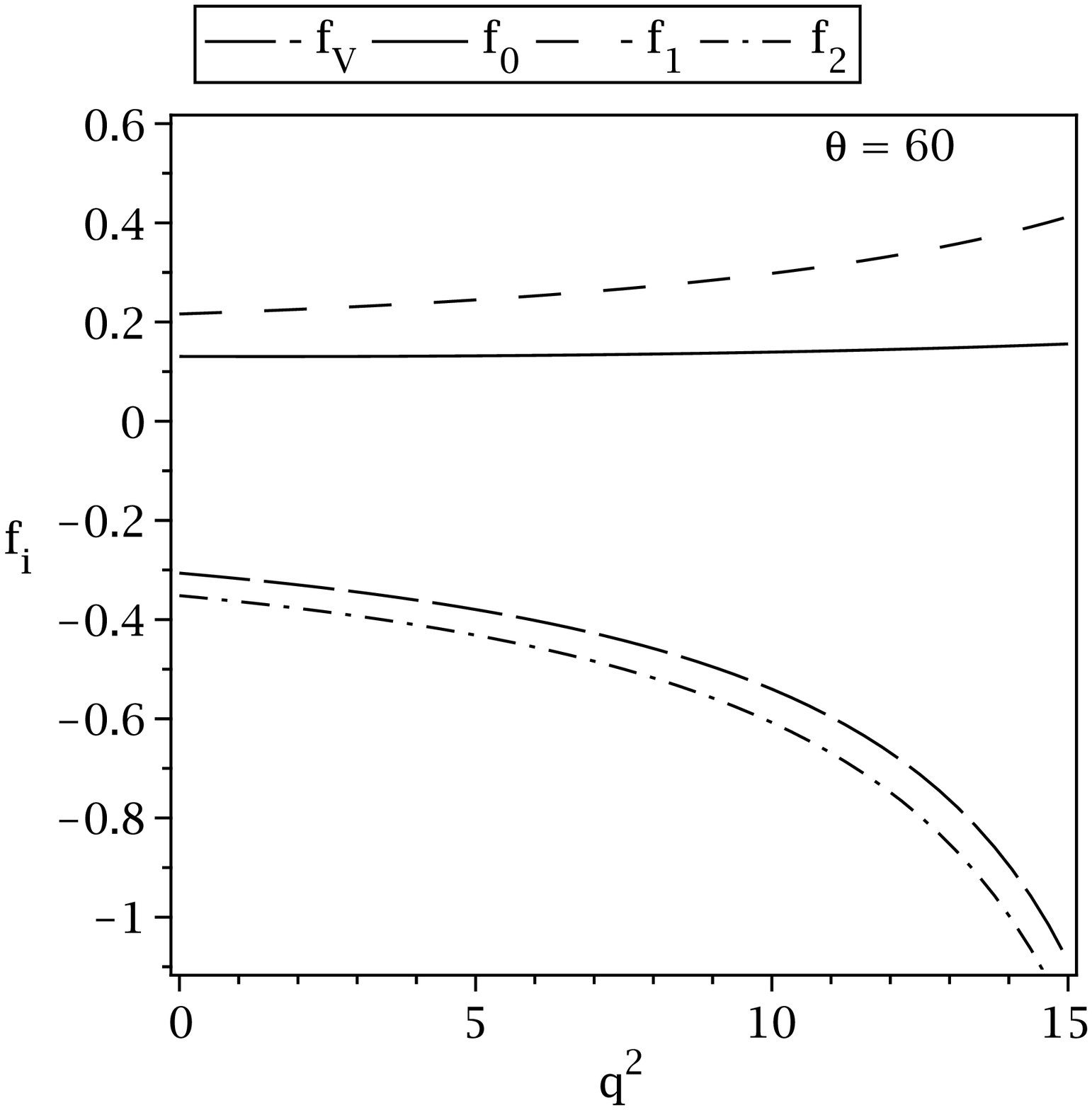}\epsfxsize=5cm \epsfbox{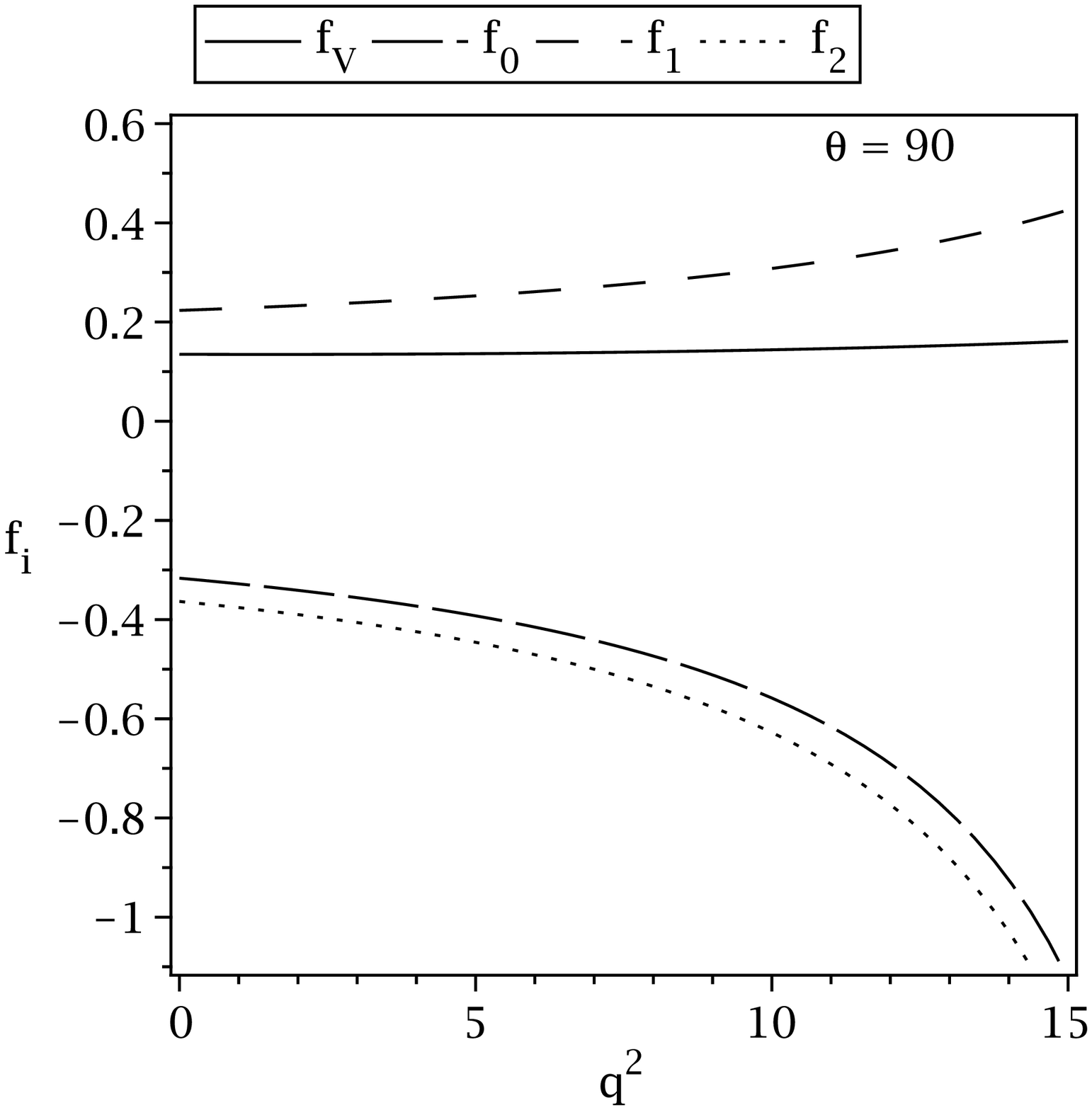} }
\end{picture}
\end{center}
\vspace*{13cm}\caption{The dependence of the transition form
factors on $q^2$ and $\theta=\pm N\pi/6,~ N= 1, 2, 3$ for the $B_c\to D_{1}^0(2420)$
transition.}\label{F7}
\end{figure}
\newpage
\begin{figure}[th]
\begin{center}
\begin{picture}(0,0)
\put(-78,-90){ \epsfxsize=5cm \epsfbox{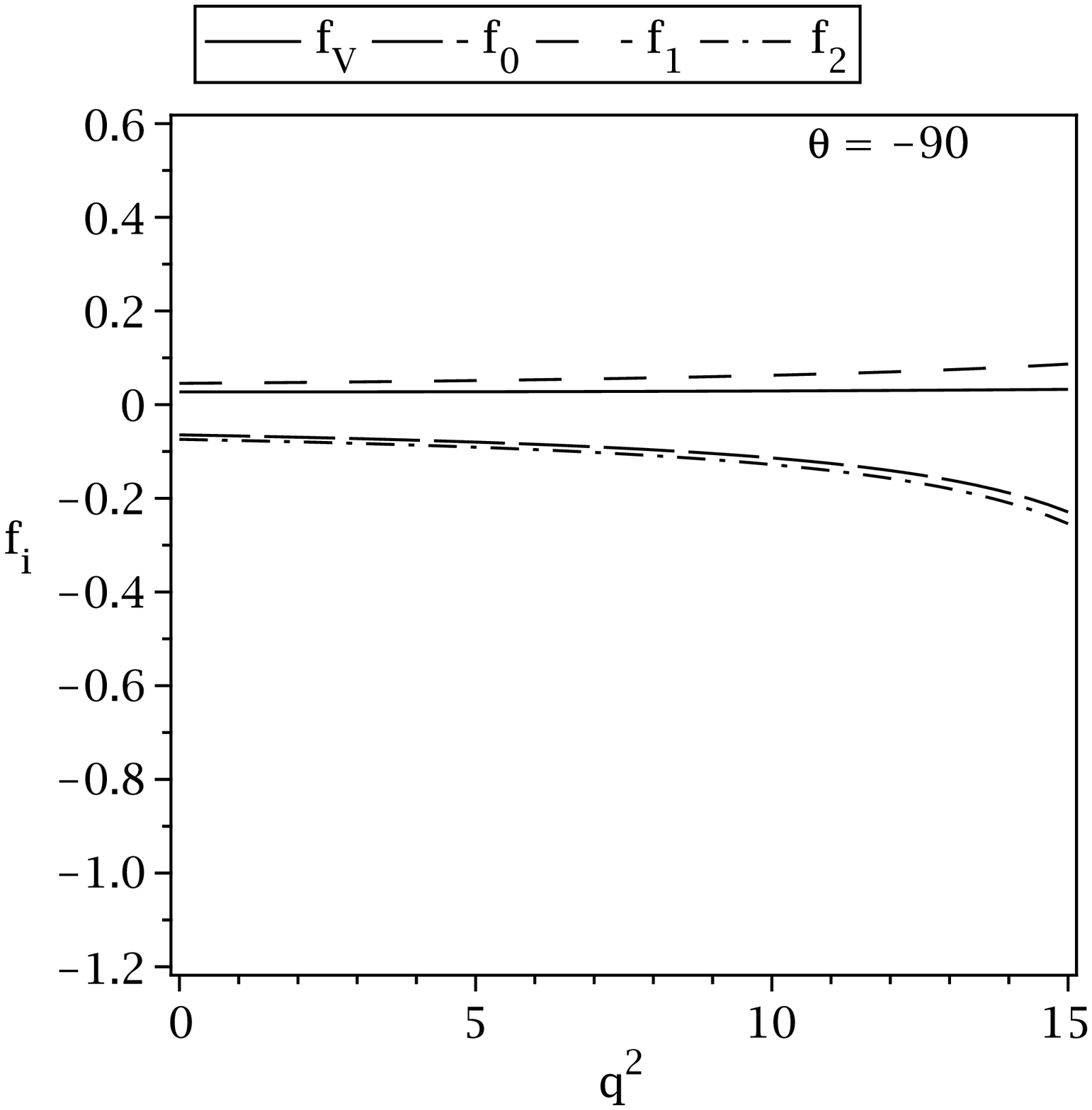} \epsfxsize=5cm
\epsfbox{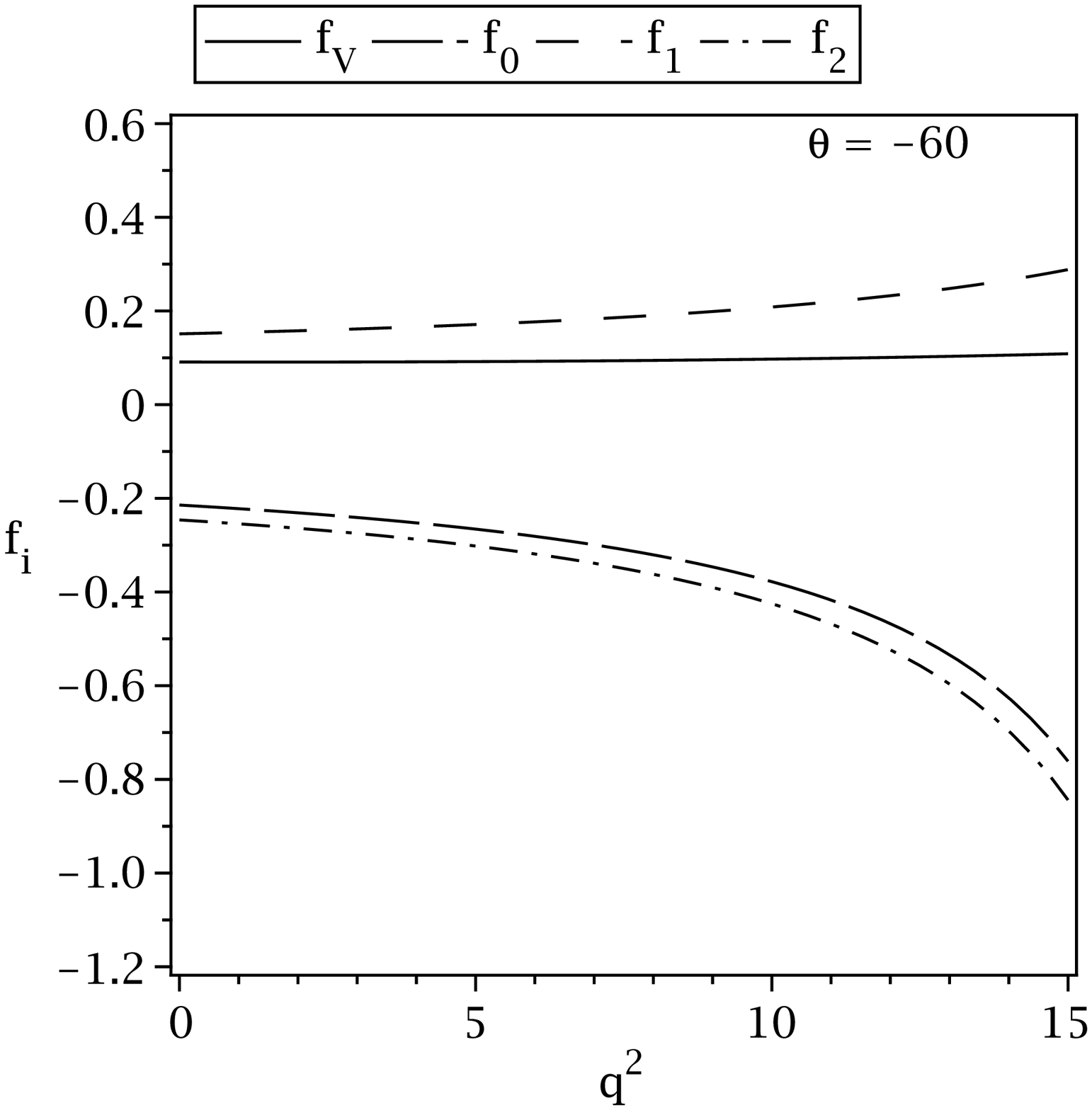}\epsfxsize=5cm \epsfbox{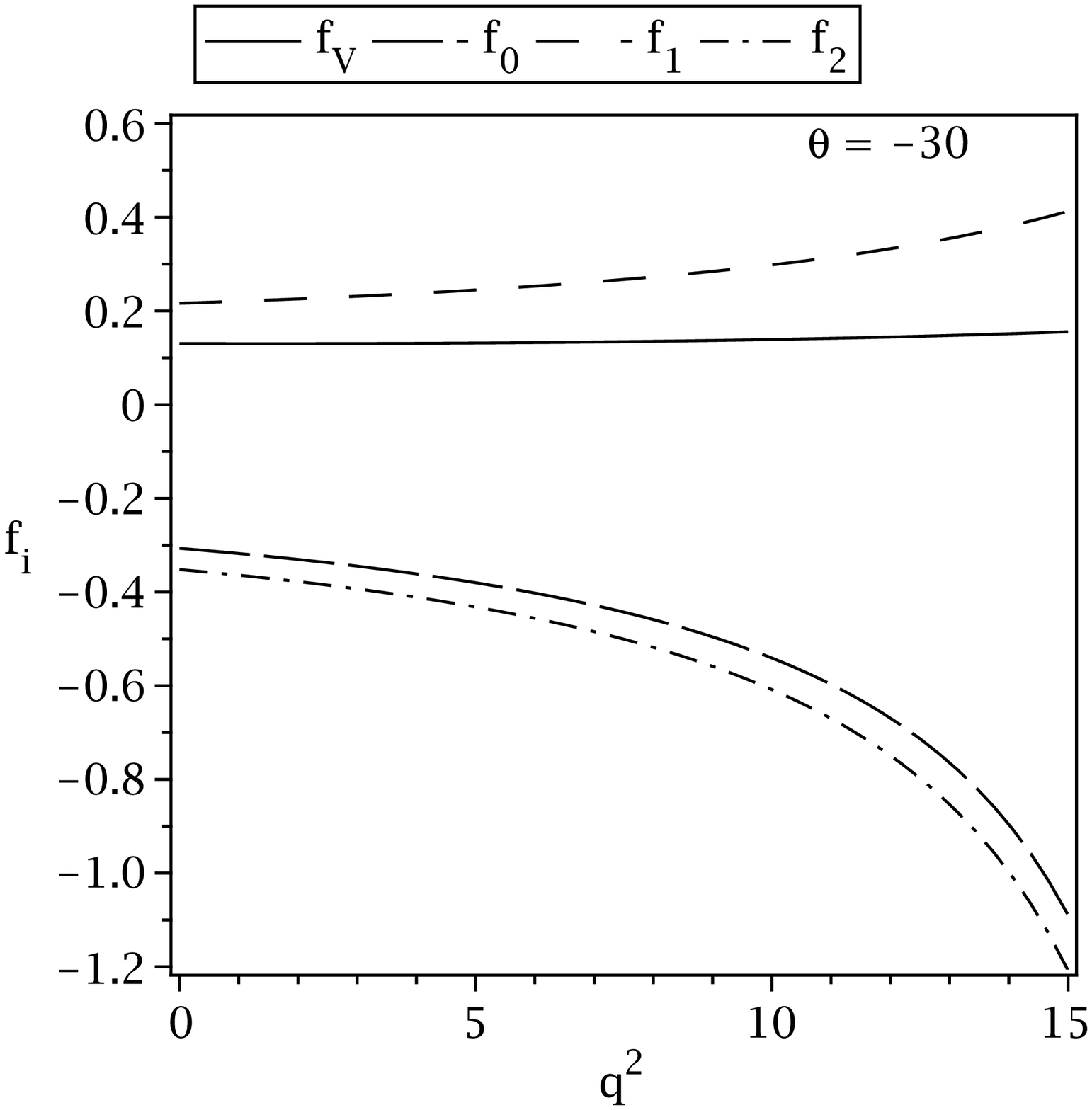} } \put(-77,-140){\epsfxsize=5cm
\epsfbox{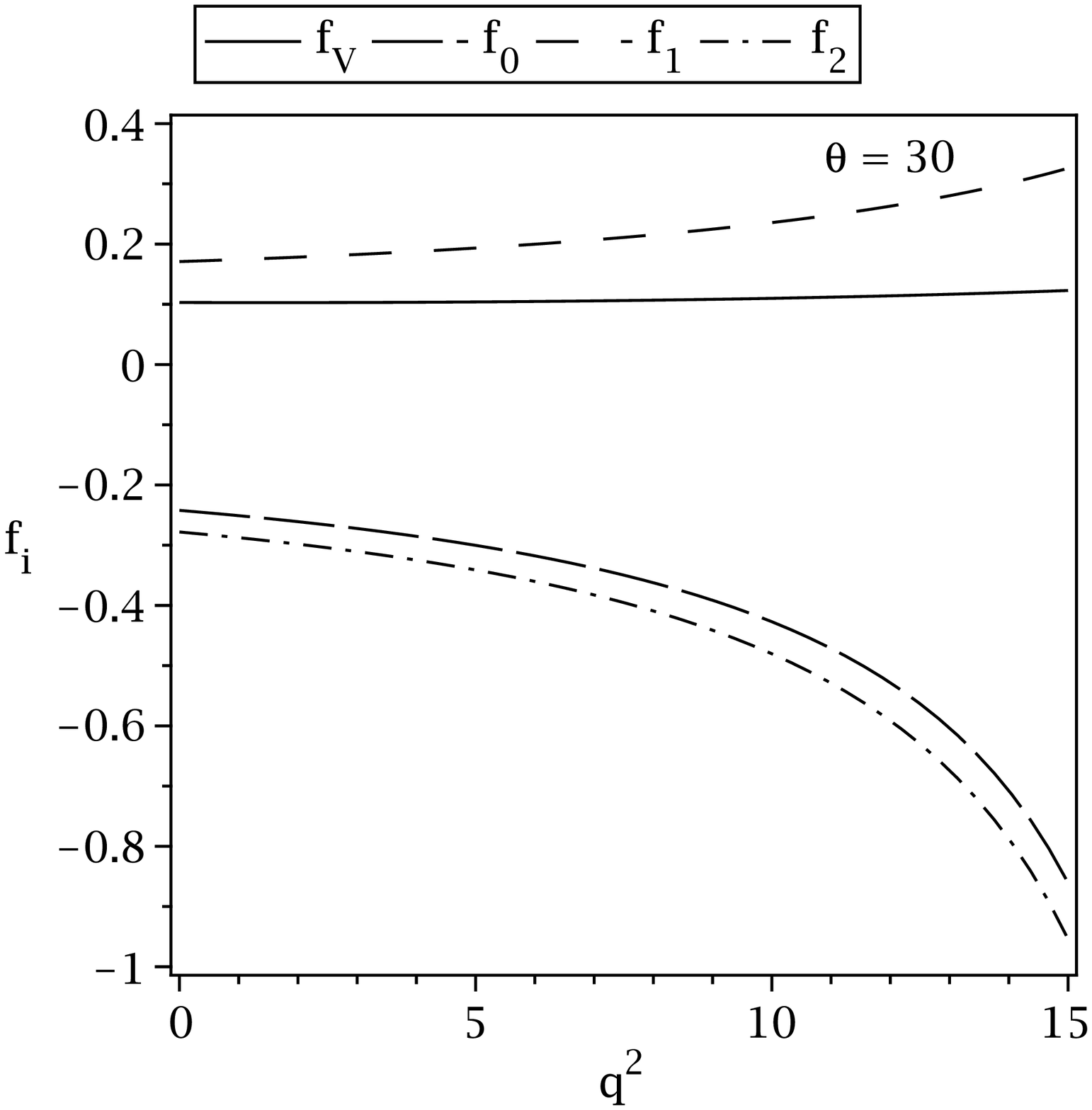} \epsfxsize=5cm \epsfbox{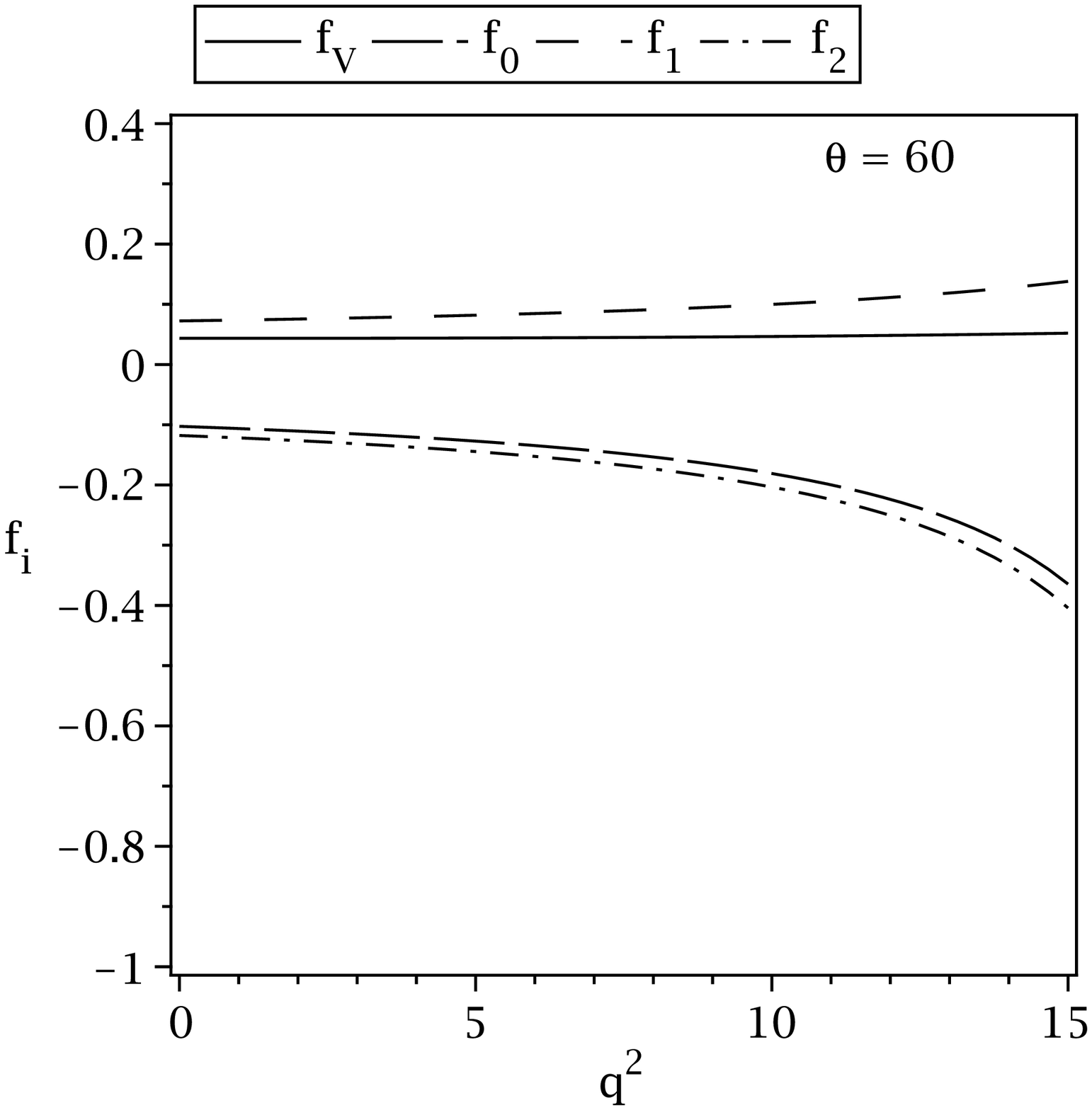}\epsfxsize=5cm \epsfbox{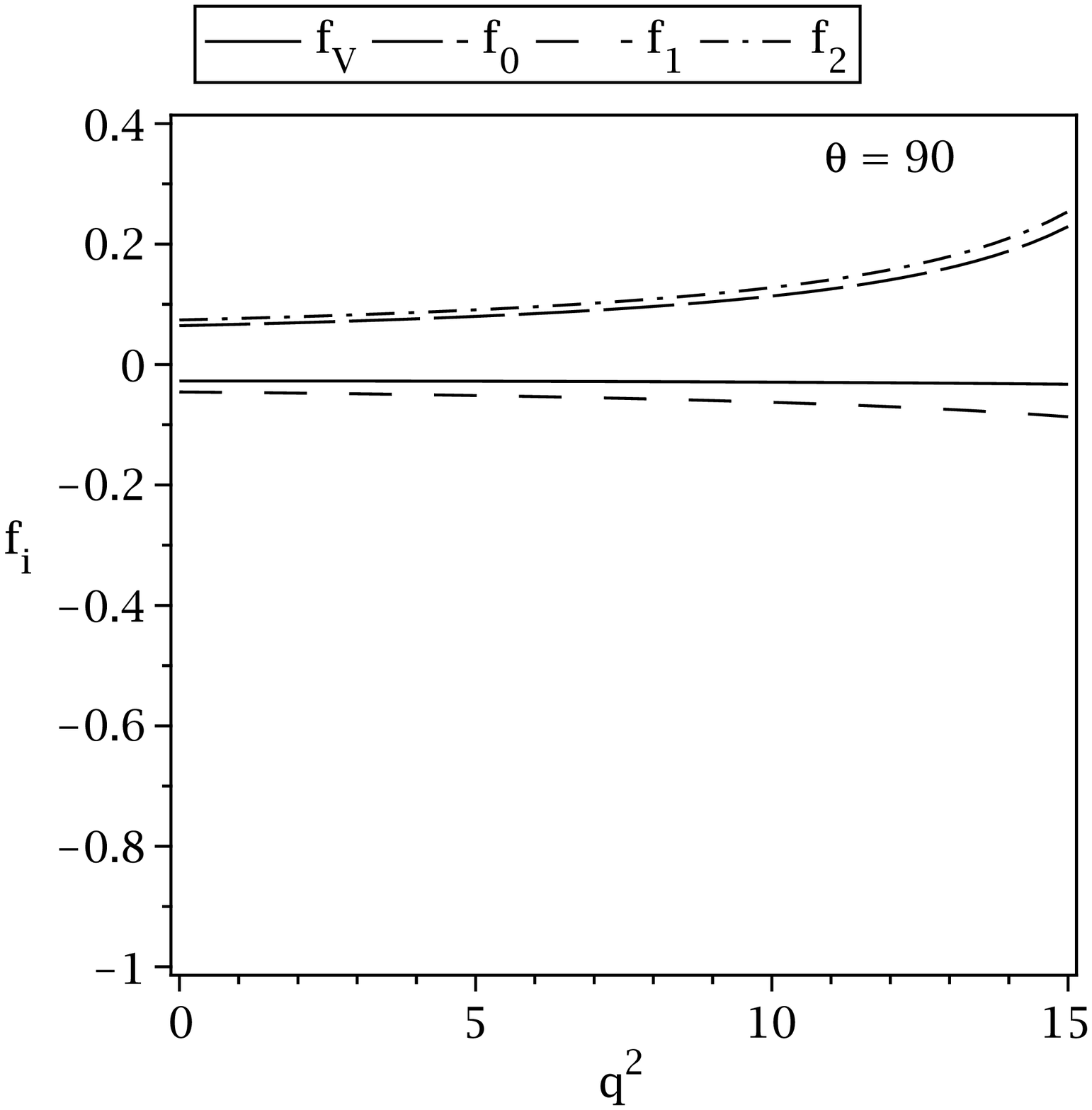} }
\end{picture}
\end{center}
\vspace*{13cm}\caption{The dependence of the transition form
factors on $q^2$ and $\theta=\pm N\pi/6,~ N= 1, 2, 3$ for the $B_c\to D_{1}^0(2430)$
transition.}\label{F8}
\end{figure}
\newpage
\newpage
\begin{figure}[th]
\begin{center}
\begin{picture}(160,100)
\put(10,20){ \epsfxsize=6cm \epsfbox{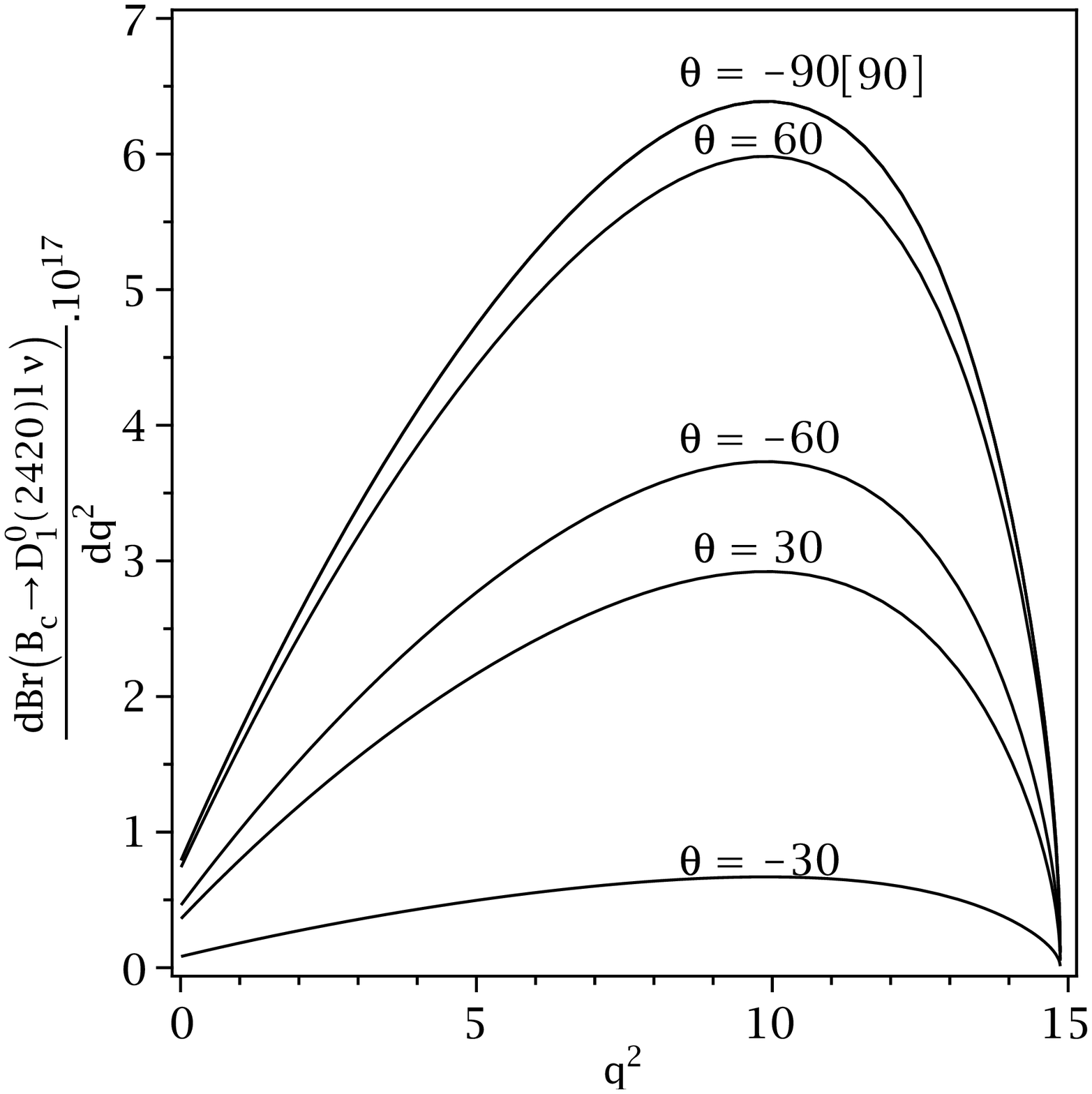}
\epsfxsize=6cm \epsfbox{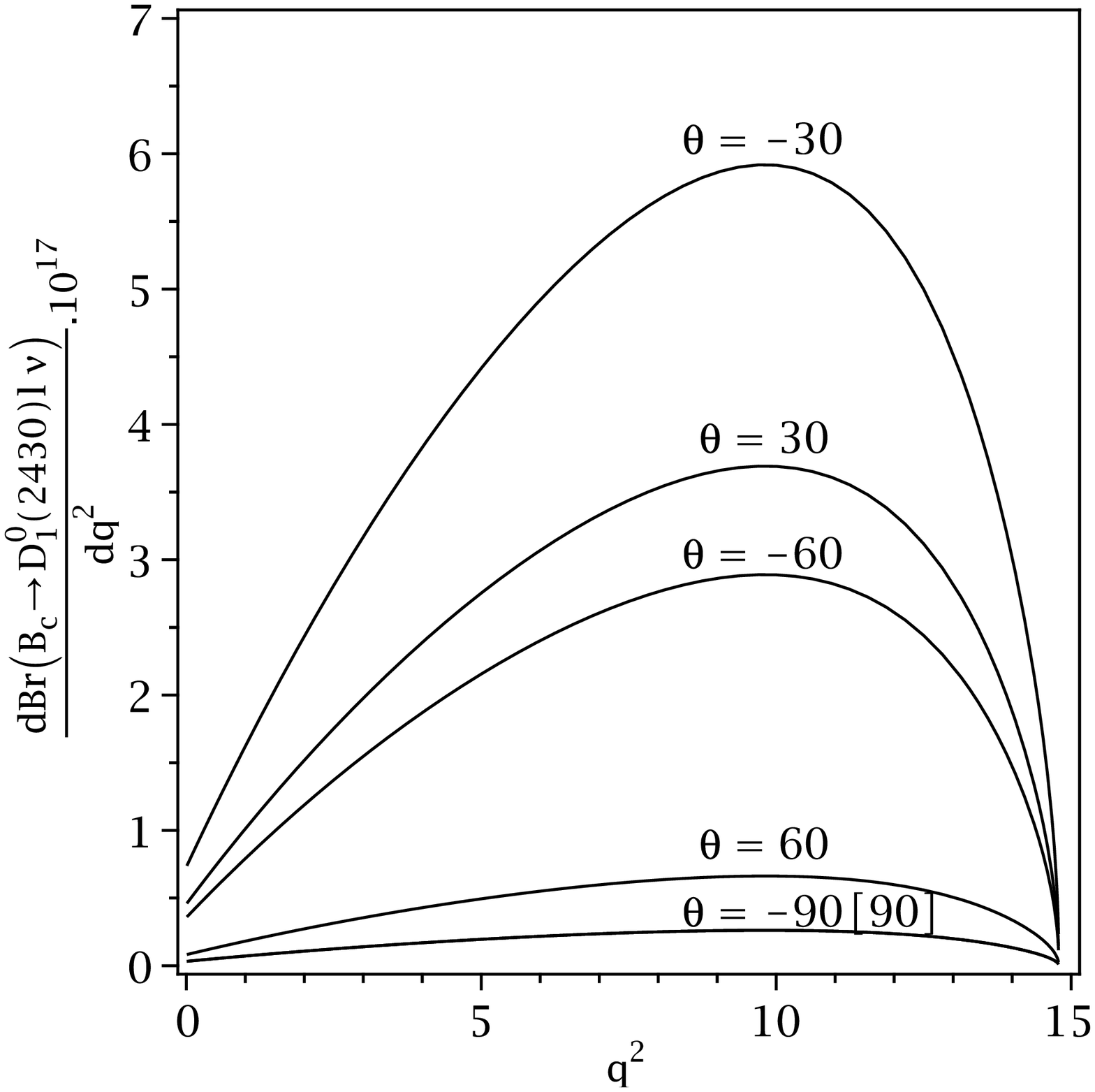}}
\end{picture}
\end{center}
\vspace*{-3cm}\caption{The decay width for $B_c\to D_{1}^0 l\nu$
with respect to $q^2$ and $\theta=\pm N\pi/6,~ N= 1, 2, 3$. }\label{F9}
\end{figure}
\normalsize
\newpage
\begin{figure}[th]
\begin{center}
\begin{picture}(160,100)
\put(-10,0){ \epsfxsize=8cm \epsfbox{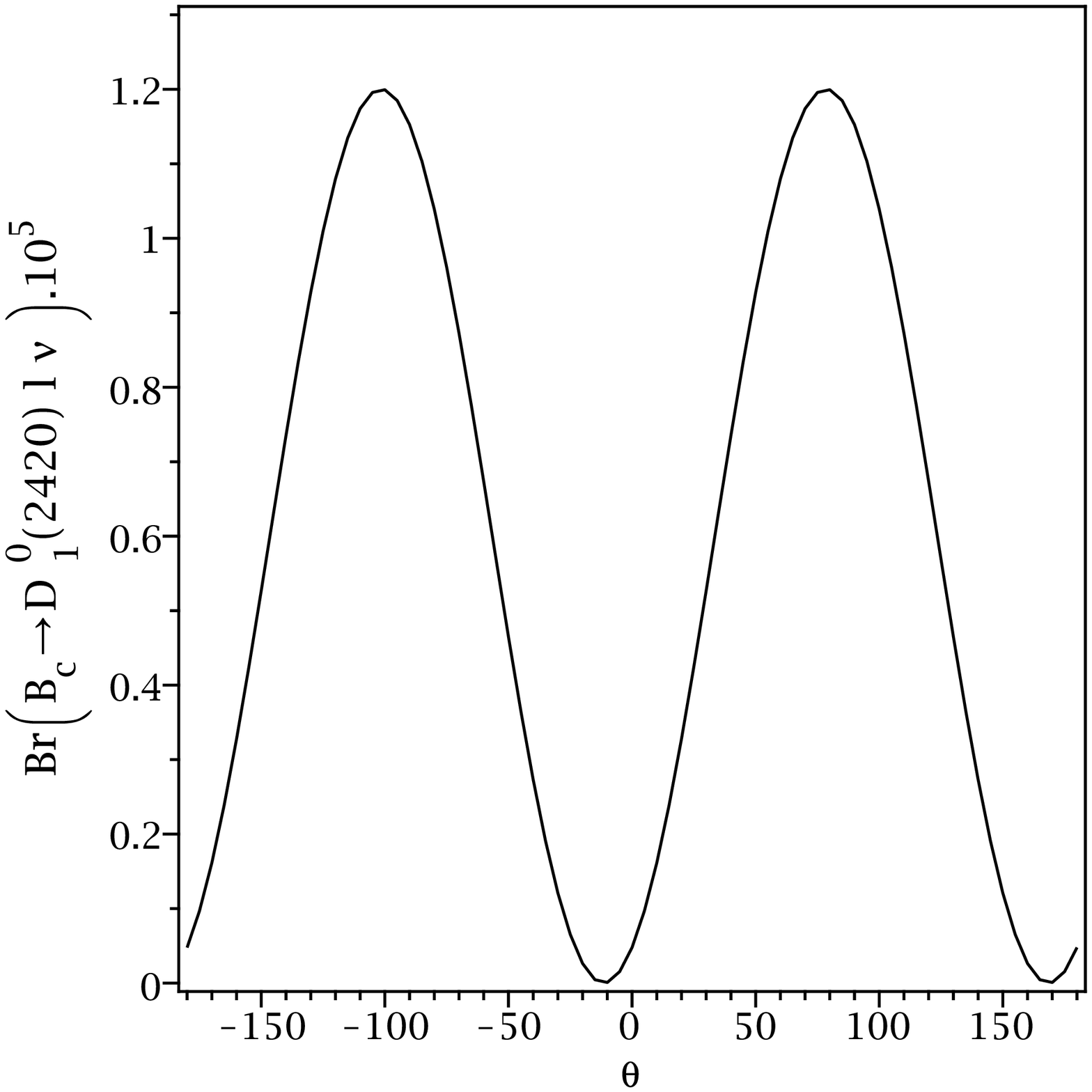} \epsfxsize=8cm
\epsfbox{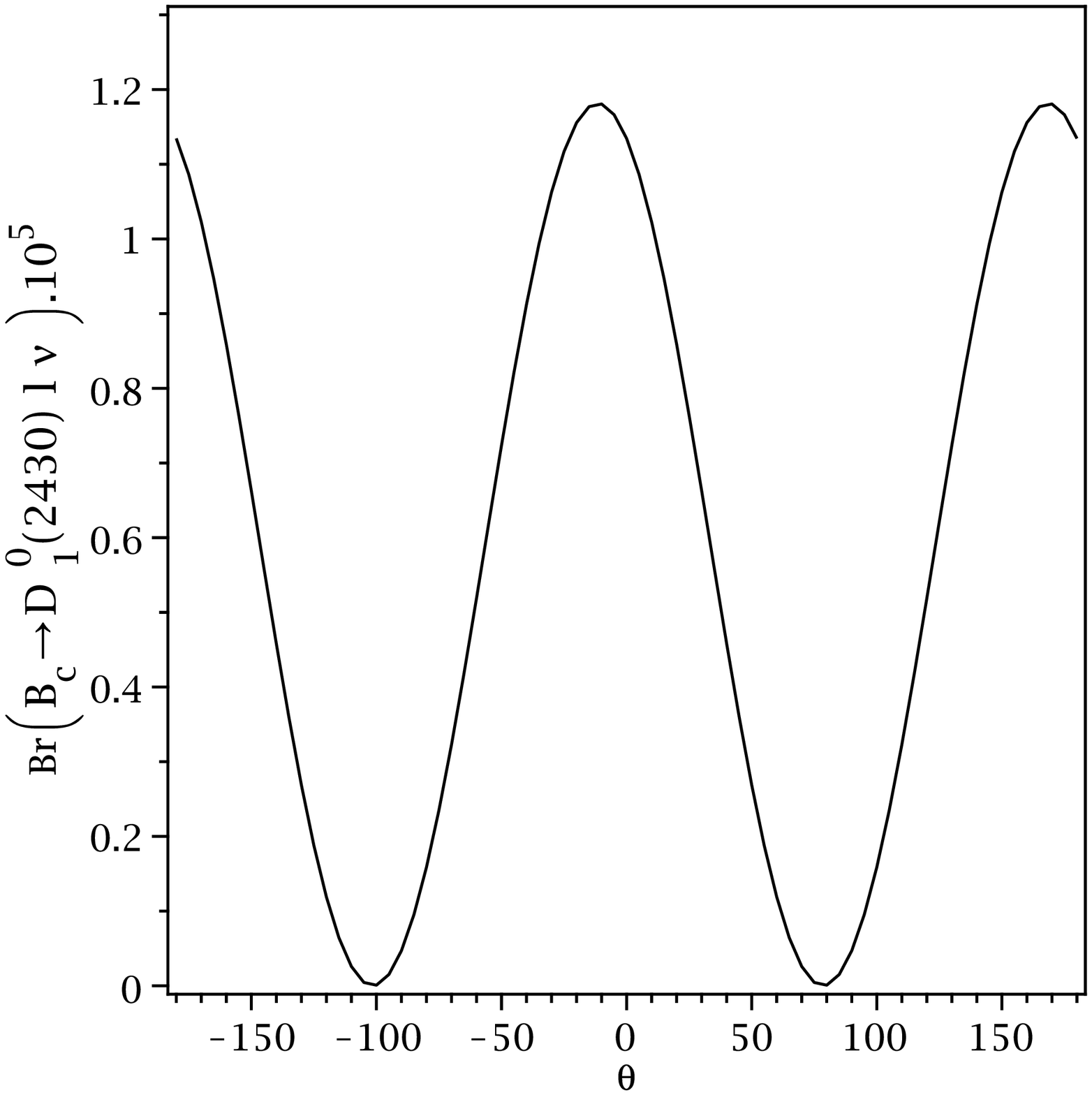}}
\end{picture}
\end{center}
\vspace*{0cm}\caption{The branching ratio functions  of the
$B_c\to D_{1}^0(2420[2430])$ with respect to $\theta$.}\label{F10}
\end{figure}
\normalsize
\newpage
\begin{figure}[th]
\begin{center}
\begin{picture}(0,0)
\put(-80,-90){ \epsfxsize=8cm \epsfbox{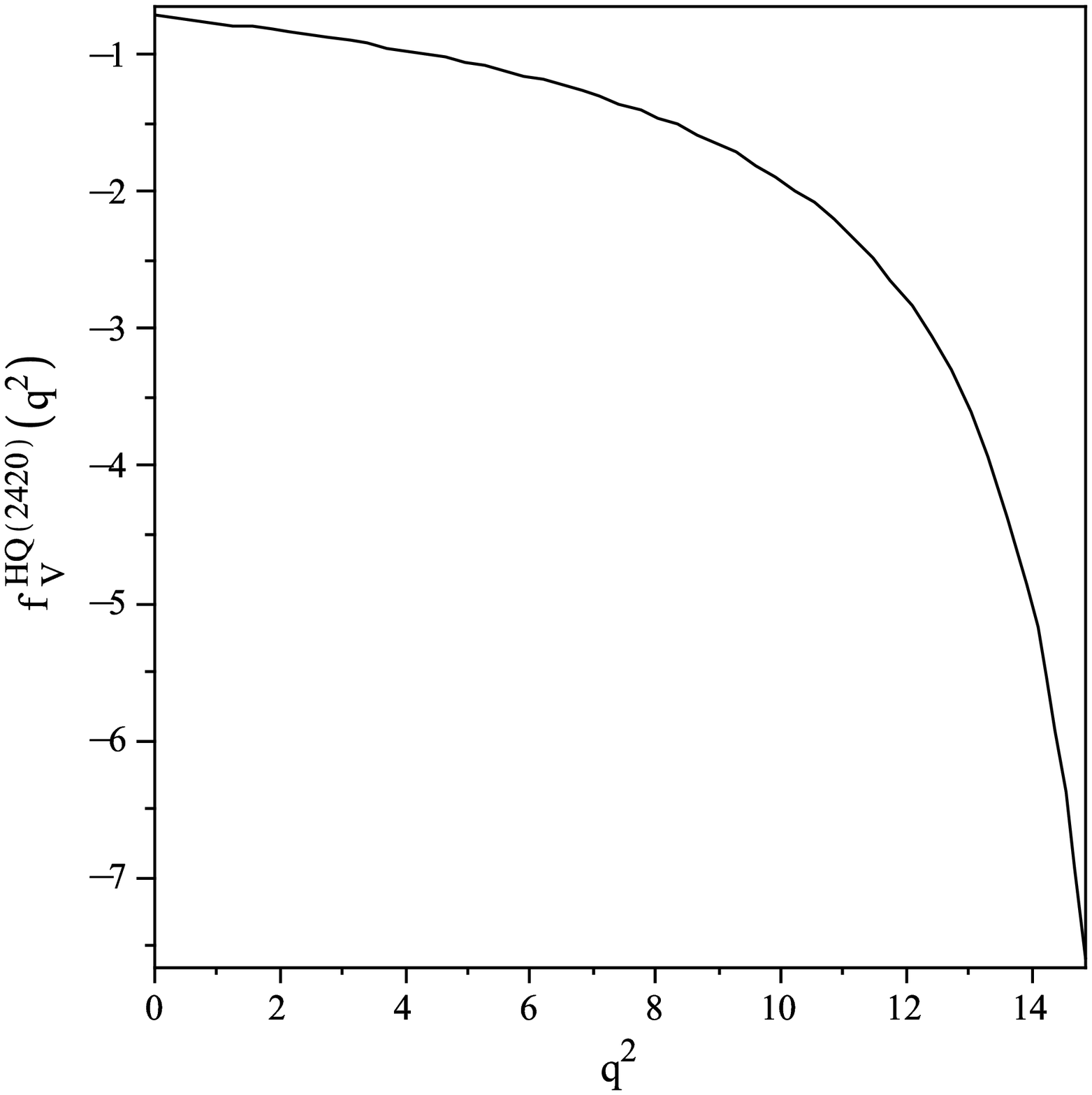} \epsfxsize=8cm
\epsfbox{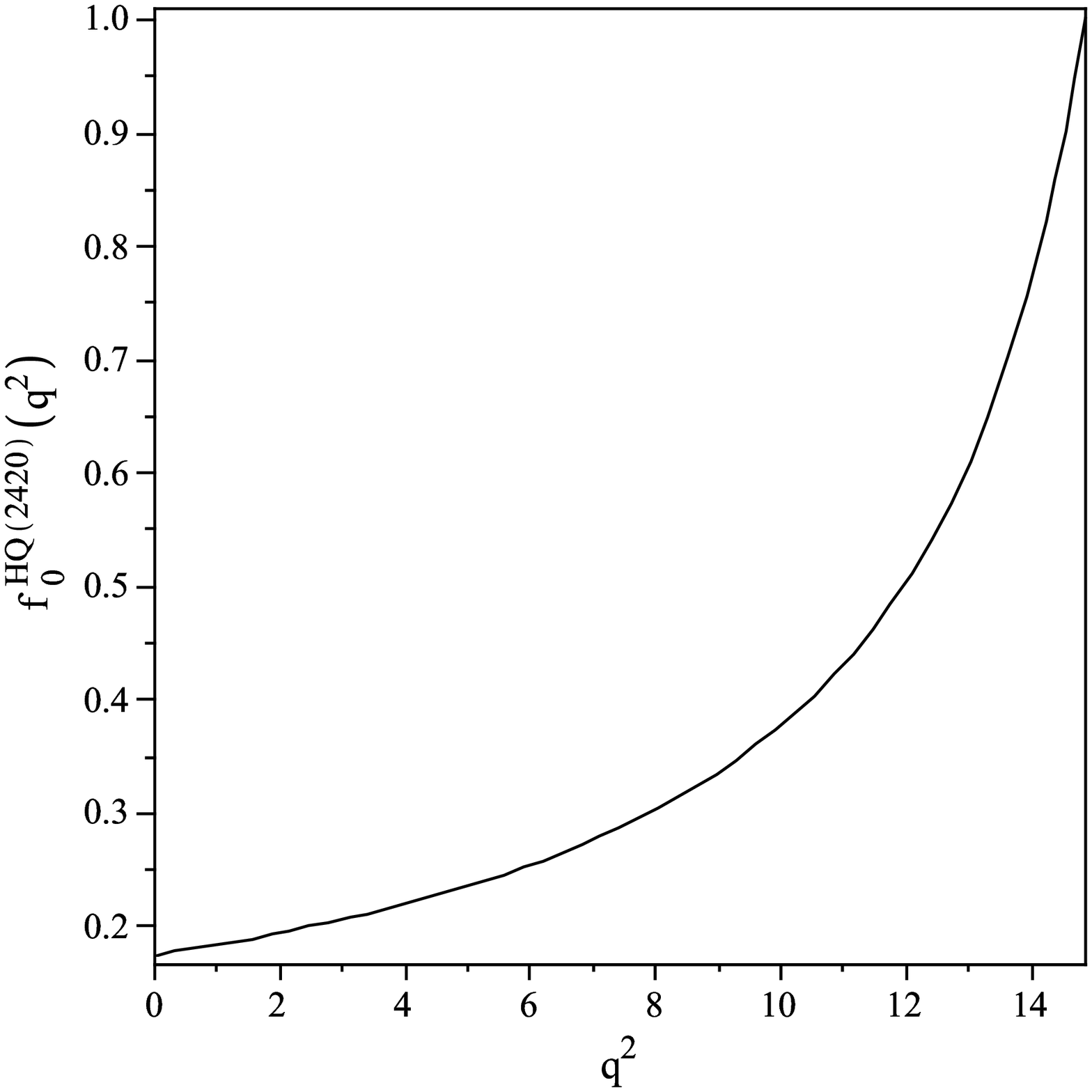} } \put(-80,-175){\epsfxsize=8cm
\epsfbox{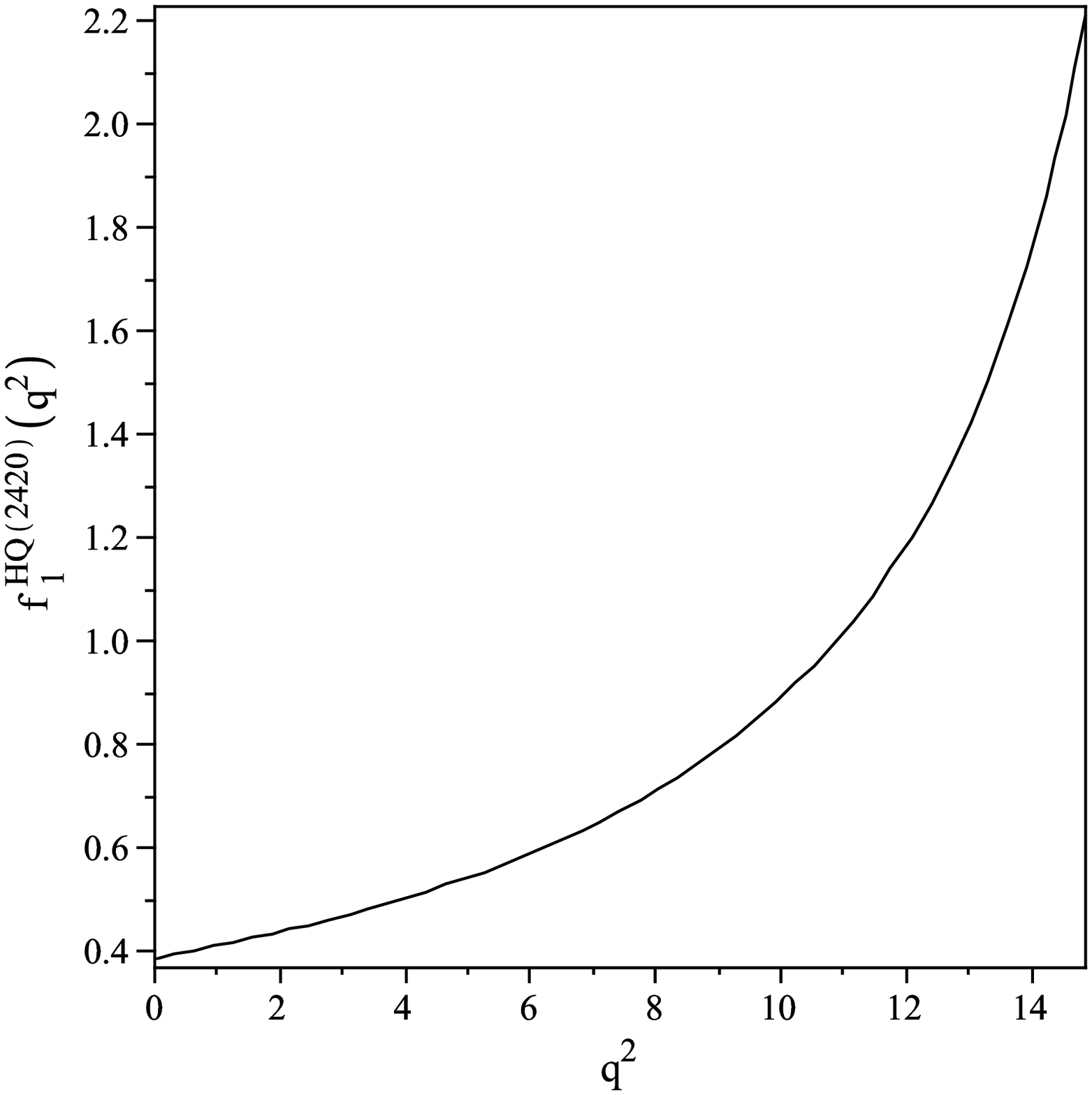} \epsfxsize=8cm \epsfbox{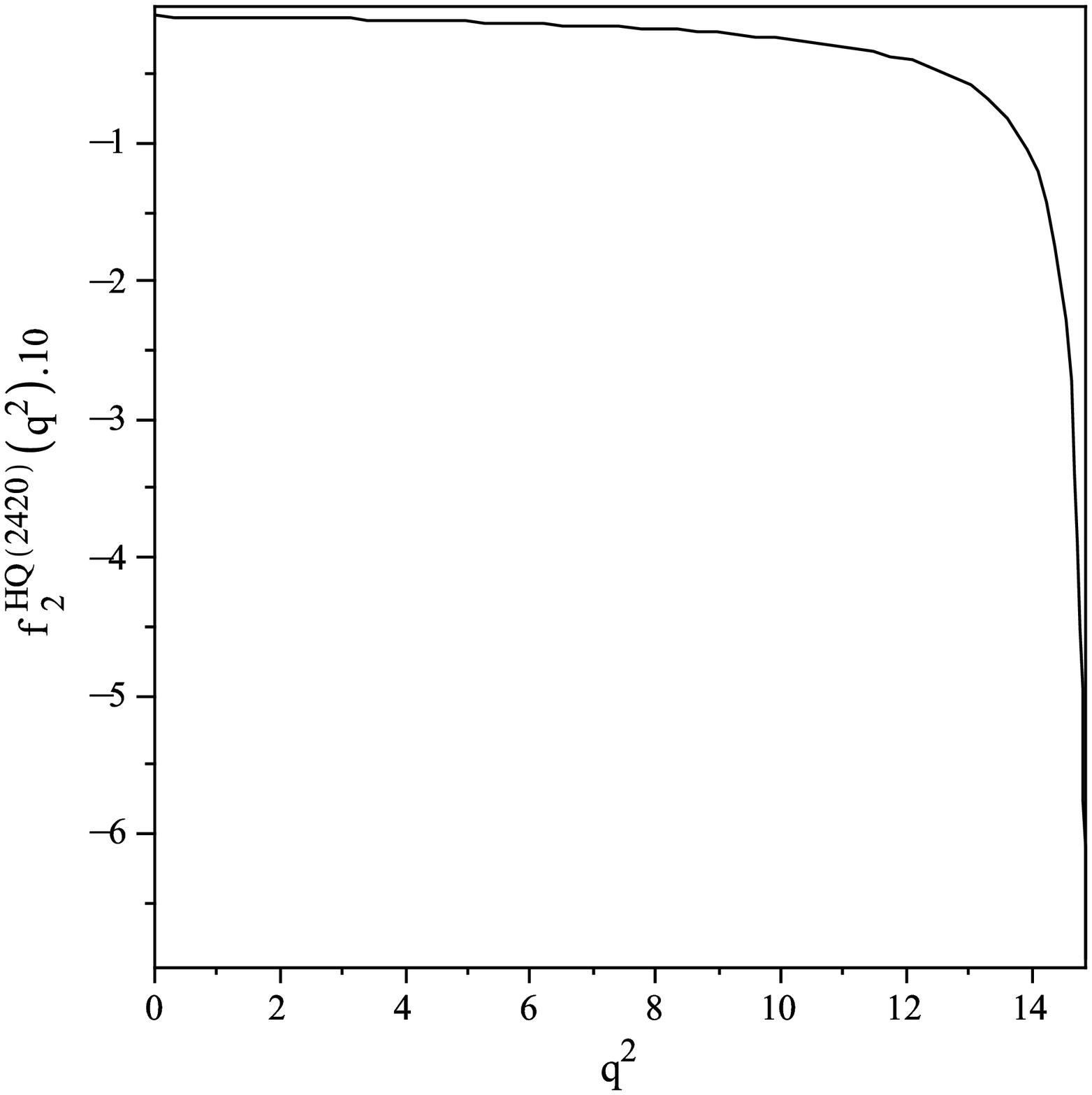} }
\end{picture}
\end{center}
\vspace*{17cm}\caption{The dependence of the HQET form factors on
$q^2$ for the $B_c\to D_{1}^0(2420)$ transition.}\label{F11}
\end{figure}
\newpage
\begin{figure}[th]
\begin{center}
\begin{picture}(0,0)
\put(-80,-90){ \epsfxsize=8cm \epsfbox{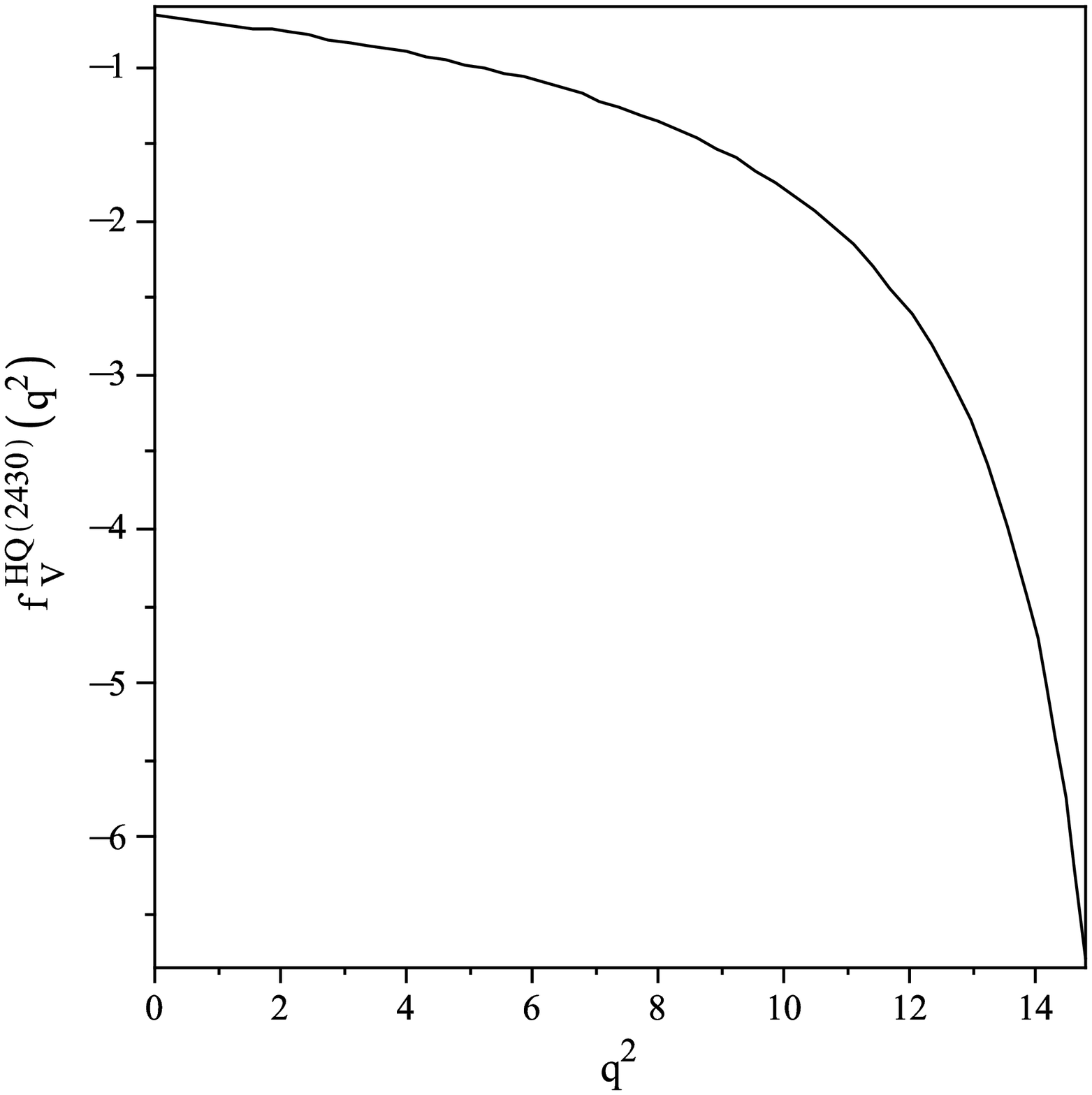} \epsfxsize=8cm
\epsfbox{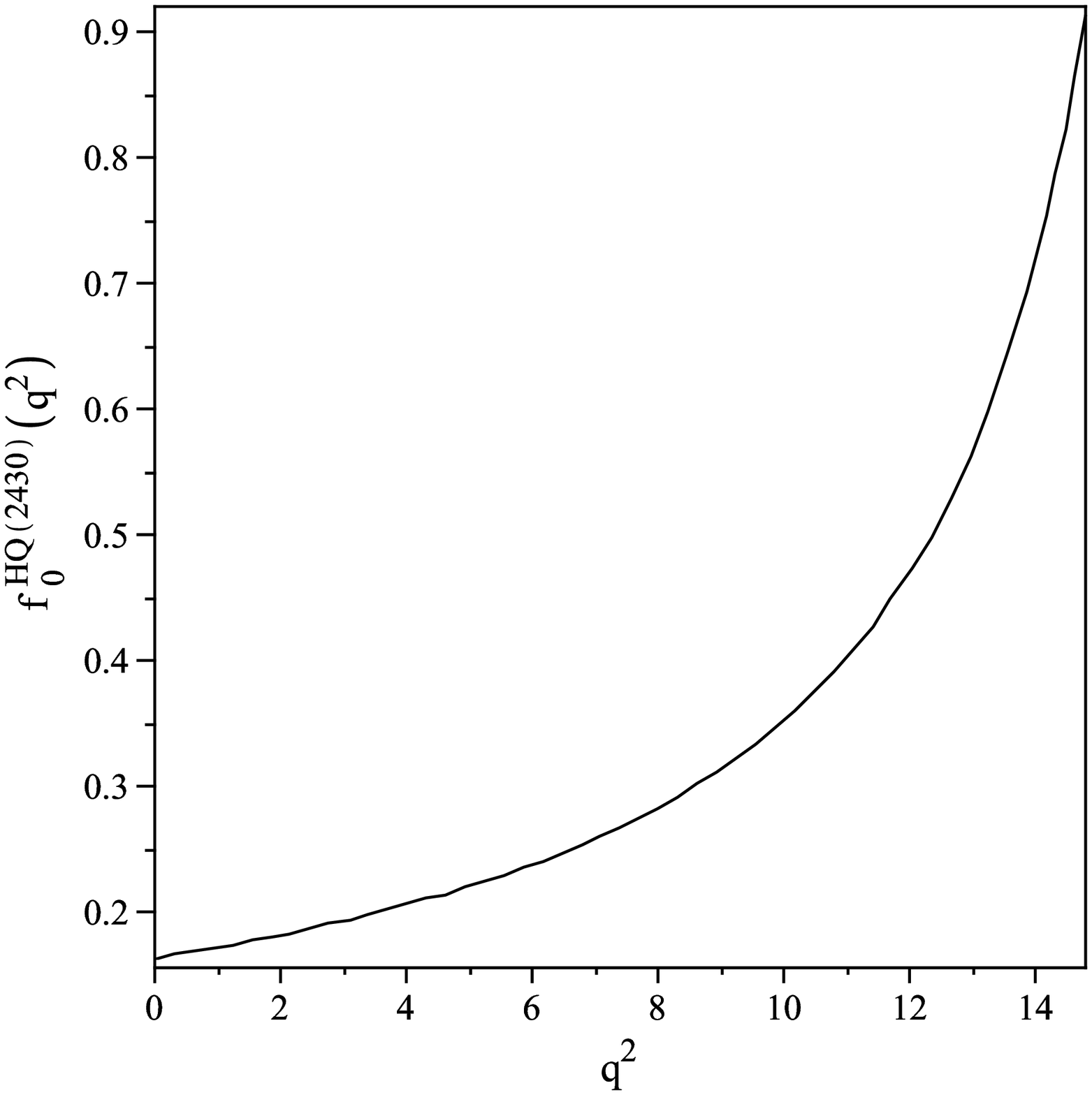} } \put(-80,-175){\epsfxsize=8cm
\epsfbox{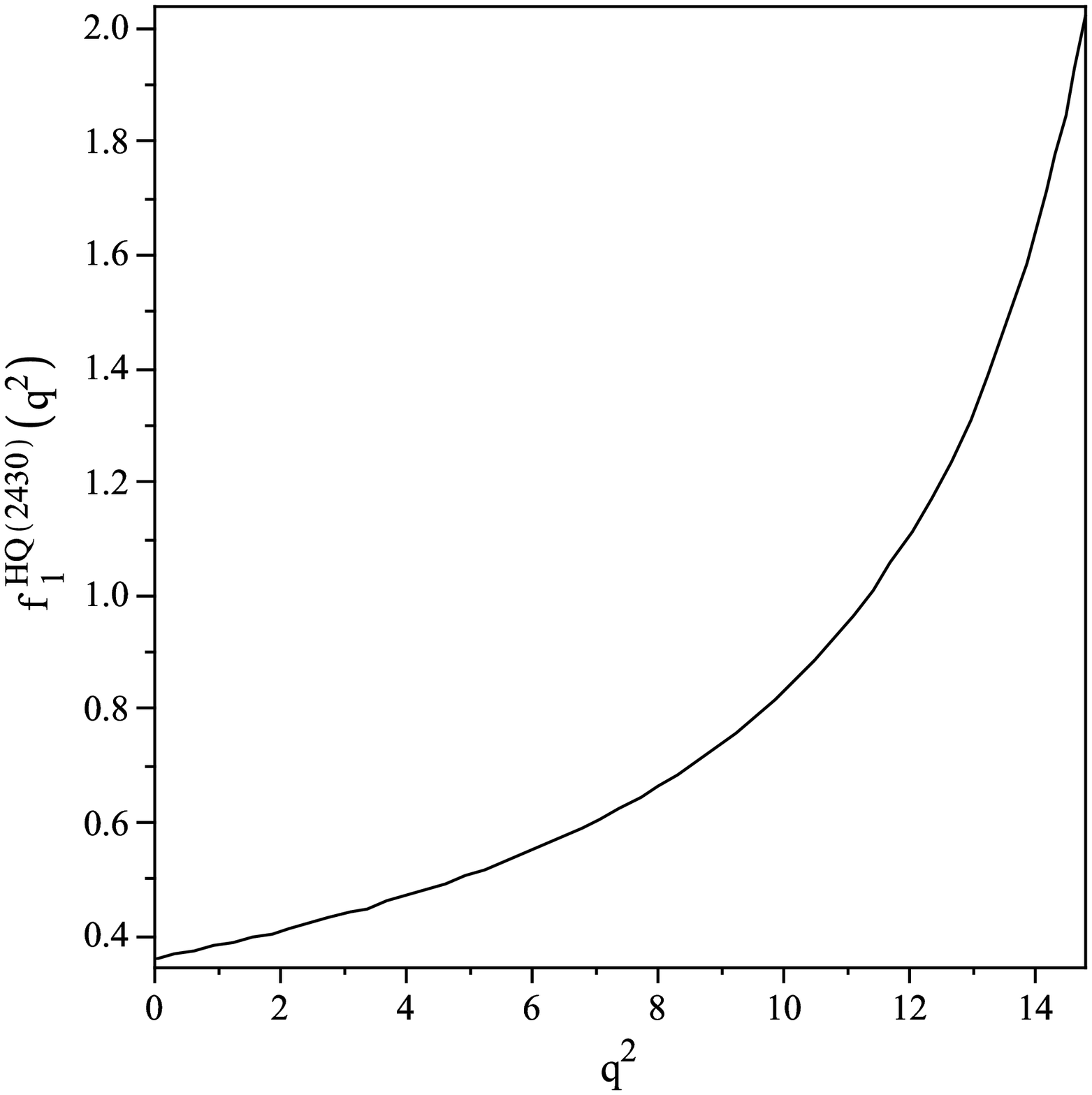} \epsfxsize=8cm \epsfbox{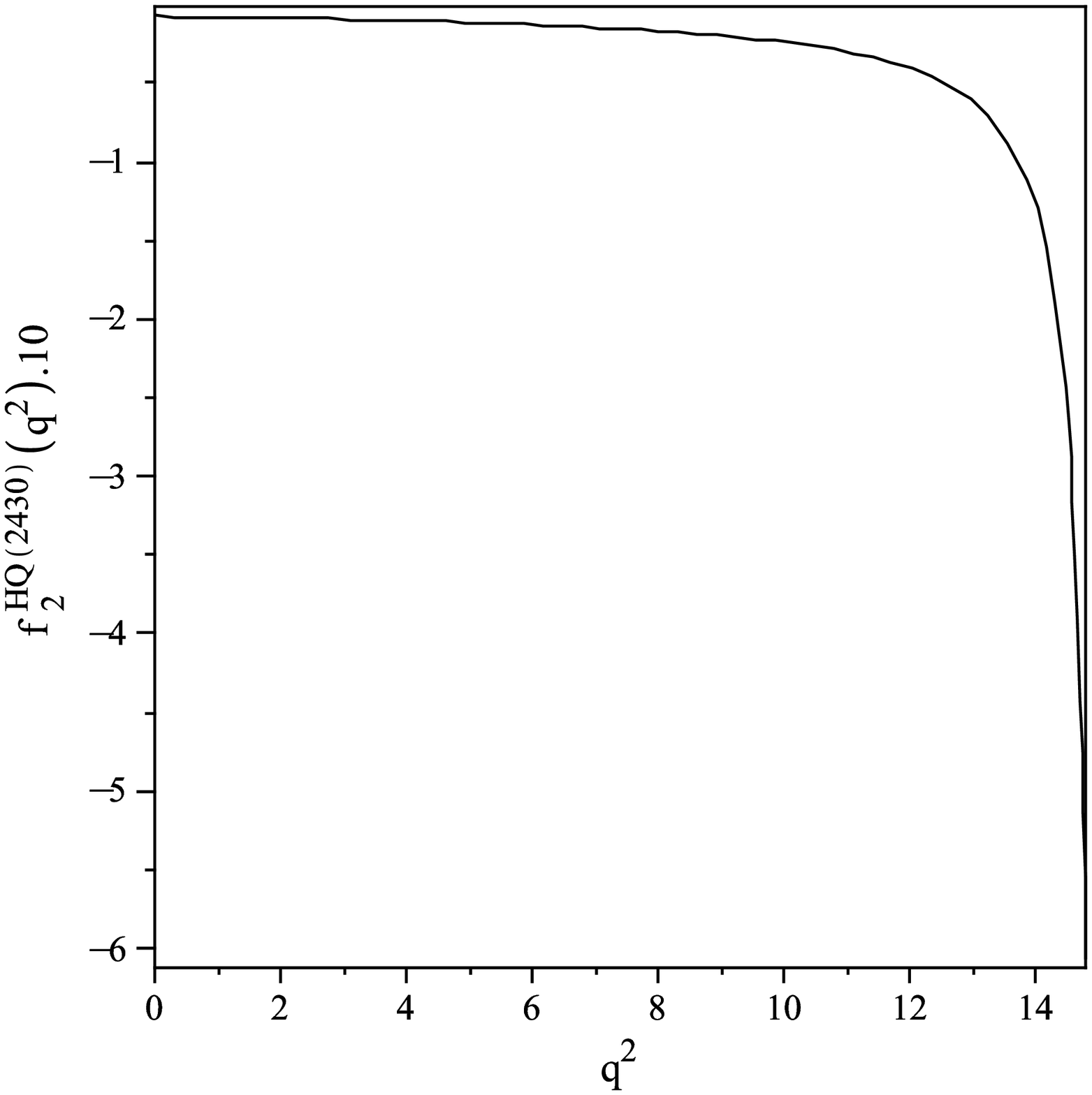} }
\end{picture}
\end{center}
\vspace*{17cm}\caption{The dependence of the HQET form factors on
$q^2$ for the $B_c\to D_{1}^0(2430)$ transition.}\label{F12}
\end{figure}
\newpage
\begin{figure}[th]
\begin{center}
\begin{picture}(160,100)
\put(-10,0){ \epsfxsize=8cm \epsfbox{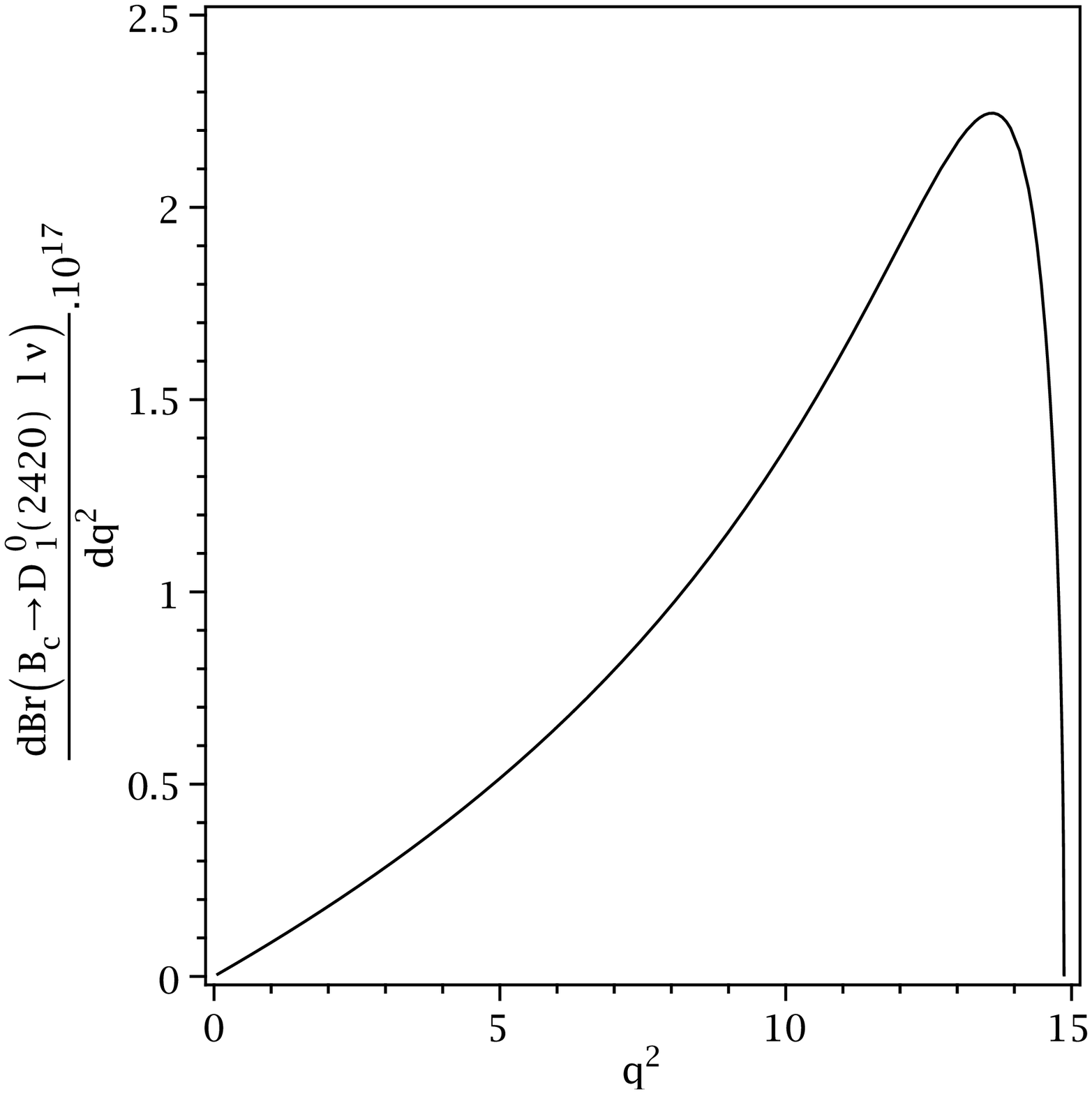}
\epsfxsize=8cm \epsfbox{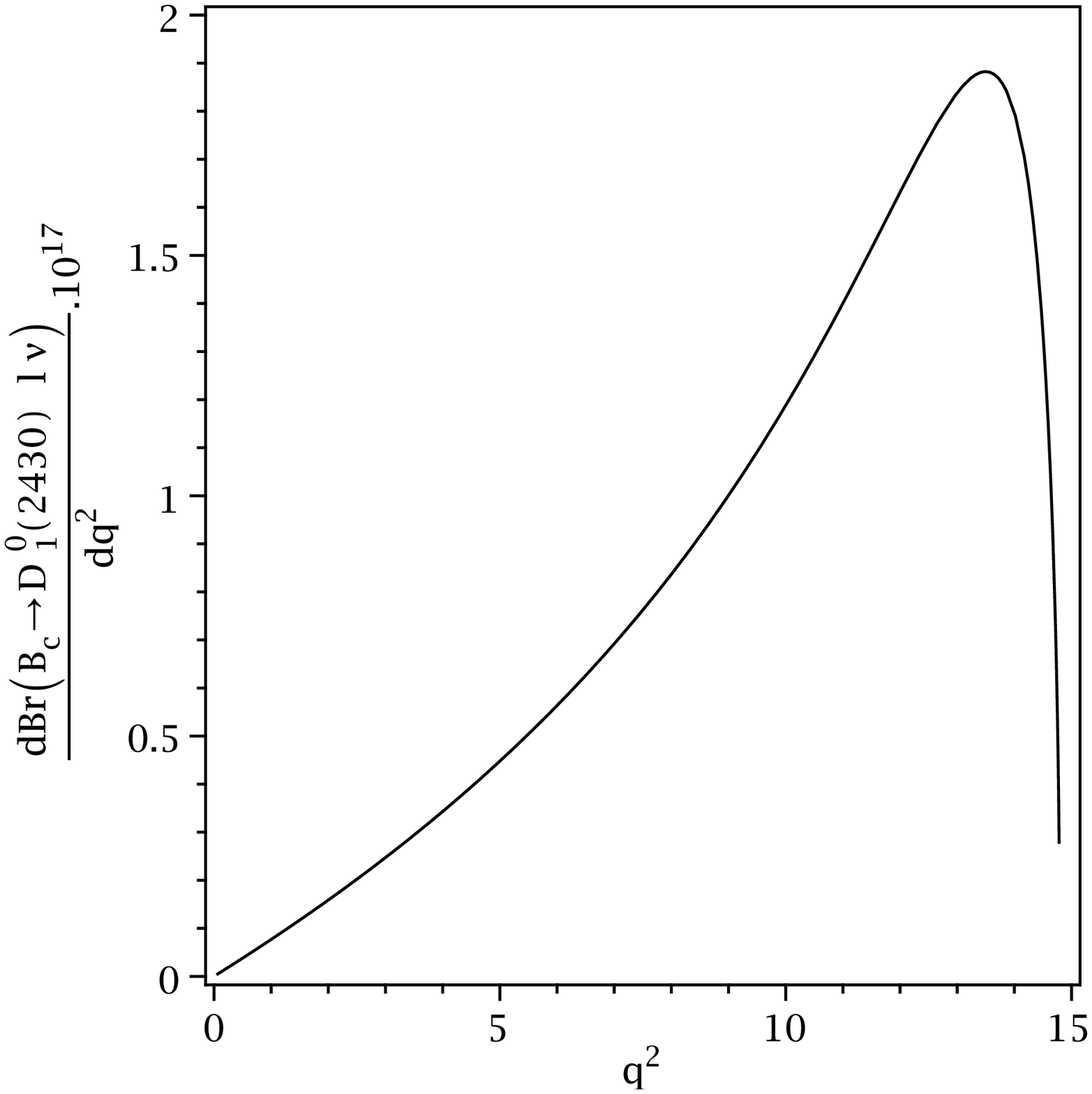} }
\end{picture}
\end{center}
\vspace*{0cm}\caption{The decay widths of the $B_c\to
D_1^0(2420[2430])$ decays in HQET approach with respect to
$q^2$.}\label{F13}
\end{figure}

\end{document}